\documentclass[%prl,
aps,
preprint %tightenlines
]{revtex4}

\usepackage[usenames]{color}
\usepackage{ulem}

\newcommand{\bea}{\begin{eqnarray}}

\newcommand{\beq}{\begin{equation}}

\newcommand{\bef}{\begin{figure}[btp]}

\newcommand{\eeq}{\end{equation}}

\newcommand{\eea}{\end{eqnarray}}

\newcommand{\eef}{\end{figure}}

\newcommand{\ket}[1]{| #1 \rangle}

\newcommand{\lket}[1]{| {\bf #1} \rangle}

\newcommand{\nnum}{\nonumber \\}

\newcommand{\op}[1]{{\hat{#1}}}

\usepackage{amssymb}
\usepackage{amsmath}
\usepackage{amsfonts}
\usepackage{graphicx}

\begin{document}
\def\ra{\rangle}
\def\la{\langle}
\def\bege{\begin{equation}}
\def\ende{\end{equation}}
\def\begarr{\begin{eqnarray}}
\def\endarr{\end{eqnarray}}
\def\ha{{\hat a}}
\def\hb{{\hat b}}
\def\hu{{\hat u}}
\def\hv{{\hat v}}
\def\hc{{\hat c}}
\def\hd{{\hat d}}
\def\no{\noindent}\def\non{\nonumber}
\def\hi{\hangindent=45pt}
\def\v{\vskip 12pt}

%\draft
\title{Optical Quantum Computation}
\author{T.~C.~Ralph\\
Centre for Quantum Computer Technology, Department of Physics,\\
University of Queensland, St Lucia 4072, Australia \\
G.~J.~Pryde \\
Centre for Quantum Computer Technology, Centre for Quantum Dynamics,\\
Griffith University, Nathan, 4111, Brisbane, Australia \\}
\date{\today}

\begin{abstract}
We review the field of Optical Quantum Computation, considering the various implementations that have been proposed and the experimental progress that has been made toward realizing them. We examine both linear and nonlinear approaches and both particle and field encodings. In particular we discuss the prospects for large scale optical quantum computing in terms of the most promising physical architectures and the technical requirements for realizing them.
\end{abstract}

\maketitle

\vspace{10 mm}

\section{Introduction}

At the most fundamental level, physical processes are described by quantum mechanics. Quantum mechanical systems possess unique properties that enable new ways of communicating and processing information (Nielsen and Chuang 2000). Large scale coherent processing of information via quantum systems is referred to as {\it quantum computation}. However, to achieve quantum computation, physical systems with very special properties are required. For example, it is essential that the quantum system evolves coherently and thus must be well isolated from the surrounding environment. Simultaneously, in order that the information stored in the system can be processed and read out, it must also be possible to produce very strong interactions between the system and classical meters and control elements. 

The invention of the laser in the early 1960's and its subsequent development led to an unprecedented increase in the precision with which light could be produced and controlled, and hence enabled the ability to systematically investigate the quantum properties of optical fields. It was soon realized that quantum optics offered a unique opportunity, not previously available to experimentalists, to test fundamentals of quantum theory (Walls and Milburn 1994) and later quantum information science (Bachor and Ralph 2004). It is natural, then, to consider quantum optics as a physical platform for quantum computation. In this article we review progress in achieving quantum information processing in optics and the prospects for building a large scale optical quantum computer.

%Each year integrated circuits become smaller and smaller, enabling more and more computing power to be packed into the devices constructed from them. If this trend continues then within the next decade or so the individual components of these circuits will approach the atomic scale. At this point further miniaturization through simple scale down of devices will no longer work as the physical principles describing their operation will segue from classical to quantum physics. As well as challenges, working at the atomic level introduces new opportunities for increasing computing power by encoding information in quantum systems. 

\subsection{Quantum Computation}

The development of computers with greater and greater power during the sixties and seventies made many more problems in quantum physics tractable to analysis. Nevertheless, it soon became clear that, in general, the simulation of quantum systems is a {\it hard} problem--- that is, the computing resources required grow exponentially with the size of the quantum system. It was Feynman (1986) who first noted that this bug might be turned into a feature. He pointed out that one quantum system could efficiently simulate another. Hence, a quantum simulator could efficiently solve problems that were intrinsically hard for a classical simulator. % \cite{FEY86}.
This raised the possibility that other computer algorithms may exist that could be more efficiently processed by quantum systems than classical systems. Although toy examples of such algorithms were suggested by Deutsch (1986) soon after, it was not till the mid-1990's that Shor (1994) showed that an important problem, the determination of the prime factors of an integer, could be solved in exponentially less time using a quantum computer. Equally importantly, it was shown shortly afterwards that {\it quantum error correction} was possible (Shor 1995; Steane 1996). This enables coherent correction of the logical errors which will inevitably creep into any calculation on a real physical device. Another influential algorithm, showing speed up for the searching of an unsorted data base, was subsequently developed by Grover (1997). These developments showed that {\it fault tolerant} quantum computers (i.e. where errors can be corrected in the presence of imperfect gate operations) were in principle possible and that such machines could solve interesting problems more efficiently than a conventional computer. This led to an explosion of interest in the field of quantum information.

Quantum information was originally framed in terms of binary systems. Consider a two-level quantum system. This could be: the spin states of an electron (up or down); two well-isolated energy levels of an atomic system; or one of many other possibilities, including various optical field states. It is clear
that such two level systems could be used to carry {\it bits} of information.
For example, we could assign the value ``zero'' to one of the states, writing it in Dirac notation (Sakurai 1985) as $|\bf 0 \rangle$,
and ``one'' to the other state, writing $|\bf 1 \rangle $. These states are typically taken to be eigenstates of the Pauli $Z$ operator with eigenvalues $\pm 1$ (Nielsen and Chuang 2000). An ordered collection of
such objects could then faithfully represent an
arbitrary bit string.

However, quantum objects offer more possible
manipulations than classical carriers of bits. In particular, not only
can we have zeros and ones, but we can also have superpositions of
zeros and ones such as the diagonal state $|+ \rangle =
(1/\sqrt{2})(|\bf 0 \rangle + |\bf 1 \rangle)$. Indeed bits can just as
effectively be encoded in these superposition basis states, for example
using $|+ \rangle$ as a zero and the anti-diagonal state $|- \rangle =
(1/\sqrt{2})(|\bf 0 \rangle - |\bf 1 \rangle)$ as a one. Following our convention, these states are eigenstates of the Pauli $X$ operator. We will refer to the $Z$ basis as the {\it computational} basis and the $X$ basis as the {\it diagonal} basis. In general we can form any superposition of the form $\alpha |\bf 0 \rangle + \beta |\bf 1 \rangle$ where $\alpha$ and $\beta$ are arbitrary complex numbers. Because of these extra
degrees of freedom we refer to information digitally encoded on quantum
systems as quantum bits or $\it qubits$ (Schumacher 1995).

A feature of qubits is their ability to span all different bit string
values simultaneously. 
%This is obviously true of a single qubit where
%the $|+ \rangle$ state, when viewed in the computational basis, $|\bf 0 \rangle$ and $ |\bf 1 \rangle$,
%equally spans the two different bit values, 0 and 1.
%This continues to be true for multi-qubit states. 
For example suppose we
start with two qubits in the state $|\bf 0 \rangle |\bf 0 \rangle$
where the first ket represents the first qubit and the second ket the second qubit and a tensor product between their two Hilbert spaces is implied. If we rotate both of them into their diagonal states we end up with
the state
\beq
{{1}\over{2}}(|\bf 0 \rangle |\bf 0 \rangle + |\bf 0 \rangle |\bf 1 \rangle + |\bf 1  \rangle |\bf 0
\rangle + |\bf 1 \rangle |\bf 1 \rangle)
\label{super}
\eeq
which is an equal superposition of all four possible two bit values.
This generalizes to $n$ qubits where the same operation of rotating
every individual qubit leads to an equal superposition of all
$2^{n}$ bit string values.

Even greater power comes from the ability to place qubits into certain special superpositions of the bit values---specifically, superpositions of correlated bit values. For example consider the two qubit state
\beq
{{1}\over{\sqrt{2}}}(|\bf 0 \rangle |\bf 0 \rangle + |\bf 1 \rangle |\bf 1 \rangle)
\label{ent}
\eeq
Now only two of the four possible combinations are present. A pair of spatially separated quantum systems are said to {\it entangled} if the state that describes the joint system cannot be factored into a product of states describing the individual systems. The state of Eq.\ref{ent} clearly cannot be factored into contributions from the individual qubits, and thus is entangled. Such a state has no classical analogue. 
%Entanglement plays the role of a resource in quantum communication protocols. For example if two distant parties share entanglement then they can communicate classical information at twice the classical rate through the technique of quantum {\it dense coding} (Bennett and Wiesner 1992). Similarly,  in the presence of entanglement, quantum information can be communicated via the exchange of classical information through the technique of quantum {\it teleportation} (Bennett, Brassard, Crepeau, Jozsa, Peres and Wootters 1993). 
Entanglement is thought to be a key ingredient in the information processing speed-up offered by quantum computation.

If we wish to perform information processing using qubits we need to introduce {\it quantum
gates}. The action of a quantum gate on a qubit state, $|\phi \rangle$, can be represented by the action of a unitary operator, $U$, on the state, i.e. $|\phi \rangle \to U |\phi \rangle$. Some quantum gates have classical counterparts, for example
the NOT gate, $X$, takes $|\bf 0 \rangle$ to $|\bf 1 \rangle$ and vice versa. Other gates have no classical analogue, such as the
Hadamard gate, $H$, for which $H|{\bf 0} \rangle = (1/\sqrt{2})(|{\bf 0} \rangle + |{\bf 1} \rangle)$
and $H|{\bf 1} \rangle = (1/\sqrt{2})(|{\bf 0} \rangle - |{\bf 1} \rangle)$. An arbitrary rotation about the $j$-axis of the Bloch sphere, where  $j=x, y, z$, is effected by the unitary $J_{\theta} = cos \theta \;I + i sin \theta\; J$ where $J = X, Y, Z$, and $I$ is the identity operator. We also require two-qubit gates such as
the controlled-NOT (CNOT) \index{CNOT}
which preforms the NOT operation on one qubit
(the {\it target}) only if the other qubit (the {\it control}) has ``one'' as its
logical value. Eventually, if large arrays
of gate operations can be implemented efficiently, and fault tolerantly, on many qubits, one could consider
performing quantum computation. 

In more recent years quantum information research has been extended to systems with Hilbert space dimensions greater than two (Bullock, OÕLeary, and Brennen 2005). For example a 3-level quantum system can encode three separate "trit" values, $|\bf 0 \rangle$, $|\bf 1 \rangle$ and $|\bf 2 \rangle$, plus any superposition of them, $\alpha |\bf 0 \rangle + \beta |\bf 1 \rangle + \gamma |\bf 2 \rangle$. Such a system is called a {\it qutrit} and, generalizing, a d-level quantum system is said to form a {\it qudit}. 
%Although it has been shown that there is no improvement in the fundamental resource scaling to be achieved using qudits \cite{qdit}, it is possible to gain significant practical advantage from using qdit levels when constructing quantum gates (Ralph, Resch and Gilchrist 2007; Lanyon, Barbieri, Almeida, Jennewein, Ralph, Resch, Pryde, OBrien, Gilchrist and White 2009).

There has also been considerable interest in infinite dimensional Hilbert spaces and the quantum information properties of continuous degrees of freedom such as position and momentum (Braunstein and Pati 2003, Braunstein and Loock 2005). It is usual to take the computational basis states to be the position eigenstates $|x \rangle$ and hence the momentum states $|p \rangle = \int dx \; e^{i x p} |x \rangle$, form the diagonal basis where, as expected, each diagonal basis state spans all computational basis values. 
%Continuous variable versions of teleportation (Braunstein and Kimble 1998) and quantum key distribution (Ralph 2000; Hillary 2000) were developed early on and many other protocols followed. 
Quantum computation proposals based on continuous variables have been developed (Lloyd and Braunstein 1999, Menicucci, Loock, Gu, Weedbrook, Ralph, and Nielsen 2006). Although these are theoretically {\it universal} in the sense that gate sets can be identified to efficiently simulate any process, a question mark exists over fully continuous schemes because of the lack of general error correction protocols. A fruitful way around this problem is to encode qubit states into the continuous spectrum (Gottesman, Kitaev and Preskill 2001; Lund, Ralph and Haselgrove 2008). 

Although considerable progress has been made, in many different physical platforms, including Ion traps (Haeffner, Roos and Blatt 2008), superconductors (Schoelkopf and Girvin 2008) and solid state (Gaebel, Domhan, Popa, Wittmann, Neumann, Jelezko, Rabeau, Stavrias, Greentree, Prawer, Meijer, Twamley, Hemmer and Wrachtrup 2006), the realization of
quantum computation experimentally still remains a long way off. Indeed it is still quite unclear what physical platforms, if any, are compatible with the task of building a full scale quantum computer. In this article we present the case for optics.

\subsection{Quantum Optics}
\label{sec:QO}

Light can be described quantum mechanically in terms of the mode {\it annihilation operator} $\op a$, its conjugate, the {\it creation operator} $\op a^\dagger$, and the electromagnetic field mode ground, or {\it vacuum state} $\ket{0}$, defined by $\op a \ket{0} = 0$ (Dirac 1958). The mode operators obey the commutation relation $[\op a, \op a^\dagger] = 1$. The action of the creation operator on the vacuum state is to create a single photon {\it number state}, in a single spatio-temporal mode, i.e. $\op a^\dagger \ket{0} = \ket{1}$. In general $\op a^\dagger \ket{n} = \sqrt{n+1}\ket{n+1}$ where $n = 0, 1, 2, ....$. Similarly the annihilation operator annihilates a single photon in a particular single spatio-temporal mode and in general  $\op a^\dagger \ket{n} = \sqrt{n}\ket{n-1}$. The number states form an ortho-normal basis convenient for representing arbitrary states of the field.  

The spatio-temporal mode operators can be further decomposed into single wave-vector operators, $\op a_k$ with the property $[\op a_{k'}, \op a_k^\dagger] = \delta(k'-k)$. For example the mode operator representing a plane wave mode propagating in the plus $x$ direction can be written
\begin{equation}
\hat a (t, x)  =  \int dk \; G(k) \;e^{i(kx-\omega_k t)}  \hat a_{k}
\label{atxp}
\end{equation}
where the optical frequency is given by $\omega_k = c |k|$. 
%Strictly, the normalization factor $N$ contains a factor of $1/\omega_k$ and should thus be under the integral. However, for realistic, finite distributions around optical frequencies, this dependence can be neglected. Canonical quantization requires that the same time commutators $[\hat a_{t,x},\hat a_{t,x'}^{\dagger}] = \delta(x-x')$. We thus choose 
$G(k)$ is a normalised spectral mode distribution
function centred around some positive wave number, $k_0$ (corresponding to an optical frequency), and is required to be zero for $k<0$. Orthogonal spatio-temporal modes are characterized by having mode operators that commute. For example if we take the mode function of Eq.\ref{atxp} and displace to a new longitudinal position $x'$ and consider its same-time commutator with the original mode we obtain
\begin{equation}
[\hat a(t,x'),\hat a(t,x)^{\dagger}] = \int dk |G(k)|^2 e^{i k (x'-x)}.
\label{atc}
\end{equation}
%
%Notice that these mode operators are explicitly global operators, depending not on $x$ and $t$, but rather on $t-x$. 
%Eq.\ref{ak} represents an idealized unphysical object, however physical modes can be represented by forming distributions of the form
For a suitably large interval $|(x-x')|$ the right hand side of Eq.\ref{atc} will go to zero and hence we will have orthogonal modes. Effectively we have created a pair of well-separated pulses. We may also create orthogonal modes as a function of transverse displacement, varying transverse or longitudinal mode shapes or differing polarizations. In this article we will simply declare suitably labelled mode operators to be orthogonal and avoid explicit decompositions.

The optical observables we will be interested in are the photon number, $\op n = \op a^\dagger \op a$, and the quadrature amplitude, $\op X^\theta = e^{i \theta} \op a + e^{-i \theta}  \op a^\dagger$. Photon number is proportional to intensity for bright fields and can be measured by photo-detectors. For dim fields individual photons can be resolved with photon counters. The quadrature amplitude of the field can be measured by beating the signal field with a bright, phase reference field at the same optical frequency, a {\it local oscillator} (LO), and then measuring it with photo-detection. This is known as {\it homodyne} detection. The angle $\theta$ is the phase difference between the signal and the LO and is usually taken to be in-phase ($\theta=0$) or in-quadrature ($\theta=\pi/2$), giving two conjugate (i.e. non-commuting) observables analogous to position and momentum.

As well as the number states, another key state in quantum optics is the coherent state (Glauber 1962). The coherent states are displaced vacuum states defined by 
\beq
\ket{\alpha} = \op D(\alpha) \ket{0}
\eeq
where the displacement operator is
\beq
\op D(\alpha) = e^{i(\op a \alpha + \op a^\dagger \alpha^\ast)}
\eeq
The coherent states are eigenstates of $\op a$ with eigenvalue $\alpha$. This leads to average values for their quadrature observables that are the same as for a classical field with the same amplitude. Hence the coherent state is often thought of as the quantum mechanical state which is the closest approximation to a classical optical field. The output of a well stabilized laser is a mixed state which can be approximately decomposed as an ensemble of coherent states with fixed magnitude but random phases (M{\o}lmer 1997). However, in situations where the phase is unimportant, or when the LO is derived from the same laser as the signal such that the phase is common mode, it is convenient to model laser output as being in a single coherent state of fixed magnitude and phase.

\section{Optical Qubits}
\label{sec:OQ}

We now consider how quantum information can be carried by light. An obvious choice is to consider photons as particles and to encode information onto some bipartite degree of freedom of individual photons such as polarization. Such an encoding always requires 2 distinct optical modes to be present, so we will refer to  particle-like encodings as {\it dual-rail} encodings (Knill, Laflamme and Milburn 2001). As the particle state is an energy eigenstate, i.e. a single photon Fock state, dual-rail qubits are stationary states that do not evolve as they propagate.

Alternatively, we could consider the field mode as the fundamental object and encode information in different field states, for example two distinct Fock states. In this encoding only one quantum optical mode is used, thus we will refer to field-like encodings as {\it single-rail} encodings (Lund and Ralph 2002). Notice, however, that now we are inevitably considering situations in which our optical modes are in superpositions of energy eigenstates and hence experience a phase evolution as they propagate. As a result, a co-propagating classical mode is implicitly needed as a phase reference for single-rail encodings. In the following, we will describe these encoding techniques in more detail and discuss a number of examples.

\subsection{Dual-Rail Encoding}
\label{subsec:DRE}

Consider two orthogonal optical modes represented by the annihilation operators $\op{a}$ and $\op{b}$ and the vacuum modes $\ket{0}_a$ and $\ket{0}_b$. For brevity we will write $\ket{0}_a \otimes \ket{0}_b \equiv \ket{00}$. We define our logical qubits as $\lket{0} = \op{a}^\dagger \ket{00} = \ket{10}$ and $\lket{1} = \op{b}^\dagger \ket{00} = \ket{01}$. That is, single photon occupation of one mode represents a logical zero, whilst single photon occupation of the other represents a logical one. This is dual-rail encoding.
%
%\begin{figure}[htb]
%\begin{center}
%\includegraphics*[width=6cm]{progoptfig1a}
%%\caption{}
%\end{center}
%\end{figure}
%
\begin{figure}[htb]
\begin{center}
\includegraphics*[width=8cm]{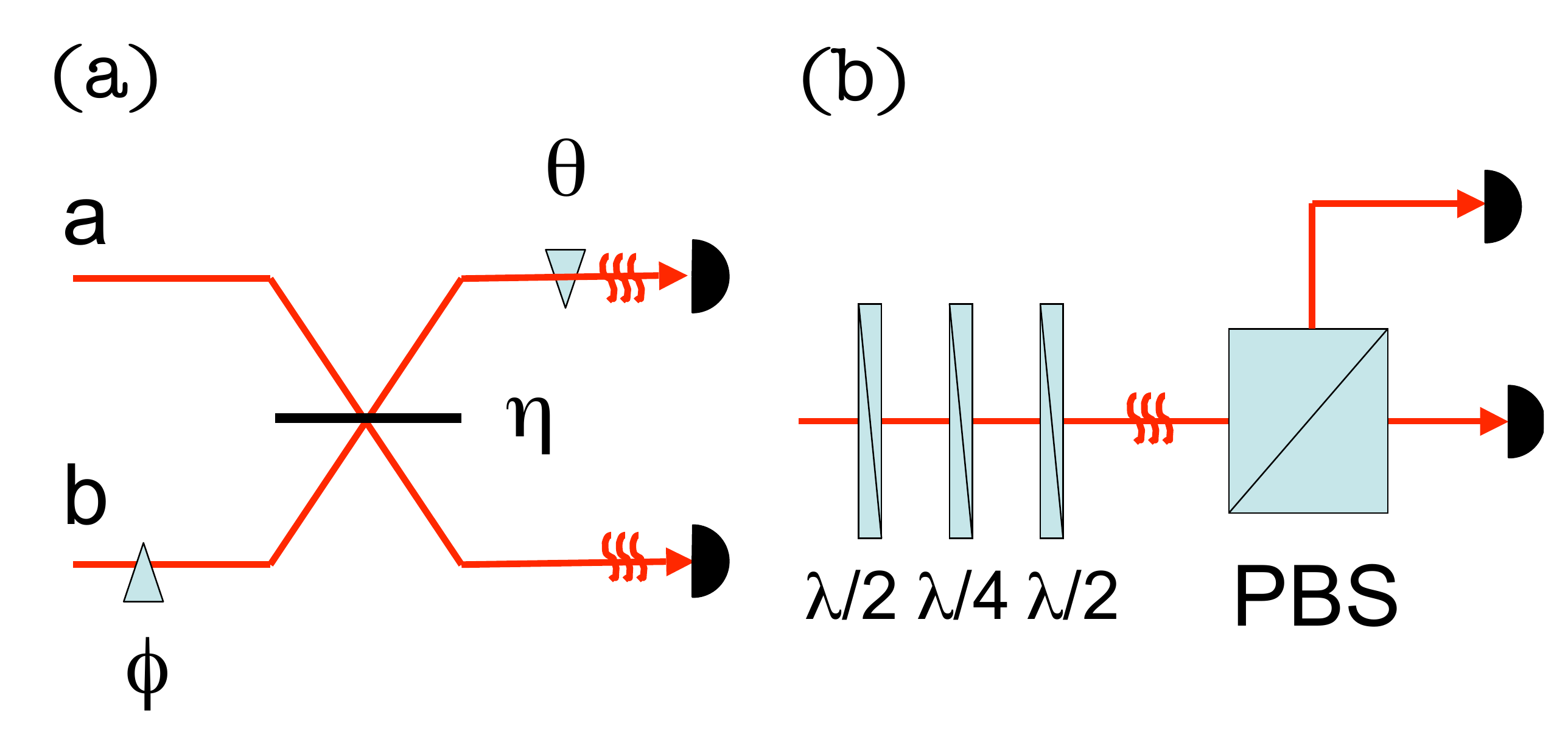}
\caption{Manipulation and detection of dual-rail qubits. (a) Beam splitter and phase-shifter circuit for producing an arbitrary single qubit evolution on a spatial dual-rail qubit. Detection in the computational basis is achieved by measuring which spatial mode holds the photon. (b) Combination of half- and quarter-wave plates oriented at particular angles achieves arbitrary single qubit evolution on a polarization dual-rail qubit. Detection in the computational basis is achieved via a polarizing beamsplitter (PBS) and photon counting. }
\label{fig1}
\end{center}
\end{figure}

For example suppose $\ket{0}_a$ and $\ket{0}_b$ are spatio-temporal modes with identical profiles, polarization and centre frequency, synchronized in time, but spatially separated in the transverse direction. Arbitrary single qubit operations can be achieved using a beamsplitter and two phase shifters as illustrated in Fig.~\ref{fig1}(a). A beamsplitter is a partially reflecting mirror that can coherently combine two optical modes in a set ratio. The interaction in the figure produces the following Heisenberg evolution of the mode operators:
\begin{eqnarray}
\op{a} & \to & \sqrt{\eta} \op{a} +e^{i \theta} \sqrt{1-\eta} \op{b} \nonumber\\
\op{b} & \to & e^{i \phi}(\sqrt{1-\eta} \op{a} -e^{i \theta} \sqrt{1-\eta} \op{b})
\label{hbs}
\end{eqnarray}
where $\eta$ is the intensity reflectivity of the beamsplitter. We have assumed the optical elements are lossless, a reasonable assumption for modern components. We have also assumed perfect mode matching between the two input modes to the beamsplitter, something rather more difficult to arrange in practice. Eq.\ref{hbs} implies the following qubit evolution (Bachor and Ralph 2004):
\begin{eqnarray}
\ket{10} & \to & \sqrt{\eta} \ket{10} +e^{i \phi} \sqrt{1-\eta} \ket{01} \nonumber\\
\ket{01} & \to & e^{i \theta}(\sqrt{1-\eta} \ket{10} -e^{i \phi} \sqrt{1-\eta} \ket{01})
\label{sbs}
\end{eqnarray}
which corresponds to an arbitrary single qubit unitary operation. Detection in the computational basis simply corresponds to measuring the photon number in each spatial mode.

More commonly two identical spatio-temporal modes but with different polarizations, say horizontal and vertical, are used as the dual rails. Then we may write $\lket{0} = \ket{10} = \ket{H}$  and $\lket{ 1} = \ket{01} = \ket{V}$. Half- and quarter-wave plates replace the phase shifters and beamsplitters in achieving arbitrary unitaries (Dodd, Ralph, and Milburn 2003). In particular, the Hadamard gate is implemented by a half-wave plate oriented at 22.5 degrees to the optic axis. Detection in any basis can be achieved via wave plates and polarizing beamsplitters, the latter of which effectively converts polarization encoding into spatial encoding (see Fig.~\ref{fig1}(b)). The ease of manipulation and phase stabilty of polarization states has made this encoding the most popular in optics (see Section\ref{subsec:LOparticle}).

Other possibilities are: temporal encodings in which the dual rails are spatio-temporal modes which are identical except for a time displacement (Stucki, Gisin, Guinnard, Ribordy, Zbinden 2002) (see example in Section \ref{sec:QO}); and  frequency encodings in which, this time, the dual-rail modes are identical except for a frequency off-set  (Huntington and Ralph 2004). 

Qubit initialization for these dual-rail schemes amounts to the ability to produce single-mode, single-photon states in a controlled way. Considerable progress has been made towards achieving this goal, as will be discussed in Section \ref{subsec:SOU}. Two qubit entangling gates turn out to be quite a challenge for dual-rail schemes, however the ease of single qubit manipulation (especially for polarization and spatial encodings) and measurement make these encodings strong contenders for large scale quantum computation.

\subsection{Single-Rail Encoding}
\label{subsec:SRE}

Single-rail encoding requires only a single quantum mode, that can be prepared in the states $\lket{0} = \ket{\phi}$ and $\lket{1} = \ket{\psi}$ or any superposition of them. The only requirement on these states $\ket{\phi}$, $\ket{\psi}$ is that they are orthogonal, i.e. that $\langle \phi \ket{\psi} = 0$. In general such qubits will be non-stationary, and therefore a good "clock" (i.e.\ a LO) is required in order to detect and manipulate them. 

Perhaps the simplest choice for $\ket{\phi}$ and $\ket{\psi}$ are the vacuum and single photon states, such that $\lket{0} = \ket{0}$ and $\lket{1} = \ket{1}$. Producing and manipulating superposition states of the form $\mu \ket{0} + \nu \ket{1}$ is not so easy, however a universal set of non-deterministic operations has been described (Lund and Ralph 2002). Also, we shall see that some two-qubit dual-rail gates are actually built from more fundamental single-rail gates of this type.
%superposition states have been produced non-deterministically in experiments (Lvovsky and Mlynek 2002; Babichev, Brezger, and Lvovsky 2004). 
%One important feature of the single-rail encoding is that it is relatively easy to produce entangled states. If a single photon is split on a 50:50 beamsplitter the resulting state is $(1/\sqrt{2})(\ket{0} \ket{1} + \ket{1} \ket{0})$ which is a maximally entangled two qubit state in the single-rail encoding. Such states can then be used as a resource for quantum processing tasks. 

Another possible choice for $\ket{\phi}$ and $\ket{\psi}$ are two different coherent states, such that $\lket{0} = \ket{\alpha}$ and $\lket{1} = \ket{\beta}$. In general such states will not be orthogonal but their overlap is given by $|\langle \alpha \ket{\beta}|^2 = \exp[-|\alpha-\beta|^2]$ which is very small for quite modest differences in the amplitudes of the coherent states. This is a continuous-variable-type encoding as we are carving out qubits from a continuous Hilbert space. A popular choice is to take $\beta = -\alpha$ (Cochrane, Munro, Milburn 1998). By choosing  $\alpha \ge 2$ a small overlap is achieved. The computational states, $\ket{\alpha}$ and $\ket{-\alpha}$ can be distinguished via homodyne detection. A useful feature of this choice  is that the equal superposition state $\ket{\alpha} + \ket{-\alpha}$ ($\ket{\alpha} - \ket{-\alpha}$) contains only even (odd) photon number terms and so these orthogonal diagonal states can be distinguished by photon counting.
% A number of groups have discussed quantum information tasks using this encoding \cite{COC98,ENK02,JEO02,RAL03}. 
As with the single photon single-rail scheme, single-qubit unitaries are difficult with this encoding. On the other hand, entanglement production is relatively easy. Splitting a superposition state like $\ket{\alpha} + \ket{-\alpha}$ many times on a beamsplitter leads to multi-mode entanglement. It turns out that this feature, i.e. easy entangling (and disentangling) operations, compensates sufficiently for the greater difficulty of performing single qubit unitaries to make the coherent state single-rail scheme a serious contender for large scale quantum computation (Lund, Ralph and Haselgrove 2008). Furthermore there has been considerable recent progress in producing diagonal-basis resouce states for this encoding, as will be discussed in Section \ref{subsec:SOU}. 

A more exotic single-rail scheme, in which the qubit states are comprised of superpositions of multiple, evenly spaced squeezed states, has also been suggested (Gottesman, Kitaev and Preskill 2001).  This scheme has the full continuous variable feature that transforming between the computational and diagonal bases is equivalent to transforming between the position and the momentum bases. In addition this structure provides a natural way for general error correction to be implemented. However it currently appears that the greater technical requirements of this approach outweigh these appealing features (Glancy and Knill 2006).

%
%Whilst production of the coherent computational states is straightforward, to date the only experimental realizations of the superposition states have come from cavity quantum electro-dynamics experiments \cite{BRU96,TUR95}, though promising schemes \cite{LUN04} and initial results \cite{WEG04} suggest small traveling wave superposition states may be possible in the near future.

\section{Universal Optical Quantum Gate Sets}
\label{UOQGS}

A universal quantum gate set enables any $n$-qubit unitary transformation to be implemented to arbitrary accuracy, for any $n$ (Nielsen and Chuang 2000). Many different universal gate sets exist. A sufficient universal gate set is comprised of arbitrary single-qubit unitary operations plus a maximally-entangling two-qubit gate. An example of the latter is the Controlled-NOT (CNOT) gate.

The CNOT gate is a two-qubit gate in which one qubit plays the role of a control and the other a target. When the control qubit is in the zero state, $\ket{\bf 0}_{c}$, the
value of the target qubit $\ket{\bf 0}_{t}$ or $\ket{\bf 1}_{t}$ is unchanged.
However, when the  control is one, $\ket{\bf 1}_{c}$, the value of the
target qubit is flipped, zero to one and vice versa. We can see that this gate is maximally entangling by considering its effect on superposition states. These can be calculated by simply making
superpositions of the aforementioned transformations. For
example if the control is in the diagonal basis we get the following
transformations
\bea
1/\sqrt{2}(\ket{\bf 0}_{c} + \ket{\bf 1}_{c}) \ket{\bf 0}_{t} & \to &
1/\sqrt{2}(\ket{\bf 0}_{c} \ket{\bf 0}_{t} + \ket{\bf 1}_{c} \ket{\bf 1}_{t}) \nnum
1/\sqrt{2}(\ket{\bf 0}_{c} + \ket{\bf 1}_{c}) \ket{\bf 1}_{t} & \to &
1/\sqrt{2}(\ket{\bf 0}_{c} \ket{\bf 1}_{t} + \ket{\bf 1}_{c} \ket{\bf 0}_{t}) \nnum
1/\sqrt{2}(\ket{\bf 0}_{c} - \ket{\bf 1}_{c}) \ket{\bf 0}_{t} & \to &
1/\sqrt{2}(\ket{\bf 0}_{c} \ket{\bf 0}_{t} - \ket{\bf 1}_{c} \ket{\bf 1}_{t}) \nnum
1/\sqrt{2}(\ket{\bf 0}_{c} - \ket{\bf 1}_{c}) \ket{\bf 1}_{t} & \to &
1/\sqrt{2}(\ket{\bf 0}_{c} \ket{\bf 1}_{t} - \ket{\bf 1}_{c} \ket{\bf 0}_{t})\nnum
\label{eq:CNOT2}
\eea
Notice that separable states are transformed into maximally entangled states. In fact, the two-qubit spanning set of entangled states on the right of Eq.\ref{eq:CNOT2} is given a special name - the {\it Bell} states. A closely related maximally-entangling two-qubit gate is the Controlled-Sign gate (CZ), for which all state components are unchanged except for $\ket{\bf 1} \ket{\bf 1} \to - \ket{\bf 1} \ket{\bf 1}$.
A CZ gate can be transformed into a CNOT gate by placing Hadamard gates before and after the CZ gate on the target qubit.
 
 In the following we will look at how universal gate sets can be realized in various ways for the two most promising optical qubits we have discussed: single photon, dual rail (spatial or polarization) and coherent state single rail.

\subsection{Nonlinear Gates}

We start our discussion with the conceptually simplest, but practically most difficult gate sets --- those based on highly nonlinear, in-line optical interactions.

\subsubsection{Kerr Nonlinearities for Dual Rail}

Arbitrary single qubit unitaries come for ``free'' with dual-rail schemes as they require only simple linear interactions. Thus our job is complete if we can implement a maximally entangling 2-qubit gate.
So how might such an interaction
between two photons be implemented? One solution is to use a $\chi^{(3)}$ nonlinear medium to induce a cross-Kerr effect between two photon modes, as first suggested by Milburn (1989). Ideally the cross-Kerr effect will produce the unitary evolution $\op U_K = \exp[i \chi \op a^{\dagger} \op a \op b^{\dagger} \op b]$, where $\op a$ represents one optical mode and $\op b$ another. Consider the schematic set-up of Fig.\ref{fig2}(a). Two polarization encoded qubits are converted into spatial dual-rail qubits using polarizing beamsplitters. One mode from each of the qubits is sent through the cross-Kerr material. The operation of this device on an arbitrary two qubit input state is given by the following evolution:
\begin{figure}[htb]
\begin{center}
\includegraphics*[width=9cm]{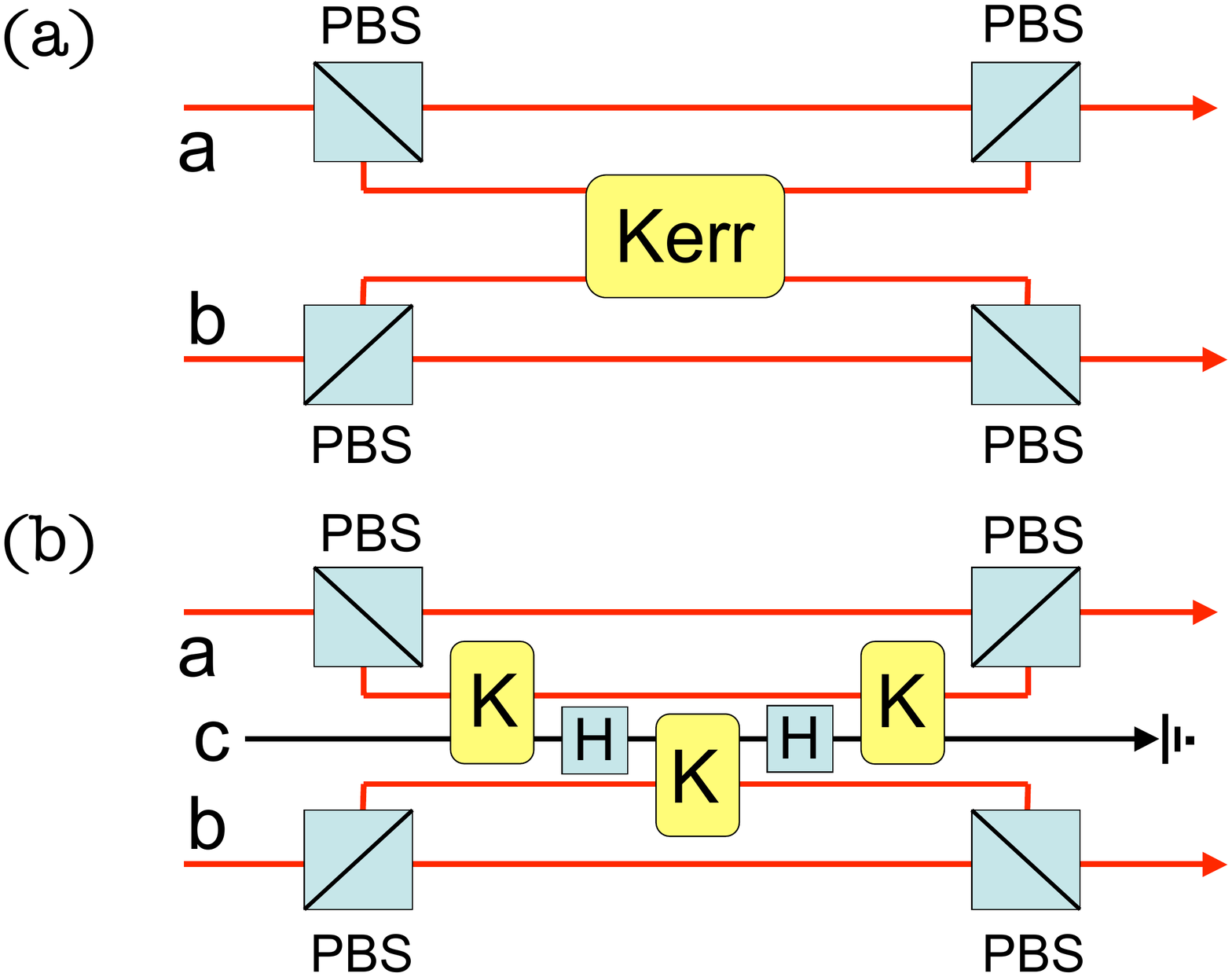}
\caption{Schematic of the implementation of an optical CZ gate between two optical polarization qubits, a and b, using a strong cross-Kerr nonlinearity: (a) using an all optical cross-Kerr nonlinearity and (b) mediated by an atomic two-level system, c. K are effective atomic/optical cross Kerr interactions, H are Hadamard rotations of the atomic two-level system and PBS are polarizing beam splitters.}
\label{fig2}
\end{center}
\end{figure}
\bea
\ket{\psi} &\to& \op U_K \ket{\psi} \nonumber\\
&=& e^{i \chi \op{a}_2^\dagger \op{a}_2 \op{b}_1^\dagger \op{b}_1}( \alpha \ket{01}_a\ket{01}_b + \beta \ket{10}_a\ket{10}_b  \nonumber\\
&& \;\;\;\;\;\;\;\;\;\;\;\;\;\;\;\;\;\;\;\;\;\;\;\;\;+ \gamma \ket{10}_a\ket{01}_b  + \delta \ket{01}_a\ket{10}_b ) \nonumber\\
&=&  \alpha \ket{01}_a\ket{01}_b  + \beta \ket{10}_a\ket{10}_b  \nonumber\\
&& \;\;\;\;\;\;\;\;\;\;\;\;\;\;\;\;\  + \gamma \ket{10}_a\ket{01}_b  + e^{i \chi} \delta \ket{01}_a\ket{10}_b  \nonumber\\
&&
\label{c-K}
\eea
Only when the modes passing through the Kerr material are both occupied is a phase shift induced. If we now choose the strength of the nonlinearity such that $\chi = \pi$, the effect is to flip the sign of one element of the superposition. Thus we directly apply a CZ gate to our dual-rail qubits, completing our universal gate set. A somewhat more complicated version of the gate uses a coherent state as a quantum bus to mediate the gate (Nemoto and Munro 2004) allowing a CZ gate to be implemented with $\chi \approx \pi/50$.

The problem with this idea in practice is that typical nonlinear materials have values of $\chi$ that are far to small. One might consider making the interaction region of the material very long in order to boost the nonlinearity, but such a strategy generally leads to very high levels of loss, which negate the desired effect. Even if loss was negligible, undesirable phase noise induced by the nonlinearity in bulk can inhibit the effect (Shapiro 2006).
%and may only be removed via sophisticated shaping of the spectral properties of the medium \cite{LEU09}. 

Non-linearities close to those required can be realized in cavity quantum electro-dynamic (QED) situations featuring single emitters in cavities of extremely high finesse and small volume. This occurs in the so-called {\it strong coupling} regime, in which the dipole coupling between the cavity field and the emitter is significantly greater than the relaxation rates of both the cavity and the dipole. These are difficult conditions to achieve as will be discussed in Section \ref{NOQCE}.
%Strong coupling has been demonstrated at optical frequencies with single alkali atoms in Fabry Perot (Turchette, Hood,
%Lange, Mabuchi and Kimble 1995; Rempe?) and micro-toroid  cavities (Aoki, Dayan, Wilcut, Bowen, Parkins, Kippenberg, Vahala and Kimble 2006) and at microwave frequencies between Rydberg atoms (Brune, Hagley, Dreyer, Maitre, Maali, Wunderlich, Raimond
% and Haroche 1996; Gleyzes, Kuhr, Guerlin, Bernu, Deleglise, Hoff, Brune, Raimond and Haroche 2007) and artificial atoms formed from superconducting {\blu Cooper-pair} boxes (Wallraff, Schuster, Blais, Frunzio, Huang, Majer, Kumar, Girvin, Schoelkopf 2004). {\blu (Do we need this microwave and Cooper-pair stuff?)}

Many schemes have been put forward for utilizing cavity QED to realize optical quantum gates. A conceptually simple scheme with several technical advantages was put forward by Duan and Kimble (2004). This scheme utilizes the ability of a single atom in the strong coupling regime to shift a cavity into or out of resonance with an optical field as a function of its internal state. An optical field reflected off a single ended cavity will acquire a $\pi$ phase shift if the cavity is on resonance but will suffer no phase shift if the cavity is off resonance. Suppose the cavity is on resonance for atomic state $| g \rangle$ but is pushed off resonance for atomic state $| e \rangle$ and an optical pulse in a superposition of vacuum and single photon states is reflected from the cavity. An effective cross-Kerr nonlinearity, with strength $\chi = \pi$ is established between the atom and the single-rail photonic qubit---that is, all state components are left the same except for $|1 \rangle |g \rangle \to - |1 \rangle |g \rangle$. Fig.\ref{fig2}(b) shows schematically how this effective Kerr effect between photon mode and cavity atom can lead to a CZ gate between photons via multiple interactions with the cavity. The major challenges with this scheme are realizing strong coupling conditions in a near ideal single-ended cavity configuration and finding parameter ranges in which spectral distortion of the photon modes is minimized.
%Many problems exist with this approach including: the difficulty of coupling photons efficiently into and out of the cavity mode; the need to isolate the cross-Kerr non-linearity from other non-linear effects and; the difficulty in maintaining a constant coupling strength between the atom and the field. A number of ingenious solutions have been suggested \cite{DUA04,NEM04} but remain unproven experimentally to date.

\subsubsection{Two Photon Absorption} 
\label{sec:zeno}

A quite different nonlinearity that can implement a CZ gate between dual-rail photons is two photon absorption, as first suggested by Franson, Jacobs, and Pittman (2004). They proposed using a pair of optical fibres weakly evanescently coupled
and doped with two-photon absorbing atoms to implement the gate. As the photons in the two fibre modes couple, the occurence of two photon state components in either of the modes is suppressed by the presence of the two-photon absorbers via the Zeno effect (Misra, Sudarshan 1977). After a length of fibre corresponding to a complete swap of the two modes a $\pi$ phase difference is produced between the $|11 \rangle$ term and the others. If the fibre modes are then swapped back by simply crossing them, a CZ gate is achieved. 

\begin{figure}[htb]
\begin{center}
\includegraphics*[width=8cm]{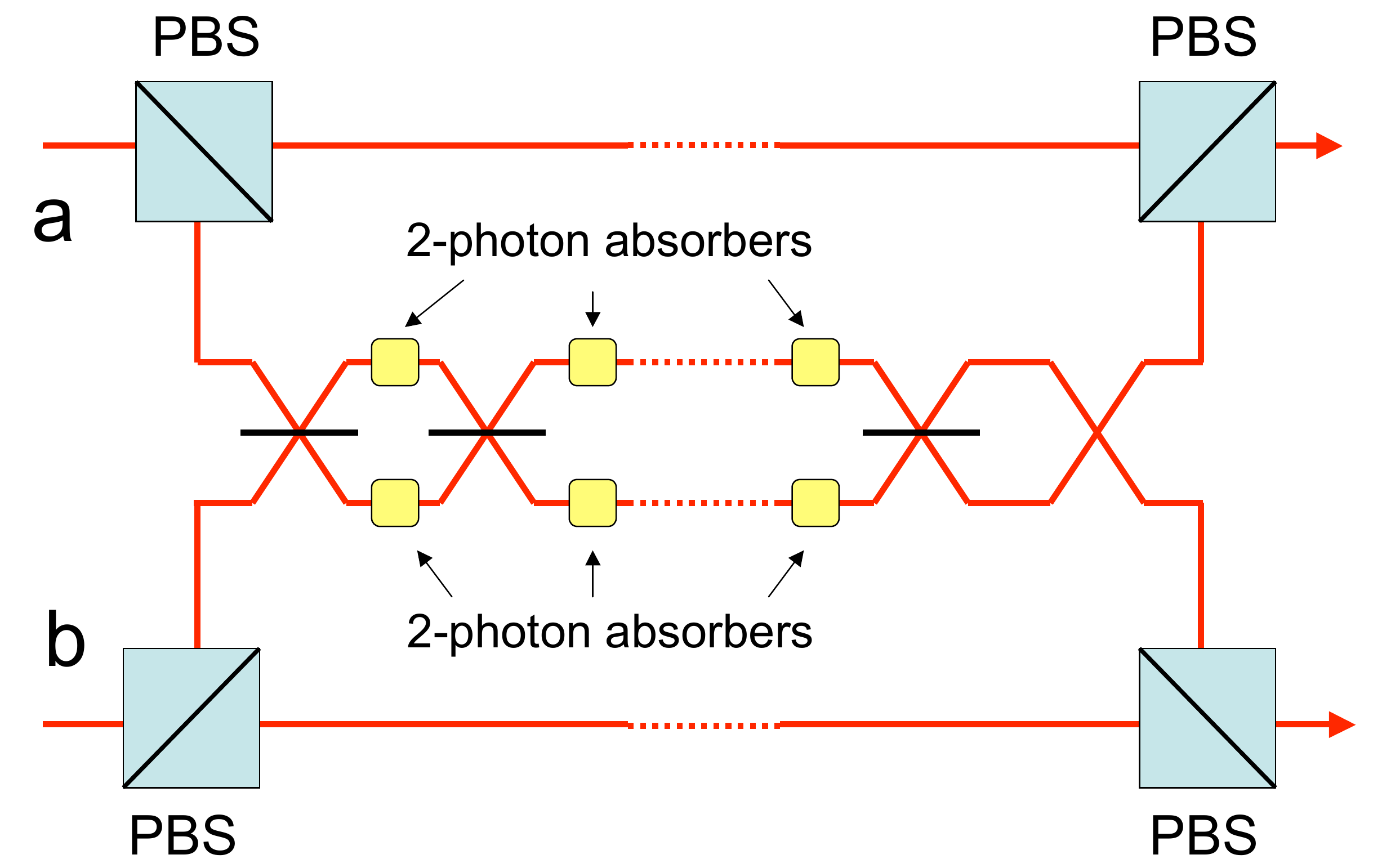}
\caption{Model of an optical CZ gate between two optical polarization qubits, a and b, using 2-photon absorption and the Zeno effect. Implementation is envisaged using evanescently coupled fibres doped with two photon absorbers. PBS are polarizing beam splitters}
\label{fig3}
\end{center}
\end{figure}

This system can be modeled as a succession of $n$ weak beamsplitters followed by 2-photon absorbers as shown in Fig.~\ref{fig3} (Leung and Ralph 2006). As $n \to \infty$ the model tends to the continuous coupling limit envisaged for the physical realization. 
%The gate operates on the single-rail encoding \cite{LUN02} for which $|0\ra_{L}=|0\ra$ and $|1\ra_{L}=|1\ra$ with the kets representing photon Fock states. Fig.~\ref{fig:CZ} shows how the single rail CZ can be converted into a dual rail CZ with logical encoding $|0\ra_{L}=|H\ra=|10\ra$ and $|1\ra_{L}=|V\ra=|10\ra$ with $|ij\ra$ a Fock state with $i$ photons in the horizontal polarization mode and $j$ photons in the vertical.

%With a series of $n$ units of weak beam splitters and two-photon
%absorbers, we can construct a Zeno CZ gate (Fig.~\ref{fig:OurCsign})
%that is in principle the same as Franson et al's fibre coupled Zeno
%gate. This CZ gate is implemented in a dual rail circuit as shown in
%Figure~\ref{fig:CZ}, where the logical basis are $|0\ra_{L}=|H\ra$
%and $|1\ra_{L}=|V\ra$. Hence if the input states are $|HH\ra$,
%$|HV\ra$, $|VH\ra$ and $|VV\ra$, then the photon number states that
%enter the CZ gate are $|00\ra$, $|01\ra$, $|10\ra$ and $|11\ra$
%respectively. The proof of our CZ gate scheme is as follow:\\
%

%\begin{figure}[h]
%\centerline{\psfig{figure=cz.ps,width=2cm,angle=90}} \caption{CZ
%gate in dual rail implementation.}\label{fig:CZ}
%\end{figure}

%The general symmetric beam splitter matrix has the form:

%%\[ e^{i\delta} \left[ \begin{array}{cc}
%%\cos\theta & \pm\sin\theta \\
%%\pm\sin\theta & \cos\theta \end{array} \right]\]

%\[ e^{i\delta} \left[ \begin{array}{cc}
%\cos\theta & \pm i\sin\theta \\
%\pm i\sin\theta & \cos\theta \end{array} \right]\]

After the PBS's the central pair of modes in Fig. \ref{fig3} are in some combination of vacuum or one-photon states. After the first beam splitter, the four possible photon number state combinations become:
\begin{eqnarray}
|00\ra & \rightarrow & |00\ra \nonumber\\
|01\ra & \rightarrow & e^{i\delta}(\cos\theta|01\ra \pm i\sin\theta|10\ra)\nonumber\\
|10\ra & \rightarrow & e^{i\delta}(\pm i\sin\theta|01\ra +\cos\theta|10\ra)\nonumber\\
|11\ra & \rightarrow & e^{i2\delta}(\cos2\theta|11\ra \pm
\frac{i}{\sqrt{2}}\sin2\theta(|02\ra+|20\ra))
\end{eqnarray}
Assuming ideal two-photon absorbers, i.e. they completely
block the two-photon state components but do not cause any single photon loss, propagation through the first pair of ideal two-photon absorbers gives the mixed state
\begin{equation}
\rho^{(1)} = P_s^{(1)} | \phi \rangle^{(1)} \langle \phi|^{(1)} + P_f^{(1)} |vac \rangle \langle vac|
\end{equation}
where $|\phi \rangle^{(1)}$ is the evolved two-mode input state obtained for the case of no two-photon absorption event and $|vac \rangle$ is the vacuum state obtained in the case a two-photon absorption event occurs. The individual components of $P_s^{(1)} |\phi \rangle^{(1)}$ transform as
\begin{eqnarray}
|00\ra & \rightarrow & |00\ra \nonumber\\
|01\ra & \rightarrow & e^{i\delta}(\cos\theta|01\ra \pm i\sin\theta|10\ra)\nonumber\\
|10\ra & \rightarrow & e^{i\delta}(\pm i\sin\theta|01\ra +\cos\theta|10\ra)\nonumber\\
|11\ra & \rightarrow & e^{i2\delta}\cos2\theta|11\ra
\label{eqn:complete}
\end{eqnarray}
Equation~(\ref{eqn:complete}) describes the transformation of
each unit, hence repeating the procedure $n$ times gives,
\begin{eqnarray}
|00\ra & \rightarrow & |00\ra\nonumber\\
|01\ra & \rightarrow & e^{in\delta}(\cos n\theta|01\ra \pm i\sin n\theta|10\ra)\nonumber\\
|10\ra & \rightarrow & e^{in\delta}(\pm i\sin n\theta|01\ra + \cos n\theta|10\ra)\nonumber\\
|11\ra & \rightarrow & e^{i2n\delta}(\cos2\theta)^n|11\ra
\label{zeno}
\end{eqnarray}
describing the transformations giving the evolved input state after $n$ units, $|\phi \rangle^{(n)}$. 
By choosing the conditions $n\theta=\frac{\pi}{2}$, $\delta=\frac{\pi}{2n}$ and going to the continuous limit $n \to \infty$, the transformation of Eq. \ref{zeno} tends to an ideal CZ gate and $P_f^{(1)} \rightarrow 0$. The main experimental challenge is to find a medium that exhibits very strong 2-photon absorption with negligible amounts of linear loss.

\subsection{Linear Optics Gates}

\label {sec:LOQC}

Although in-line nonlinear interactions are an efficient way to implement optical quantum computation in principle, we have already noted that there are many difficulties with this approach. If \textit{linear} optics could be used, many of these difficulties would be reduced. In fact any quantum unitary can be simulated using linear optics and a single photon (Reck, Zeilinger, Bernstein and Bertani 1994). However, this is achieved using a \textit{unary} encoding in which the number of modes required to simulate an $n$ qubit circuit grows as $2^n$. As a result this scheme has an exponential resource overhead and cannot in general be used for quantum computation \footnote{A special case is Grover's algorithm as discussed in Ralph (2006).}.

The first researchers to show that a scaleable linear optics scheme was possible were Knill, Laflamme and Milburn (2001). We will refer to their scheme as KLM. The standard dual-rail qubit encoding was used, but arbitrary processing was predicted to be possible without in-line nonlinearity or an exponential overhead. Instead, the KLM toolbox comprises: linear optical elements; single photon sources; photon-counting detectors; and electro-optic feed-forward. The introduction of single-photon ancilla and their subsequent measurement leads to {\it measurement induced nonlinearities} being applied to the qubits. Another way of viewing this is that KLM trades in-line nonlinearities for off-line nonlinearities in the form of single photon sources and detectors. KLM led to a surge of research activity in dual-rail linear optical schemes. We refer to such schemes in general as {\it linear optical quantum computing} (LOQC) (Kok, Munro, Nemoto, Ralph, Dowling, Milburn 2007).

\subsubsection{KLM}
\label{sec:KLM}
We begin by reviewing the original KLM scheme. The KLM scheme can be broken up into three tiers: non-deterministic entangling gates; non-deterministic teleportation gates; and error encoding against teleportation failure.
%
%\begin{enumerate}
%\item Non-deterministic two qubit gates which can be used to produce entangled resource states.
%\item Non-deterministic teleportation gates which are driven by entangled resource states and fail by accidentally measuring the value of the qubit.
%\item Error correcting codes that protect the qubits from accidental measurement during the application  of the teleportation gates and hence allow scale up of universal circuits without an exponential overhead.
%\end{enumerate} 
%%
%We now discuss each of these tiers in turn.

\begin{figure}
\includegraphics*[width=9cm]{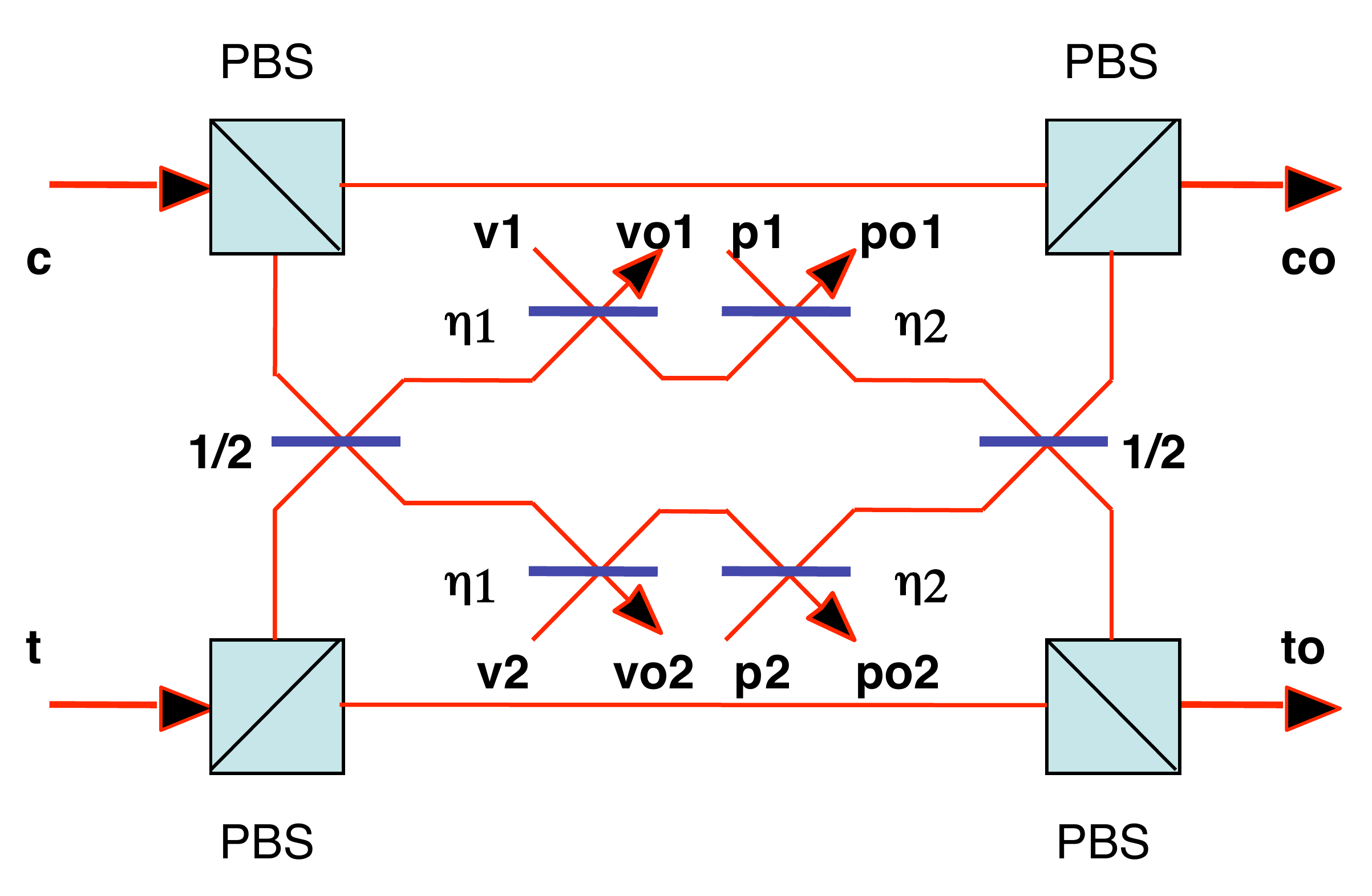}
\caption{Schematic representation of a non-deterministic CZ gate.
Polarization encoded qubits are injected at $c$ and $t$. Acilla photons are injected at $p1$ and $p2$. Successful operation is heralded by the detection of no photons at
outputs $vo1$ and $vo2$ and the detection of one and only one photon
at each of outputs $po1$ and $po2$. PBS are polarizing beam splitters.}
\label{fig4}
\end{figure}

{\it Non-Deterministic Entangling Gates:} At the first level, KLM introduced two-qubit gates that could take separable, single photon inputs, and produce entangled outputs. In particular KLM showed how to make a CZ gate that was
non-deterministic, but heralded. That is, the gate does not always
work, but an independent signal heralds successful operation. A
somewhat simplified version of this gate is shown in Fig.\ref{fig4} (Ralph, White, Munro and Milburn 2001).
In addition to the single-photon polarisation
qubits incident at ports $c$ (control) and $t$ (target), the gate also
has ancilla inputs comprising
two vacuum input ports, $v1$ and $v2$, and two single photon input
ports, $p1$ and $p2$. The beamsplitter reflectivities are given by
$\eta_{1}=5-3 \sqrt{2}$ and $\eta_{2}=(3-\sqrt{2})/7$. It can be shown
that when no photons are detected at outputs $vo1$ and $vo2$, and one
and only one photon is detected at each of $po1$ and $po2$, then the gate
has succeeded and the photon qubits exiting through $co$ and $to$ have
had the CZ transformation applied to them. The probability of
successful operation is $\eta_{2}^{2} \approx 0.05$.
% Recently it has been proved by Eisert \cite{EIS05} that $1/16$ is the upper bound for success probability for a gate of this type (as achieved by the original KLM proposal).
%

Even at this first level the technical requirements
are demanding. Four photons must simultaneously enter the circuit.
The detectors at $po1$ and $po2$ have to distinguish between zero,
one or two photons. Any inefficiency in the production or detection of
photons will lead to mistakes and rapidly erase the operation of the
gate. High-visibility single photon interference and two photon interference (Hong, Ou, Mandel 1987)
are required simultaneously: as a result excellent mode-matching and
photon indistinguishability are essential. Since KLM, several different suggestions for non-deterministic linear optical CZ (or CNOT) gates have been made (Ralph, Langford, Bell and White 2002; Hofmann and Takeuchi 2002;  Pittman, Fitch, Jacobs, and Franson 2003, Knill 2003; Bao, Chen, Q.Zhang, Yang, H.Zhang, Yang, and Pan 2007). There has also been recent work on numerically identifying optimal probabilities of success for LOQC gates (Uskov, Kaplan, Smith, Huver, and Dowling 2009). 

{\it Teleportation Gates:} We now discuss the second tier of the KLM scheme. Although the gates discussed in the previous section give us access to non-trivial two-qubit operations and small scale circuits, they are ultimately not scaleable in their own right. A cascaded sequence of
such non-deterministic gates would be useless for quantum computation because
the probability of many gates working in sequence decreases
exponentially. In order to make a scaleable system we must move to {\it teleportation gates}. 
%This
%problem may be avoided by using teleportation. 

The
idea that teleportation can be used for universal quantum computation was
first proposed by Gottesman and Chuang (1999). Teleportation uses an entangled Bell pair as a resource to transfer a qubit state from one mode to another (Bennett, Brassard, Crepeau, Jozsa, Peres and Wootters 1993). To achieve this, a two-mode measurement in the Bell basis (see section \ref{UOQGS}) is made between the input state and one of the particles from the Bell pair. The result of the Bell measurement is used to make bit flip and/or phase flip corrections to the other member of the pair which is transformed into the input state. In gate teleportation, the desired gate transformation is performed on the entangled state. The transformed entangled state is then used to teleport the qubit state which subsequently acquires the transformation (modulo certain commutation requirements between the transformation and the measurement corrections).  The main point
is that the application of the gate need not be deterministic. Non-deterministic gates can be used in
a trial-and-error manner to produce the required entangled state, which is then used to teleport the gate onto the qubit(s). 
%From this point of view the gates of the previous section can be regarded as entanglement factories - producing entangled states for use in teleportation. Alternatively we can note that some photon sources, such as parametric down-conversion, can produce entangled photons directly.
%Figure. New figure to be supplied
%
%\begin{figure}
%\begin{center}
%\includegraphics*[width=10cm]{progfigF}
%\caption{ Schematic representation of gate operation via
%teleportation. Figures (a) and (b) are equivalent, yet in (b) a
%non-deterministic CNOT gate is sufficient as failure only destroys the
%entanglement: the operation can be repeated till successful without
%losing the qubit.}
%\label{figF}
%\end{center}
%\end{figure}

\begin{figure}
\begin{center}
\includegraphics*[width=9cm]{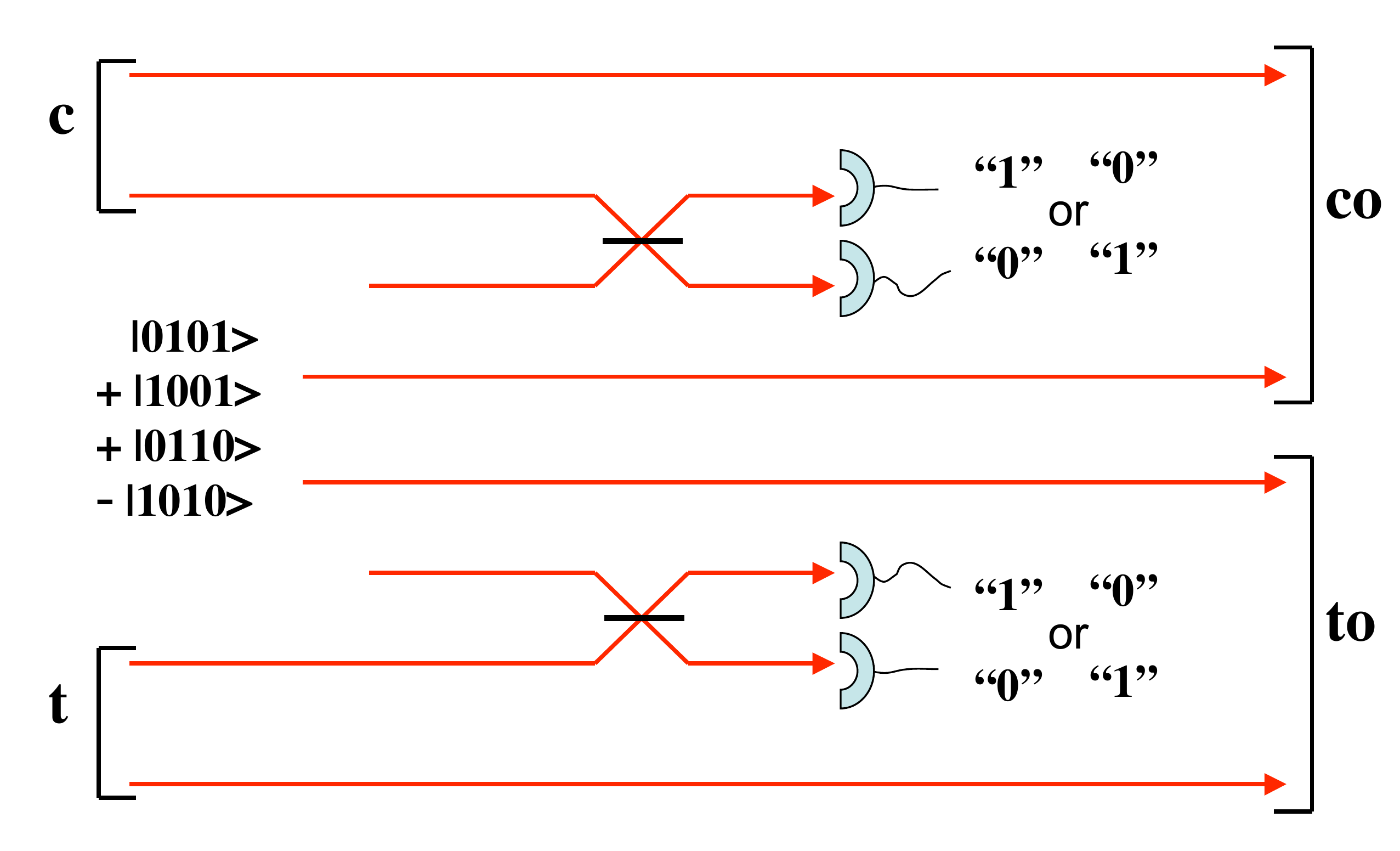}
\caption{ Schematic representation of optical CZ gate operation via
teleportation. Success is heralded by a single photon being detected at each of the two pairs of detectors in one of the patterns shown. Depending on the particular detection patterns, Pauli corrections may be neccessary. If zero (two) photons are detected at one of the detector pairs then the corresponding qubit has been measured to be in the zero (one) logical state and the gate has failed. The probability of success of the gate is 25\%.}
\label{fig5}
\end{center}
\end{figure}

The simplest teleportation gate is shown in Fig.\ref{fig5}. The heart of the gate is a teleported single-rail CZ gate (see section \ref{subsec:SRE}). 
%C operation on single-rail qubits can equally well be used to produce CS operation on dual-rail qubits simply by adding additional rails which do not participate in the interaction (Fig.\ref{figG}). 
The entangled resource is the state 
\beq
{{1}\over{2}}(\ket{0101} + \ket{0110} + \ket{1001} - \ket{1010})
\eeq
which can be interpreted as two single-rail Bell states which have had a CZ gate applied between. Indeed the circuit of Fig.\ref{fig4} could be used to produce this state from separable single photon inputs. Alternatively one can recognize this state as the dual-rail Bell state $\ket{0101} + \ket{1010}$ with a Hadamard gate applied to the second qubit. Such a state can be generated directly (but still non-deterministically) by parametric down conversion (see section \ref{subsubsec:BellPS}).  This latter interpretation is due to Pittman, Jacobs and Franson (2001). 

Unfortunately, Bell measurements, as required for the teleportation, can only be carried out non-deterministically with linear optics (L\"utkenhaus, Calsamiglia, and Suominen 1999). For the simplest scheme (the beamsplitters in Fig. \ref{fig4}) these fail 50\% of the time, thus the total probability of success of this gate is 25\%. KLM showed how to increase this probability of success using more complicated entangled states in the teleporter, however only modest improvements are practical due to the rapidly increasing overheads needed to produce these entangled states.

{\it Error Encoding Against Teleportation Failure:} We have seen that teleportation gates can be implemented which have higher probability of success than the first tier non-deterministic gates. A key feature of the teleportation gates is that failure results in the measurement of the logical values of the qubits. KLM introduced an error correction code  to protect against such computational basis measurements (Z-measurements) of the qubits. A logical qubit can be encoded across 2 physical qubits as (Knill, Laflamme and Milburn 2001) 
\begin{equation}
    \ket{\phi}^{(2)} = \alpha (|\bf 0 \rangle 
    |\bf 0 \rangle + |\bf 1 \rangle |\bf 1 \rangle) + \beta (|\bf 0 \rangle 
    |\bf 1 \rangle + |\bf 1 \rangle |\bf 0 \rangle)
    \label{parity1}
\end{equation}
This is a parity encoding---the ``zero'' state is represented 
by an equal superposition of all the even parity combinations of the 2 qubits, whilst the ``one'' 
state is represented by all the odd parity combinations. Notice that if a 
Z-measurement is made on either of the physical qubits of the state in 
Eq.\ref{parity1} and the result ``0'' is obtained, then the state 
collapses to an unencoded qubit, however the superposition is 
preserved. Similarly if the measurement result is ``1'' a bit-flipped 
version of the unencoded qubit is the result, but again the 
superposition is preserved so the qubit can be recovered. 

This encoding thus enables recovery from teleportation gate failure and so improves the probability of success of the gate by allowing second attempts. 
%An in principle demonstration of this encoding was made by O'Brien {\it et al} \cite{OBR05} using the two photon CNOT gate discussed in section \ref{sec:NDEG} to produce the required parity encoded states, where the CNOT gate takes an unencoded qubit as its target input and a diagonal state as its control input. It was shown that measurement of either physical qubit led to the expected unencoded qubit being projected onto the remaining photon to an accuracy of greater than 90\% fidelity. 
%
However, notice that a two-qubit (and thus non-deterministic) gate is needed to produce the parity encoding. It is not immediately obvious that producing encoded states non-deterministically which then can be used to improve the performance of more non-deterministic gates, is a winning strategy. KLM showed however, that provided you start with teleporters with a probability of success greater than 50\%, this strategy does improve gates success. For example a $2/3$ teleporter used with the parity encoding leads to a CZ gate success probability of about 58\% (as opposed to 44\% without encoding). In order to further improve the probability of success KLM concatenates the two qubit parity code. For example, using the nomenclature of Eq. \ref{parity1}, the next level up logical qubit is given by 
\begin{eqnarray}
    \ket{\phi}_{L4} &=& \alpha (|{\bf 0} \rangle^{(2)} 
    |{\bf 0}  \rangle^{(2)}  + |{\bf 1} \rangle^{(2)}  |{\bf 1}  \rangle^{(2)} ) \nonumber\\
    &&\;\;\;\;\;\;\;\;+ \beta (|{\bf 0}  \rangle^{(2)}  
    |{\bf 1}  \rangle^{(2)}  + |{\bf 1}  \rangle^{(2)}  |{\bf 0}  \rangle^{(2)} )
    \label{eq2}
\end{eqnarray}
High probabilities of success are obtained after a few levels of concatenation, leading to the claim of a scalable system. 

\subsubsection{Parity States}
\label{sec:PS}

The KLM result was a major step forward both in opening the door to small-scale demonstrations of optical quantum circuits, and in pointing the way towards a scalable system. However, in its original form the resources required for scale-up were exorbitant. For example, one can estimate that tens of thousands of Bell pairs are needed to implement a single CZ gate with 95\% probability of success using the original KLM approach. We now discuss a modification of KLM that massively reduces this overhead.
%Fortunately, considerable progress has been made in recent years in reducing this overhead \cite{YOR03, NIE04, HAY04} with the most efficient approaches requiring of order 100 Bell pairs for a CS with $> 95\%$ success \cite{BRO05,GIL05}. Two related but distinct approaches have emerged which we now discuss.

An alternative way to scale up the parity states, introduced by Hayes, Gilchrist, Myers and Ralph (2004), is not to concatenate the code as per Eq.\ref{eq2}, but instead to increase it incrementally. Hence a 
logical qubit can be encoded across $n$ qubits by representing logical 
``zero'' by all the even parity combinations of the $n$ qubits and 
logical ``one'' by all the odd parity combinations. This code retains the feature that if the 
logical qubit is encoded across $n$ physical qubits then a 
computational basis measurement on any one of the qubits reduces the state to a logical 
qubit encoded across $(n-1)$ physical qubits (with the possible need for a bit-flip). 
Specifically, this parity encoding is given by
\begin{eqnarray}
\label{parity}
\ket{{\bf 0}}^{(n)} & \equiv & (\ket{+}^{\otimes n}+\ket{-}^{\otimes
n})/\sqrt{2}\nonumber \\  
\ket{{\bf 1}}^{n} & \equiv & (\ket{+}^{\otimes n}-\ket{-}^{\otimes
n})/\sqrt{2},
\end{eqnarray}
where  $\ket{\pm} = (\ket{{\bf 0}} \pm \ket{{\bf 1}})/\sqrt{2}$.  

There are two operations which are easily performed on parity encoded
states: an arbitrary $X$ rotation, i.e. $X_\theta=\cos(\theta/2) I + i\sin(\theta/2)X$, 
which can
be performed by applying that operation to any of the physical qubits and;  a $Z$ operation, which can be performed by applying $Z$ to
all the physical qubits (since the odd-parity states will acquire an
overall phase flip).  

%A key operation we will use is the partial Bell state
%measurement \cite{Wei94,Brau95}.  This consists of mixing two physical qubits
%on a polarising beam splitter followed by measurement in the
%diagonal-antidiagonal basis.  A successful event occurs when a photon is
%counted at each out put of the beamsplitter. An unsuccessful event occurs when
%both photons appear at one of the outputs. When successful it projects onto the
%Bell states $\ket{00}+\ket{11}$ and $\ket{00}-\ket{11}$. When unsuccessful it
%projects onto the separable states $\ket{01}$ and $\ket{10}$, thus measuring
%the qubits in the computational basis. 
The teleportation gates are reduced to just partial single-rail and dual-rail Bell-state measurements. A dual-rail Bell measurement can be used to add $n$ physical qubits to a parity encoded state using a resource of $\ket{{\bf 0}}^{(n+2)}$. This is referred to as type-II fusion ($f_{II}$)  (Browne and Rudolph 2005). The result of $f_{II}$ is
\begin{equation}
	f_{II}\ket{\psi}^{(m)}\ket{{\bf 0}}^{(n+2)}\rightarrow\left\{ \begin{array}{cl}
		\ket{\psi}^{(m+n)} & \mbox{(success)}\\
		\ket{\psi}^{(m-1)}\ket{{\bf 0}}^{(n+1)} & \mbox{(failure)}
	\end{array}\right.
	\label{encoding}
\end{equation}
When successful (with probability $1/2$), the length of the parity qubit is
extended by $n$. A phase flip correction may be necessary depending on the
outcome of the Bell-measurement.  If unsuccessful, a physical qubit is removed
from the parity encoded state, and the resource state is left in the state
$\ket{{\bf 0}}^{(n+1)}$ (which may be recycled).  This encoding procedure is equivalent
to a gambling game where we either lose one level of encoding, or gain $n$
depending on the toss of a coin. Clearly, if $n \ge 2$ this is a winning game. The required resource states can be built from Bell pairs using a combination of single-rail Bell measurements (type I fusion) and $f_{II}$. The remaining gates required to achieve a universal gate set (a $Z_{90} = I + i Z$ and a
\textsc{cnot} gate) can be efficiently performed using these fusion techniques (Gilchrist, Hayes and Ralph 2007). The resource overhead for performing gates in this way is of order 100 Bell pairs per gate.

\subsection{Coherent State Gates}

Linear optical protocols are also possible for coherent state qubits. In fact they are arguably the simplest linear optics schemes known. The off-line resources in this case are {\it cat states}, i.e. superpositions of distinct coherent states. The greater difficulty involved in producing these states somewhat off-sets the greater simplicity of these schemes. As with KLM, photon resolving measurements and feedfoward are also required. We will refer to schemes of this type as Coherent State Quantum Computing (CSQC).

\subsubsection{Coherent State Qubits with Large Amplitudes}

We will first consider the scheme of Ralph, Munro and Milburn (2002) in which it is assumed that very large amplitude cat states are available as resources. The qubits are coherent states with amplitudes zero (i.e. the ground or vacuum state) and $\alpha$. These qubits are not exactly orthogonal, but the approximation of
orthogonality is good for even moderately large $\alpha$ as
$\langle \alpha |0\rangle =e^{-\alpha^2/2}$.
In this section it is assumed that $\alpha \gg 1$ so that qubits have negligible overlap.

As we have seen, two-qubit gates represent a formidibile challenge in single-photon dual-rail schemes. Surprisingly, for this coherent state encoding, a
non-trivial two-qubit gate can be implemented using only a single
beamsplitter. Consider the beamsplitter
interaction given by the unitary transformation
%
%\begin{equation}
$U_{BS}=\exp[i\theta (a b^{\dagger}+a^{\dagger} b)]$,
%\end{equation}
%
where $a$ and $b$ are the annihilation operators corresponding to two
coherent state qubits $|\gamma \rangle_{a}$ and $|\beta
\rangle_{b}$, with $\gamma$ and $\beta$ taking values of $\alpha$ or
$0$. It is well known that the output state
produced by such an interaction is
\begin{eqnarray}
U_{BS} |\gamma \rangle_{a} |\beta \rangle_{b}=|\cos \theta \gamma+i
\sin \theta \beta \rangle_{a} |\cos \theta \beta+
i \sin \theta \gamma \rangle_{b} \nonumber\\
 \label{Ho}
\end{eqnarray}
where $\cos^{2} \theta$ ($\sin^{2} \theta$) is the reflectivity
(transmissivity) of the beamsplitter.
%Now consider the overlap between the output and input states. Using
%the relationship \cite{WAL94} $\langle \tau|\alpha \rangle =
%\exp[-1/2(|\tau|^{2}+|\alpha|^{2})+\tau^{*} \alpha]$ we find
%%
%\begin{eqnarray}
%&&\langle \gamma |_{a}\langle \beta |_{b} |\cos \theta \gamma+i
%\sin \theta \beta \rangle_{a} |\cos \theta \beta+i \sin \theta \gamma
%\rangle_{b} \nonumber\\
%&&\;\;\;\;= \exp[-(\gamma^{2}+\beta^{2})(1-\cos \theta)+2 i \sin
%\theta \gamma \beta]
%\label{state}
%\end {eqnarray}
%
Now assume that $\theta$ is sufficiently small, so that
$\theta^{2} \alpha^{2}\ll1$ but $\alpha$ is sufficiently large, so
that $\theta \alpha^{2}$ is of order one. Physically this corresponds
to large coherent states impinging on an almost perfectly reflecting beamsplitter. %Eq. \ref{state} then
%approximately becomes
%%
%\begin{eqnarray}
%&&\langle \gamma |_{a}\langle \beta |_{b} |\cos \theta \gamma+i
%\sin \theta \beta \rangle_{a} |\cos \theta \beta+i \sin \theta \gamma
%\rangle_{b}\nonumber\\
%&& \;\;\;\; \approx \exp[2 i \theta \gamma \beta]
%\label{astate}
%\end {eqnarray}
%%
%Eq.\ref{astate} shows that the only difference between the input
%and output states of the beamsplitter is a phase shift proportional
%to the amplitudes of the input qubits, that is:
Under these conditions it can be shown that to high fidelity
\begin{eqnarray}
U_{BS} |\gamma \rangle_{a} |\beta \rangle_{b}
& \approx & \exp[2 i \theta \gamma \beta] |\gamma \rangle_{a}
 |\beta \rangle_{b}
 \label{H3}
\end{eqnarray}
If we further
require that $\theta \alpha^{2}=\pi/2$ then this transformation produces
a CZ gate. % \cite{note2}. 
%That is if either or both of the qubits
%are in the logical zero state ($\gamma=0$ and/or $\beta=0$)
%the transformation produces no
%effect on the state. However if both modes are initially in the logical one
%state (i.e $\gamma=\beta=\alpha$) then a sign change is produced.
%Such a gate is a non-trivial two qubit gate.

For universal computation we require, in addition to the above maximally-entangling two-qubit gate, the ability to perform arbitrary single-qubit unitaries. 
Arbitrary unitaries can be constructed given the ability to do arbitrary rotations around the z-axis ($Z_{\phi}$), bit-flips, plus the Hadamard gate (Nielsen and Chuang 2000).
A bit flip (or $X$ gate) is equivalent to a displacement of
$-\alpha$ followed by a $\pi$ phase shift of the coherent amplitude.
%$X=U(\pi) D(-\alpha)$,
%where $U(\pi)=exp[i \pi a^{\dagger}a]$ is physically just a
%half-wavelength delay, whilst a displacement can be implemented by
%mixing a very strong coherent field with the qubit on a highly
%reflective beamsplitter \cite{WAL94}.
%
The action of the $Z$ rotation gate is
$ Z_{\phi}(\mu|0 \rangle_L+\nu|1 \rangle_L)=\mu|0
    \rangle_L+e^{i \phi}\nu|1 \rangle_L$.
It can be implemented, to a good approximation, by imposing a
small phase shift on the qubit:
\begin{eqnarray}
    U(\epsilon)|\alpha \rangle & = & e^{i \epsilon a^{\dagger}a}|\alpha
\rangle \nonumber\\
& \approx & e^{i \epsilon \alpha^{2}}|\alpha
\rangle  =  Z_{\phi}|\alpha \rangle
\label{phase}
\end{eqnarray}
with $\phi=\epsilon\alpha^2$ and as before we require 
$\epsilon^{2}\alpha^{2}\ll1$ .

\begin{figure}
\begin{center}
\includegraphics*[width=9cm]{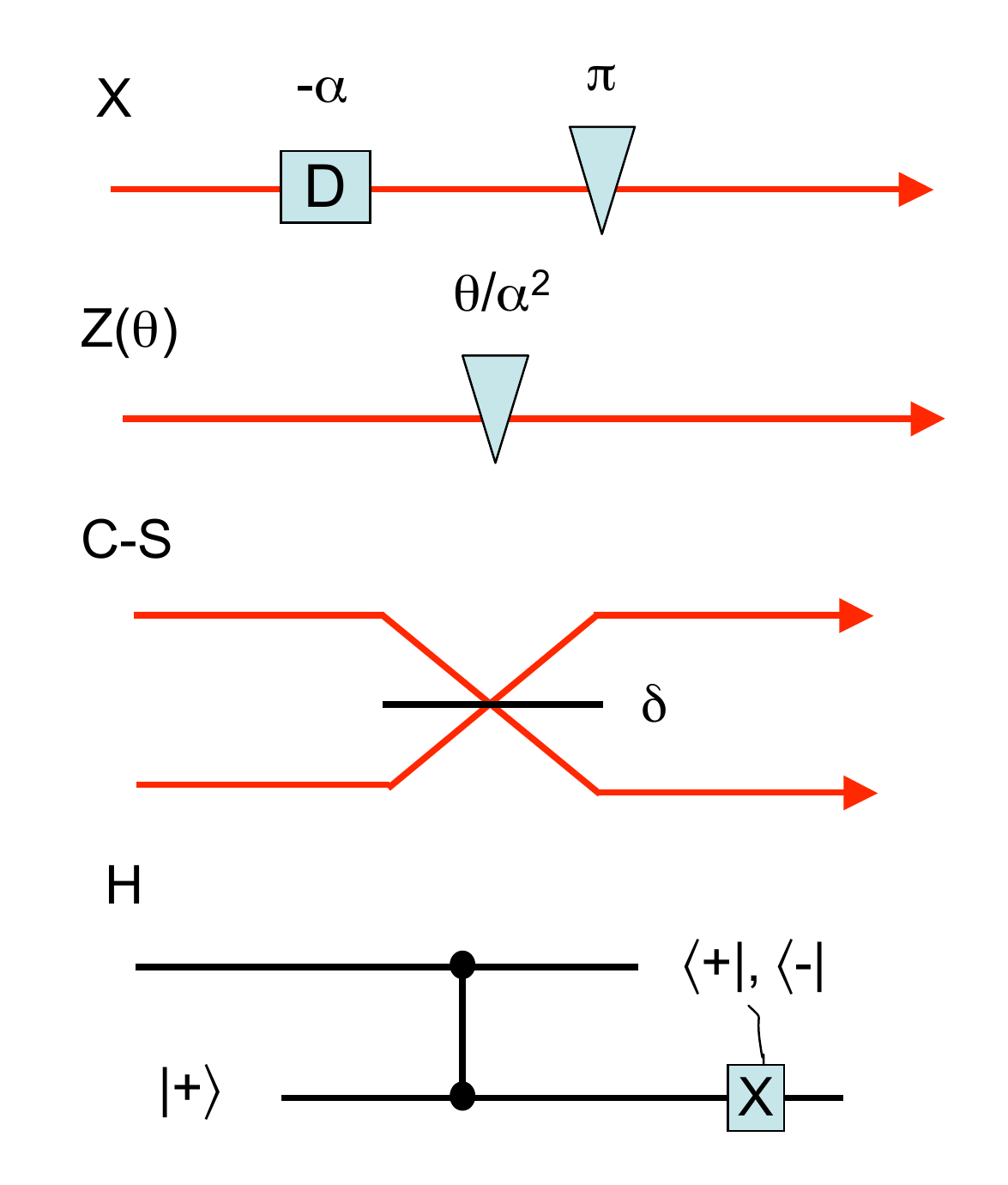}
\caption{ Schematic summary of linear optics coherent state universal gate set of Ralph, Munro and Milburn (2002). $D$ is a displacement and triangles represent phase shifts. The beamsplitter reflectivity is $\delta = cos^2 \theta$ where $\theta \alpha^{2}=\pi/2$. The gates only operate correctly if $\alpha\gg\pi/2$.}
\label{fig6}
\end{center}
\end{figure}

In addition to these gates, a Hadamard gate is required in order to achieve an
arbitrary qubit rotation. 
This can be achieved using an ancilla cat state, $1/\sqrt{2}(|0\rangle+|\alpha \rangle)$, the CZ gate, a diagonal (or cat) basis measurement and a possible $X$ correction.
A cat basis measurement
can be implemented by first displacing by
$-\alpha/2$. This transforms our ``0'', ``$\alpha$'' superposition into
``$\alpha/2$'', ``$-\alpha/2$'' superposition. These new states are parity eigenstates. A photon number measurement determines the sign of the superposition but not the sign of the amplitude as required for a diagonal basis measurement.
The complete gate set is summarized in Fig.\ref{fig6}. Its simplicity is clear, however this comes at a price. Recall that we require the conditions $\theta^{2} \alpha^{2}\ll1$ whilst $\theta \alpha^{2} = \pi/2$ in order to realize a CZ gate with high fidelity. This in turn implies $\alpha\gg\pi/2$. For high fidelity Ralph, Munro and Milburn (2002) showed that $\alpha > 20$ is required. Not only does this present great difficulties in terms of producing the required cat states but also places very high technical requirements on the detectors and linear manipulations.

\subsubsection{Coherent State Qubits with Small Amplitudes}

The requirement for large coherent amplitudes in the Ralph, Munro and Milburn (2002) scheme can be relaxed by introducing gate teleportation in a similar way to the KLM scheme. Gate teleportation can be carried out directly with high probability of success in the coherent state scheme, thus we end up with a two-tier scheme that is still significantly simpler than the three-tier KLM scheme. This was first shown by Ralph, Gilchrist, Munro, Milburn and Glancy (2003) where it was still assumed that $\alpha \ge 2$ so that any overlap between the coherent state qubits could be neglected. Lund, Ralph and Haselgrove (2008) further generalized the scheme, so that the gates would work non-deterministically for coherent amplitudes of {\it any} size, and with sufficient probability of success to be scaleable provided $\alpha > 1.2$. We now describe this scheme.

\begin{figure}
\begin{center}
\includegraphics*[width=9cm]{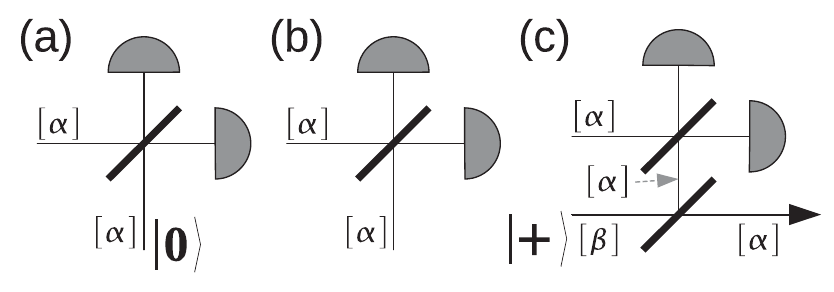}
\caption{ Schematic summary of (a) unambiguous qubit measurement, (b) unambiguous Bell measurement and (c) teleportation, for coherent state qubits. In (a) one input is an arbitrary qubit and the other is in the zero logical state (both with absolute values of their amplitudes [$\alpha$]). The arbitrary qubit is measured to be in $|\alpha \rangle$ if counts register at the top detector and $|-\alpha \rangle$ if counts register at the bottom detector. In (b) both inputs are arbitrary qubits. The four Bell states are determined by which detector fires and whether the number of counts registered is even or odd. In (c) the top input is an arbitrary qubit but the lower input is in the logical plus state with absolute value of its amplitude $\beta = \sqrt{2} \alpha$. The arbitrary qubit is teleported to the output modulo corrections dependent on which detector fires and whether the counts are odd or even. In all three cases the measurement/operation fails if no counts are registered at the detectors.}
\label{fig7}
\end{center}
\end{figure}

These later papers used the more familiar $\alpha, -\alpha$ encoding such that an arbitrary coherent state qubit is represented by the state 
\begin{equation}
N_{\mu,\nu}(\alpha) (\mu |\alpha \rangle + \nu |-\alpha \rangle).
\end{equation}
As noted before, transformation between this basis and the $0$, $\alpha'=\sqrt{2} \alpha$ basis requires only a displacement. We now allow for small values of $\alpha$ by including the exact normalization 
\begin{equation}
N_{\mu,\nu}(\alpha) = (|\mu|^2 + |\nu|^2 + e^{- 2 \alpha^2} (\mu \nu^* + \nu \mu^*))^{-1/2}.
\end{equation}
As with KLM, measurements play a key role. As this scheme does not restrict the size of $|\alpha|$, the Z-basis and Bell-state measurements must, in general,
distinguish as best as possible between nonorthogonal states. This can be achieved using linear optics and photon counting. The computational or Z-basis measurement is shown in
Fig.\ref{fig7}(a) and the Bell state measurement is shown in
Fig.\ref{fig7}(b). For the measurement
to be unambiguous and error-free, it must have
a failure outcome (Ivanovic 1987). This occurs in both measurements
when no photons are detected. The probability of failure
tends to zero as $|\alpha|$ increases.

The procedure for teleporting coherent state qubits was first described by van Enk and Hirota (2002) and Jeong, Kim and Lee (2001) and is shown in Fig.\ref{fig7}(c). The teleporter
uses unambiguous Bell state measurements which have 
5 outcomes. Four outcomes correspond
to successfully identifying the respective Bell states, and the remaining one is the failure mode.
When the appropriate Pauli corrections are made, the input
qubit is successfully transferred to the output. The fifth
outcome corresponds to the measurement failure whose
probability again decreases to zero as $|\alpha|$  increases. Upon
failure, the output of the teleporter is unrelated to the input
and hence the qubit is erased. Gate teleportation, as described below, can be used to produce a universal set of gates. The ability to unambiguously
teleport the qubit value, in spite of the fact that the
basis states are nonorthogonal, allows gates to be implemented for all values of $|\alpha|$.

{\it Single Qubit  Gates}: The  Phase Rotation Gate ($Z_{\theta}$) can be achieved using the resource state 
\begin{eqnarray}
\ket{Z_{\theta}} &=& \exp[i \theta]  \ket{\alpha, \alpha} + \exp[-i \theta]  \ket{\alpha, -\alpha} \nonumber\\
&+& \exp[i \theta]  \ket{-\alpha, \alpha} + \exp[-i \theta]  \ket{-\alpha, -\alpha}. 
\end{eqnarray}
whilst the Hadamard gate (H) can be achieved using the resource state
\begin{equation}
\ket{had} = \ket{\alpha, \alpha} + \ket{\alpha, -\alpha} + \ket{-\alpha, \alpha} - \ket{-\alpha, -\alpha}. 
\end{equation}
It is straightforward to show that if a Bell-state measurement is made between an arbitrary qubit state $\ket{\sigma}$ and the first qubit of $\ket{Z_{\theta}}$ ($\ket{had}$) then the operation $Z_{\theta} \ket{\sigma}$ ($H \ket{\sigma}$) is performed, where, dependent on the outcome of the Bell-measurement a bit-flip correction, a phase-flip correction, or both may be necessary. If the outcome is zero photons in both arms the gate fails and the qubit is erased. The resource states $\ket{Z_{\theta}}$ and $\ket{had}$ can be produced non-deterministically from cat state resources, linear optics and photon counting.

{\it Controlled-Sign Gate}: To complete the universal gate set the CZ gate can be achieved using the resource state 
\begin{eqnarray}
\ket{CZ} &=& \ket{\alpha, \alpha, \alpha, \alpha} + \ket{\alpha, \alpha, -\alpha, -\alpha} \nonumber\\
&+&  \ket{-\alpha, -\alpha, \alpha, \alpha} -  \ket{-\alpha, -\alpha, -\alpha, -\alpha}. 
\end{eqnarray}
If a Bell-state measurement is made between an arbitrary qubit state $\ket{\sigma}$ and the first qubit of $\ket{CZ}$ and a Bell-state measurement is made between an arbitrary qubit state $\ket{\phi}$ and the last qubit of $\ket{CZ}$ then the operation $CZ \ket{\sigma}\ket{\phi}$ is performed, where again, dependent on the outcomes of the Bell-measurements bit-flip and/or phase-flip corrections may be necessary. If the outcome is zero photons in both arms of either of the Bell measurements the gate fails and the respective qubit is erased. The resource state $\ket{CZ}$ can be produced non-deterministically from cat state resources, linear optics and photon counting.

{\it Correction of Phase-flips}: After each gate it was noted that bit flip and/or phase flip corrections may be necessary. Bit flips can be implemented easily by simply delaying a qubit with respect to the local oscillator. However phase-flips are more difficult and require further teleportation. However Jeong and Ralph (2007) have pointed out that only active correction of bit-flips is necessary. 
%This is because phase-flips commute with the phase rotation gate and the control sign gate but are converted into bit flips by the Hadamard gate. They suggest the following strategy: After each gate operation any bit-flips are corrected whilst phase-flips are noted. After the next Hadamard gate the phase flips are converted to bit-flips which are then corrected and any new phase-flips are noted. By following this strategy only bit-flips need to be corrected actively, with, at worst, some final phase-flips needing to be accounted for in the final measurement of the circuit.

This CSQC scheme requires only small amplitude cats and remains significantly more resource efficient than dual-rail schemes, needing around 12 cats per gate on average. Of course this assumes that cat states and dual-rail Bell states are of approximately equal difficulty to produce. Progress in producing both these key resource states will be discussed in Section \ref{subsec:SOU}.

\section{Cluster States}
\label{CIL}

So far we have been implicitly discussing quantum computation in terms of the circuit model, familiar from classical computation, in which qubits are prepared in some fiducial state, acted on sequentially to produce logic operations, and then measured in their computational bases to obtain the answer to the computation. Raussendorf and Briegel (2001) have suggested an alternative way of performing quantum computing, distinct from the usual circuit model, called cluster-state quantum computation. It is based on measurement induced quantum evolution and so is sometimes referred to as ``one-way'' quantum computation. In principle, cluster state computation can be carried out on any physical platform. However, the emphasis in many optical quantum computation architectures on measurement induced nonlinearities and off-line resources turns out to be particularly compatible with the cluster state approach.

In Raussendorf and Briegel's protocol, a large entangled state of a particular form---called a cluster state---is constructed first. Quantum computation is then carried out by making a series of measurements in diagonal ($X$), and phase-rotated-diagonal ($Z_{-\theta}XZ_{\theta}$) bases on the cluster state. For example any evolution of a single qubit can 
be simulated by: (i) preparing a string of qubits all in the 
$\ket{+}$ state; (ii) linking each nearest neighbour by CZ 
gates (this forms a linear cluster state); and (iii) measuring the single qubits in the string in sequence.   
The measurement basis chosen for each qubit depends on the single 
qubit unitaries one wishes to simulate and the result of the 
measurement of the preceding qubit. In particular, a qubit measurement in the basis $\ket{R1(\theta)}=\ket{\bf 0}+ e^{\pm i \theta} \ket{\bf 1}$, 
$\ket{R2(\theta)}=
-\ket{\bf 0}+e^{\pm i \theta} \ket{\bf 1}$ simulates 
the unitary evolution $H  Z_\theta$ on the adjoining qubit. The ``$+$'' (``$-$'') sign in the phase factor is chosen if the outcome of the previous measurement was $R1$ ($R2$). The last remaining qubit in the chain is the output of the evolution and can be measured in the computational basis.
An arbitrary single qubit unitary can be simulated using a four-qubit 
cluster state and three measurements.
\begin{figure}
\begin{center}
\includegraphics*[width=9cm]{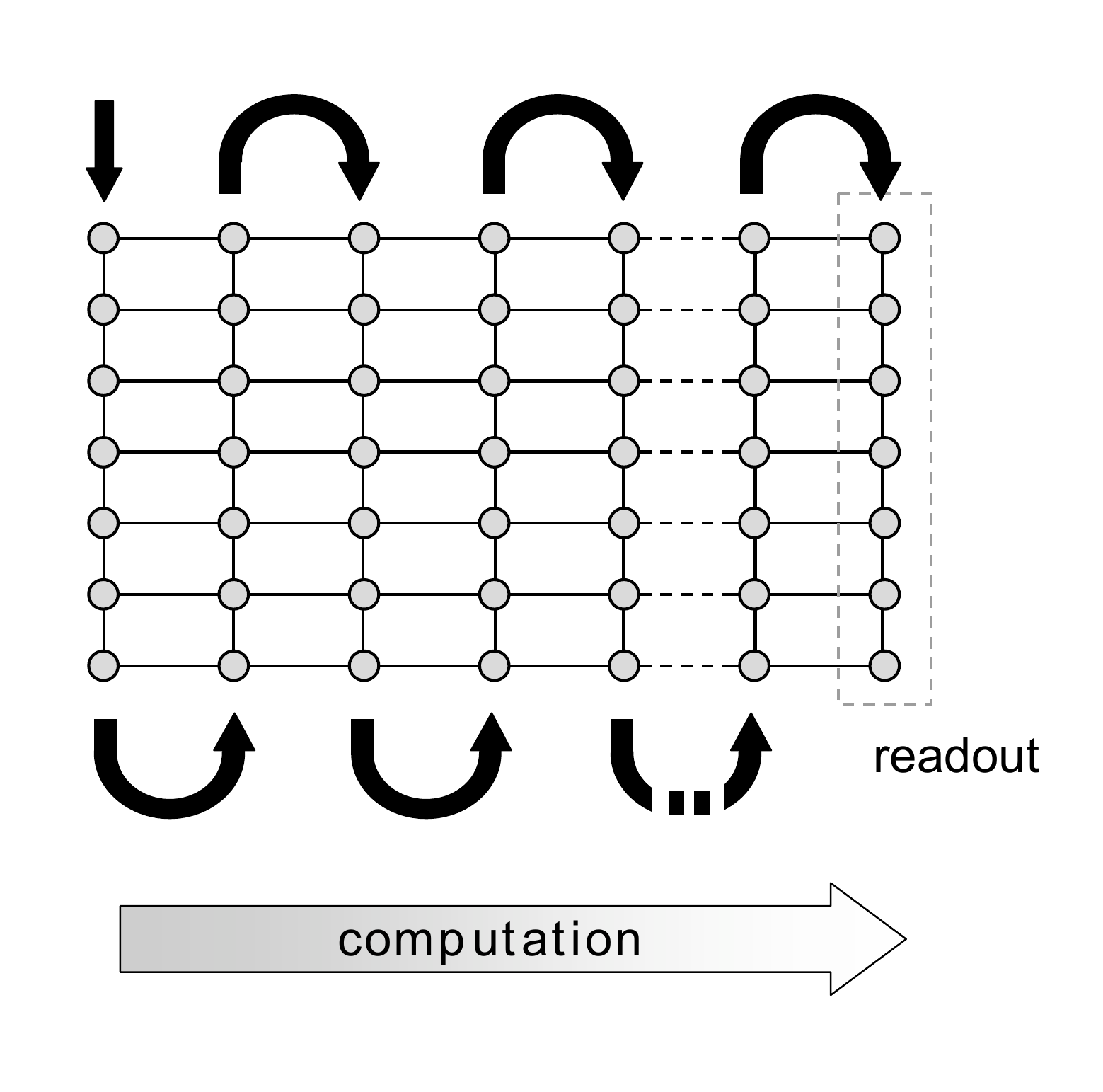}
\caption{ Conceptual diagram representing cluster state quantum computing. Each circle represents a qubit, and the lines connecting them represent entanglement in the form of CZ operations. In this 2D square cluster, the computation starts with a measurement on the first column of qubits (black arrow)---each qubit is measured seperately (i.e.\ the measurement is separable). Based on the outcomes, feedforward is used to change the measurement bases for measurements on the next column. Measurement on the final column yields he outcome of the computation.}
\label{fig8}
\end{center}
\end{figure}

By joining linear chains with CZ gates to create 2-dimensional cluster states, two qubit gates can be built into the cluster, enabling universal quantum computation (see Fig.\ref{fig8}). The first suggestion that measurement-based quantum computation could help to reduce the resources in an optical system was made by Yoran and Resnik (2003). Subsequently Nielsen (2004) adapted the complete cluster state approach to LOQC. He showed that cluster states could be efficiently built up using the KLM teleportation gates. This follows from the fact that the cluster states are able to recover from computational basis measurements in a similar (but not identical) way to that of the parity states. The application of the fusion techniques we described for parity states in section \ref{sec:PS}, that were in fact initially developed by Browne and Rudolph (2005) for cluster state production, further reduces the resource overhead. In this approach ``mini-cluster'' states are built up non-deterministically and then fused on to the main cluster in a similar way to that already described for parity states. This is perhaps the most efficient of the linear optical dual-rail schemes, requiring approximately 60 Bell pairs per two-qubit gate, though the exact meaning of "per gate" in the cluster state paradigm is not as obvious as in circuit models. Work has also been done on cluster state construction methods based on percolation approaches which can dramatically reduce the amount of conditional optical operations required (Kieling, Rudolph and Eisert 2007). In-principle optical demonstrations of one-way quantum computation using dual-rail encoding have now been achieved as will be discussed in section \ref{subsec:LOparticle}.

\section{Fault Tolerance}
\label{sec:FT}

The discussions up to this point have only considered errors that occur due to the physics of the fundamental interactions, such as teleportation gate failures. However, in any realistic implementation there will be additional experimental imperfections in the devices that may lead to additional problems. Typically, such non-ideal interactions will lead to random errors being introduced. Even if these errors are small, when large scale quantum processing is considered we have to worry about their propagation during gate operations. If uncorrected, such errors would grow uncontrollably and make the computation useless. The answer to this problem is fault tolerant error correction (Shor 1995; Steane 1996). In the following, after briefly reviewing the basic principles of error correction, we will look at particular schemes that have been developed and evaluated for several of the optical protocols we have discussed. 

The idea of error correction is self-explanatory, though the description of its application to quantum systems requires some care. Classically we might consider using a redundancy code such that (for example) $0 \to 0,0,0$ and $1 \to 1,1,1$. If a bit flip occurs on one of the bits we might end up with $0,1,0$ or $1,0,1$, but we can recover the original bit value by taking a majority vote. At first it may seem that such a code cannot be used for quantum mechanical systems because: (i) the no-cloning theorem (Wootters and Zurek 1982) means we can not make copies of an unknown qubit; and (ii) taking the majority vote is a measurement that will collapse our quantum superposition. It turns out however that a quantum analog is possible. An example of a quantum redundant encoding is $\alpha \lket{0} + \beta \lket{1} \to \alpha \lket{000} + \beta \lket{111}$ where we have created an entangled state rather than copies. It is then possible, using two CNOT gates and two ancillas, to identify an error without collapsing the state, by reading out the parity of pairs of qubits. For example a bit-flip error might result in the state $\alpha \lket{001} + \beta \lket{110}$. The parity of the first two qubits will be zero whilst the parity of the second two qubits will be one, thus unambiguously identifying that an error has occured on the last qubit. Because we are measuring the parity, not the qubit value, the superposition is not collapsed. Such codes can be expanded to cope with the possibility of more than one error occurring between correction attempts and to cope with multiple types of errors. Of course the CNOT gates being used to detect and correct the errors may themselves be faulty. An error correction code is said to be {\it fault tolerant} if error propagation can be prevented even if the components used to do the error correction introduce errors themselves. Typically this is only possible if the error rate per operation is below some level known as the {\it fault tolerant threshold}. 

\subsection{Loss Tolerance}

Presently the dominant source of errors in optical quantum processing is photon loss---in components, detectors and sources. As a result, initial codes for error correction in optical quantum computing were aimed specifically against loss. KLM estimated a threshold of about 1\% for {\it loss tolerance}, i.e. fault tolerance where the only error considered is loss. Remaining with the original KLM gate approach, Silva, Roetteler and Zalka (2005) were able to show that the loss threshold might lie as high as 11\%. Using the parity state approach and assuming that sources and detectors each had an equal loss of $x \%$, Ralph, Hayes and Gilchrist (2005) numerically obtained a loss threshold of $x=17\%$. A roughly equivalent value was obtained by Varnava, Browne and Rudolph (2008) for loss tolerance of cluster states. 

Readers with some knowledge of fault tolerant codes will notice that these threshold values are very high. The reason for this is that all the codes so far mentioned are dual-rail codes for which loss constitutes a {\it locatable} error. That is, as dual-rail codes are particle-like codes, and loss destroys particles, so loss acts to remove the qubit from the qubit space. Finding out if all your qubits (photons) are present can be achieved without encoding via a quantum non-demolition measurement (QND) (Gleyzes, Kuhr, Guerlin, Bernu, Deleglise, Hoff, Brune, Raimond and Haroche 2007). Thus an error---a photon missing---is locatable without encoding. In contrast, a bit flip leaves the qubit in the qubit space and can only be found using codes such as the one discussed at the beginning of this section. Such errors are described as being {\it unlocatable}. Coding is still required in order to recover from locatable errors, but the easier detection of such errors makes thresholds considerably higher than for unlocatable errors. For single-rail codes, loss tends to produce unlocatable errors. As a result, targeting loss for single-rail systems holds no particular advantage.

Unfortunately, the loss only codes so far described have been shown to be incompatible with more general codes. Specifically, these loss codes tend to amplify unlocatable errors such as bit flips, strongly decreasing the effective thresholds for their correction (Rohde, Ralph and Munro 2007) and hence making them impractical in the presence of even very small rates of unlocatable errors. Although it can be shown that there is no fundamental reason why codes that optimally correct locatable errors must increase the rate of unlocatable errors (Haselgrove and Rohde 2008), presently no explicit examples of such codes have been demonstrated. Instead, codes have been developed that simultaneously correct both types of error, though with loss thresholds significantly lower than for the loss-only codes. In the following we will discuss such a code, developed specifically for optical systems.

\subsection{Telecorrection}

Telecorrection is a modified version of the Steane error correction protocol (Steane 1996) developed by Dawson, Haselgrove and Nielsen (2005). Although developed in the context of dual-rail optical cluster states (see section \ref{CIL}), the code is well suited to optical schemes in general and has been applied to several different protocols, enabling a consistent comparison between them.
\begin{figure}
\begin{center}
\includegraphics*[width=8cm]{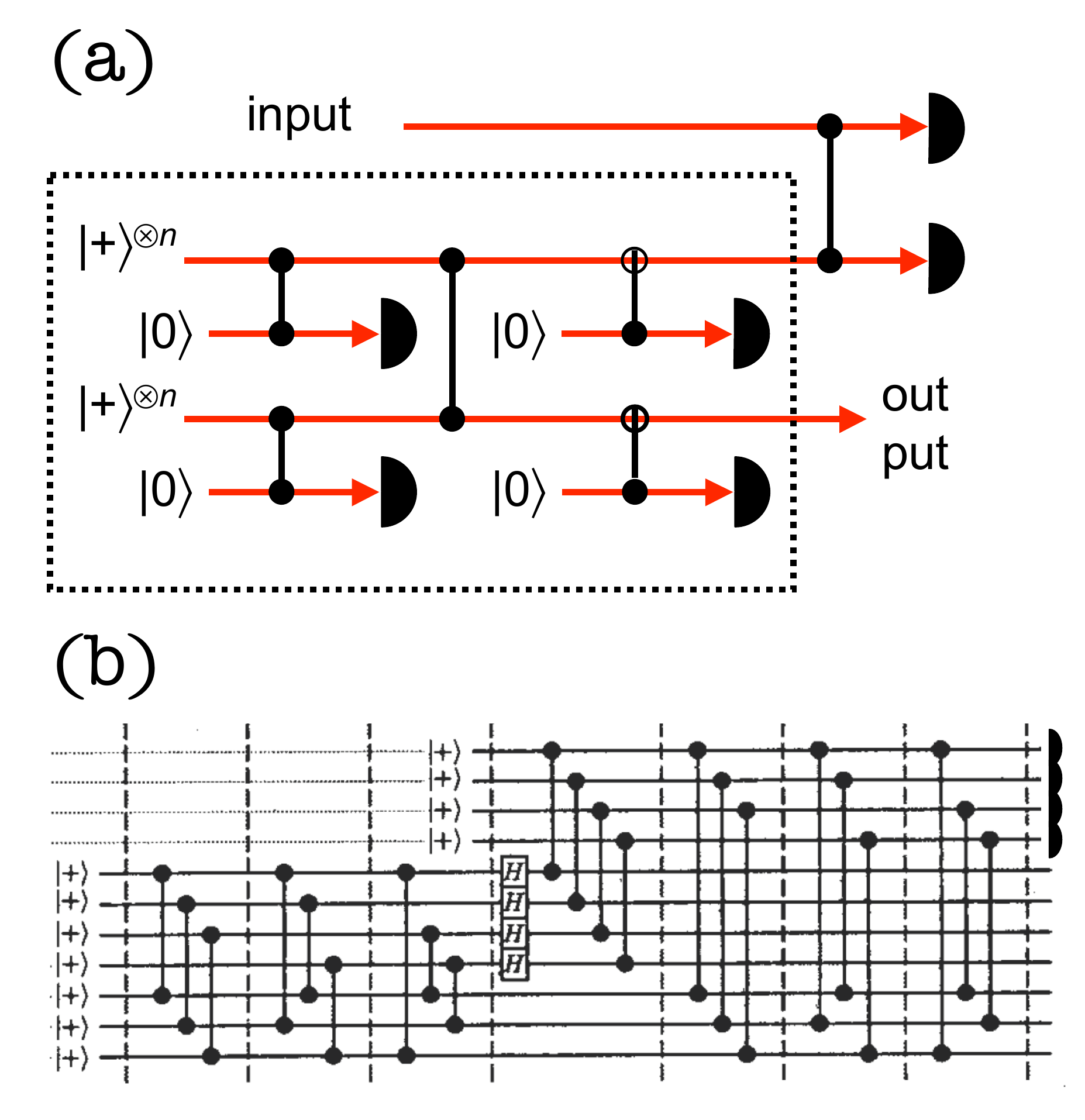}
\caption{ Schematic of (a) telecorrector circuit at the level of Steane code logical qubits and (b) ancilla circuit for fault-tolerantly producing the $|\bf 0 \rangle$ ancilla state from unencoded qubits.}
\label{fig9}
\end{center}
\end{figure}

Logical qubits are encoded across seven physical qubits according to the Steane code. A feature of this code is that logical Clifford gate operations can be implemented by applying the desired gate at the physical qubit level on (or between, for two-qubit gates) the individual physical qubits making up the Steane logical qubit(s). The Hadamard, Pauli and CZ gates are examples of Clifford gates. In order to implement a universal set of gates on the logical qubits, at least one non-Clifford gate will be required (e.g. $Z_{\theta}$). Such logical non-Clifford gates are more complicated to apply, requiring teleportation with a specially prepared resource state.

\begin{figure}
\begin{center}
\includegraphics*[width=9cm]{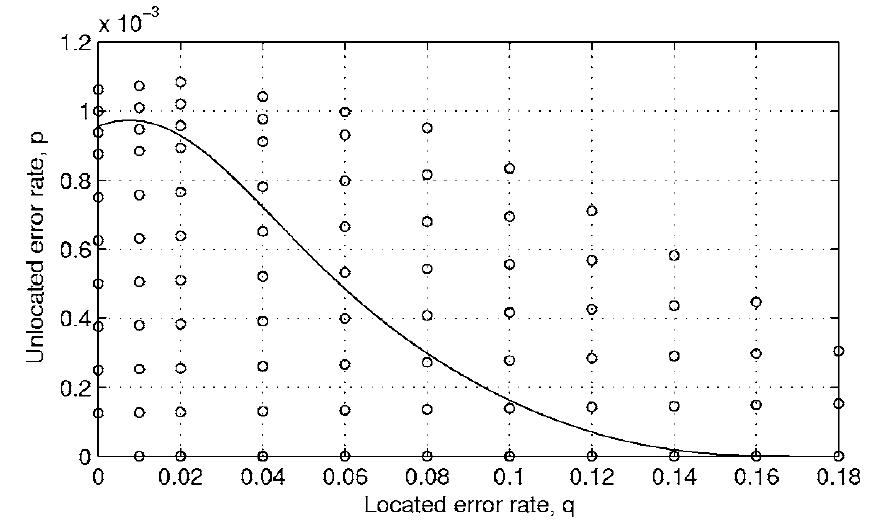}
\caption{ Threshold graph for 7-qubit Steane code based telecorrection, plotted as trade-off between locatable and unlocatable errors. Fault tolerant operation is possible for parameters in the region below the solid line. Circles indicate the parameter values for which the simulations were run. Reproduced from Dawson, Haselgrove and Nielsen (2006).}
\label{fig10}
\end{center}
\end{figure}

Rounds of error correction are also applied via teleportation, where now the resource state is the telecorrector state.  The logical circuit for creating the telecorrector state is depicted inside the dotted box of Fig.\ref{fig9}(a) with the physical qubit level circuit for creating the Steane code logical zero states, that are fed into the logical circuit, depicted in Fig.\ref{fig9}(b). Notice that there are several places in the preparation procedure for the telecorrector state at which measurements are made. These correspond to syndrome measurements in the original Steane protocol (Steane 1996). These measurements are made in the process of creating the telecorrector state, prior to its interaction with the logical qubits. If unwanted measurement outcomes occur, the telecorrector state can be rejected and a new one prepared without affecting the logical qubits. The logical circuit for implementing the teleportation is shown in Fig.\ref{fig9}(a). Numerical simulations show (Dawson, Haselgrove and Nielsen 2006) that if the physical error rates of the components used to construct the telecorrector state are below a certain threshold, then the teleported logical qubit will have a smaller error rate than the input qubit, thus implementing a round of error correction. The threshold level depends on the relative distribution of errors between locatable and unlocatable, with higher thresholds obtained when the errors are predominately locatable. The trade-off is shown in Fig.\ref{fig10} for a generic system.

\subsection{Thresholds and Resource Counts for Optical Schemes}

Converting the generic error rate thresholds shown in Fig.\ref{fig10} into threshold levels for actual physical parameters requires the creation of physical models for particular gate implementations and their analysis with respect to error rates given particular levels of imperfection. Once such models are in place, it is possible to estimate the number of resources required to create the telecorrector state and hence the resources required for one round of error correction. 

This process has now been carried out (at least as a function of loss) for four different optical schemes, allowing a direct comparison between them. 

{\it LOQC Cluster States}: Dawson, Haselgrove and Nielsen (2006) applied their telecorrector code to the linear optical cluster state protocol developed by Nielsen (2004) and Browne and Rudolph (2005). They obtained numerical thresholds as a function of the physical loss rates and depolarization rates of their dual-rail photons and estimated the number of operations and hence the number of photonic Bell pairs required to implement a round of error correction. As well as results specific to optical cluster state, results for generic codes were also presented that could be used to estimate thresholds for more general platforms.

{\it Non-linear Zeno Gates}: Leung and Ralph (2007) used the generic telecorrector code (Dawson, Haselgrove and Nielsen 2006) to estimate thresholds for a modified version of the nonlinear Zeno protocol developed by Franson, Pittman and Jacobs (2003). The thresholds were estimated as a function of the physical loss rates in the nonlinear medium and the detectors and the level of mode matching of the optics. Resources can also be estimated from the number of operations in the generic code.

{\it CSQC}: Lund, Ralph and Haselgrove (2008) generalized the CSQC protocol developed by Ralph, Gilchrist, Munro, Milburn and Glancy (2003) and applied the telecorrector code. They obtained numerical thresholds as a function of the physical loss rates of the source, detectors and memory. They obtained the number of operations required and hence the average number of cat state resources consumed in one round of error correction.

{\it LOQC Parity States}: Hayes, Haselgrove, Gilchrist and Ralph (2009) applied the telecorrector code to the LOQC parity state protocol (Hayes, Gilchrist and Ralph 2005; Gilchrist, Hayes and Ralph 2007). They obtained numerical thresholds as a function of the physical loss rates and depolarization rates of their dual-rail photons. They obtained the number of operations required and hence the average number of Bell state resources consumed in one round of error correction.
\begin{figure}
\begin{center}
\includegraphics*[width=9cm]{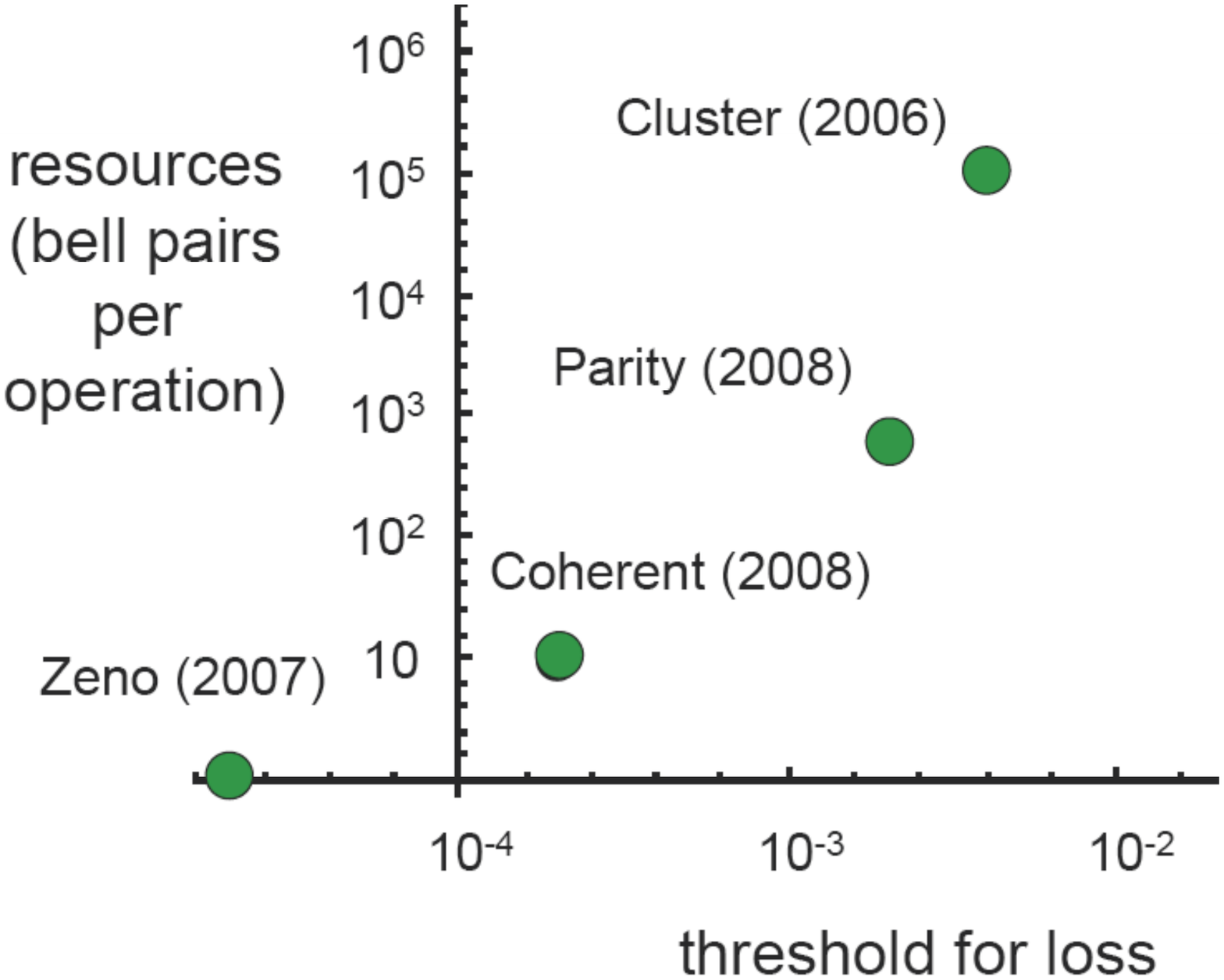}
\caption{ Threshold versus resources for various optical implementations of the telecorrector-Steane code. The code is fully fault tolerant and thus can correct all noise types  however, for the purpose of this graph, it is assumed that loss is the dominant noise type and the threshold with respect to it is plotted. Resources are the number of Bell pairs (cat states for the coherent scheme) needed per physical operation when running the fault tolerant code (typically one round of 1st level error correction requires 1,000 operations). The different points are based on: Cluster, Dawson Haselgrove, Nielsen (2006); Zeno, Leung and Ralph (2007); Coherent, Lund, Ralph, and Haselgrove (2008); and Parity, Hayes, Haselgrove, Gilchrist and Ralph, (2009). }
\label{fig11}
\end{center}
\end{figure}

In Fig.\ref{fig11} we show a comparison of these results, showing resources versus threshold (assuming the dominant error source is loss) for the four protcols. An interesting trade-off between threshold and resources emerges from the plot. Choosing the best architecture then depends on whether the technological bottlenecks lying ahead favour higher thresholds or lower resource requirements. Based on the present analysis the approaches offering the best compromise for medium scale quantum computing (100s of logical operations) appear to be the CSQC scheme and the Parity State scheme. For large-scale quantum computing, the overheads for all implementations, optical or otherwise, still remain very high.

\section{Experimental Demonstrations}

% Geoff's new section

There have been a wide range of experimental demonstrations of optical quantum logic, applications and resource generation. The majority of optical experiments have used linear optics methods, with either the KLM approach or the use of exisiting entanglement (e.g. cluster states) to add the required nonlinearity. These linear optical schemes can be divided into those where the qubits are encoded in \textit{particles}, and those with a \textit{field} encoding.

\subsection{Linear optics, particle encodings}
\label{subsec:LOparticle}

In the particle encoding, a qubit is typically encoded onto the state of a single photon. This is a dual-rail qubit---the logical state corresponds to the occupation of one or another mode by the photon. Although path, frequency and other encodings have been used, the most common approach is to represent the logical states by orthogonal polarization states of the photon's electric field. Polarization is especially practical because logical modes are both degenerate and co-propagating. This means that most sources of phase noise in an optical circuit (e.g. vibrations of components) are common-mode, and the qubit is robust against dephasing. %Indeed all optical qubits are robust against external sources of noise in free space (the state of the free-space bath is vacuum).
 It has been noted that polarization is preserved over long distances in space---light from the Crab Nebula, some 6500 light years away, is still significantly polarized in some regions of the spectrum (Oort and Walraven (1956)). In dielectric media such as glass optical fibers, the polarization state is subject to changes in birefringence (or refractive index and dispersion for other encodings). As the interactions with these media are essentially non-resonant at optical frequencies, changes due to locally fluctuating electric and magnetic fields are small and the predominant effect is from thermal changes in the material, something that happens on a time scale much longer than the time taken for the qubits to propagate through the device. Therefore, even in dielectric media, phase noise on optical (and especially common-mode polarization) qubits are generally negligible for a single shot. 

Additionally, linear optical transformations on a single qubit are generally easy and can be performed with high precision (Peters, Altepeter, Jeffrey, Branning, and Kwiat (2003)). Polarization transformations, for example, can be achieved with fidelities of $> 99\%$, using waveplates. Transformations in path encodings may be achieved using a combination of beam splitters and phase shifters, and in some cases using the extra polarization degree of freedom can assist in implementing operations with high fidelity. For example, high visibility interferences of over $99\%$ can be achieved in polarizing interferometers.  

The standard polarization encoding requires, not surprisingly, the ability to make single photon states of definite polarization. Since the late 1980's, the solution of choice has been parametric down conversion in a $\chi^{(2)}$ medium (Ghosh and Mandel (1987)).  Weak degenerate parametric down conversion results in the spontaneous conversion of single pump photons at the harmonic frequency into pairs of photons at the fundamental. If the down conversion is spatially non-degenerate then, in the Schr\"odinger picture, initial vacuum inputs are transformed according to
\beq
|0 \rangle_{a} |0 \rangle_{b} \to 
(|0 \rangle_{a} |0 \rangle_{b} + \chi' |1 \rangle_{a} |1 \rangle_{b}
+ \chi'^{2} |2 \rangle_{a} |2 \rangle_{b} + \ldots)
\eeq
where $\chi'$ is an effective non-linear interaction strength, proportional to the pump power. If we now allow $\chi'$ to be very small (which is not hard to arrange experimentally) then the 
state produced is given to an excellent approximation by
\beq
|\psi \rangle_{ab} =
|0 \rangle_{a} |0 \rangle_{b} + \chi' |1 \rangle_{a} |1 \rangle_{b}.
\label{EQ:2PHOTON}
\eeq
%
%In contrast to a coherent state, the state in equ.(\ref{EQ:2PHOTON}) has
%only the desired two photon term to first order in $\chi'$.
 If we
postselect only those events from the detection record in which 2
photons are detected ``simultaneously'', or in coincidence
(within some preset time window), then
we will only record the part of the state which is due to the pairs of photons.
Thus by using the combination of parametric down-conversion, the 
polarization degree of freedom
and postselection,  we can perform two-qubit experiments. 
Experiments carried out this way are sometimes referred to as
coincidence experiments and we will discuss various examples in later sections.
However, note that this source is still spontaneous, i.e.\
successful events are rare, random and
we do not know if they have occurred until after the fact. Although 3- and 4-qubit experiments have been achieved by a simple generalization of the techniques just outlined, the cost is an exponential drop in the probability of success.
%, and so direct scaling to much higher qubit numbers is impractible. 
Therefore, experiments carried out in coincidence can demonstrate the basic physics of particular systems, and prove the priniciples behind optical quantum information techniques, but are not intrinsically scaleable to large-scale quantum information processing.
Progress in
producing sources without this drawback is discussed in the next
section.

%As mentioned in section \ref{ggg}, 
The strength of linear optics in implementing single qubit gates is balanced by the main challenge of linear optical quantum computing, which is implementing two-qubit entangling gates. The key to implementing these with linear optics is the measurement-induced nonlinearity of the KLM and related schemes, or the pre-existing entanglement of the cluster states schemes \footnote{Measurement-induced nonlinearity is required in these schemes too, for building the cluster initially}. We will now consider some of the experimental demonstrations of two-qubit gates using linear optics in a photonic encoding, as well as some of the small-scale applications of these gates.

Ralph, Langford, Bell and White (2002), and independently Hofmann and Takeuchi (2002), proposed a simplification of the KLM scheme in which simple two-photon gates could be implemented nondeterministically with two photonic qubits and no ancillas. This scheme provided for a powerful proof-of-principle of measurement-induced nonlinearity gates without the need for complicated architectural overheads. 
%However, for these gates to be made scaleable, they need extra ancilla-based circuitry (QND detectors and teleporters), because they also necessarily destroy the signal and control photons as part of the measurement-induced nonlinearity \footnote{The group of Andrew White has therefore called these ``external ancilla'' gates.}. 
Although, being coincidence gates they are not immediately scaleable to large systems, nevertheless in many cases it is possible to chain logic gates together---prior to photon detection---making this scheme suitable for small circuit demonstrations. 
\begin{figure}
\begin{center}
\includegraphics*[width=9cm]{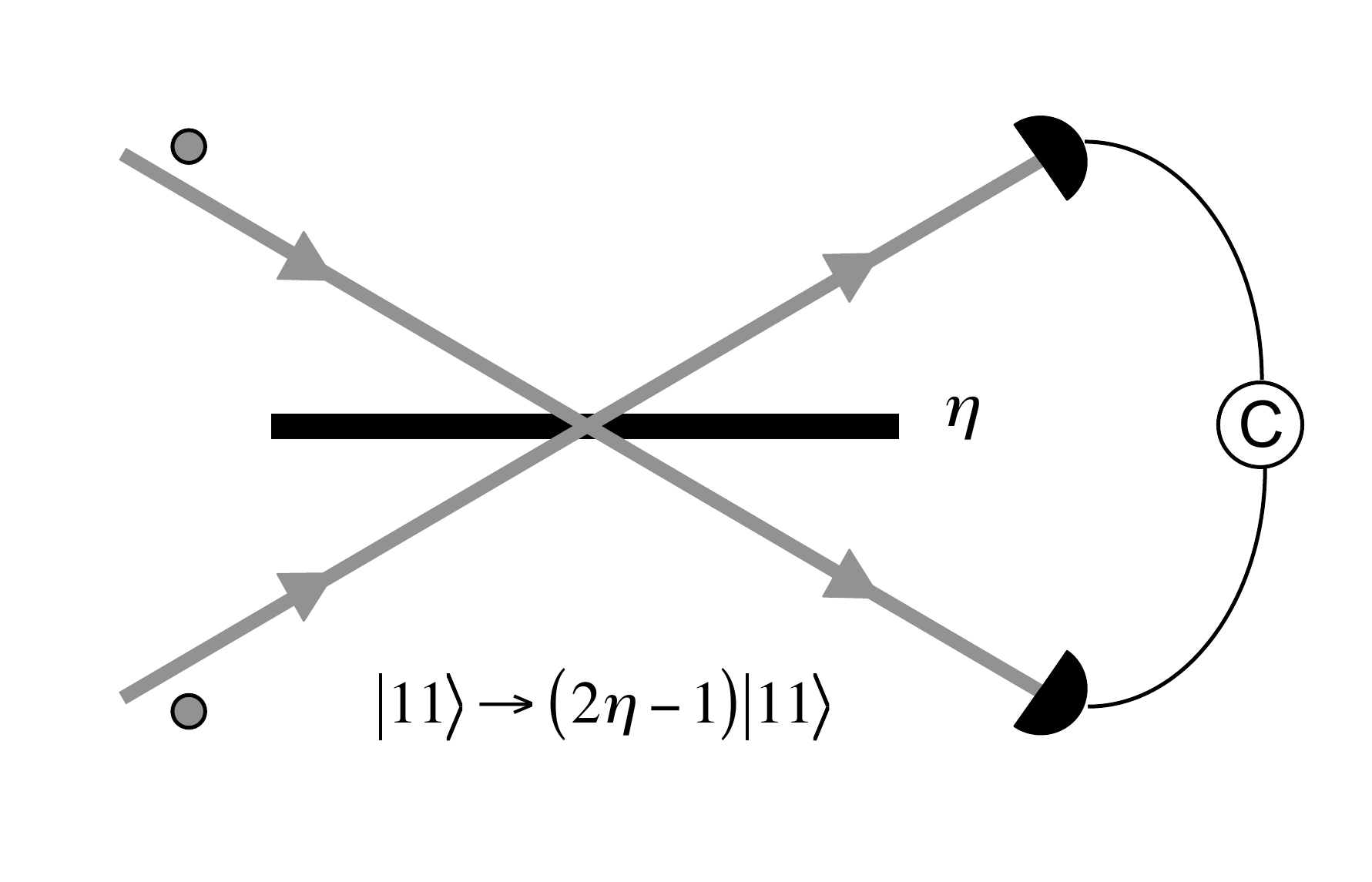}
\caption{ Partial Hong-Ou-Mandel interference between two single photons on a beam splitter of reflectivity $\eta$. The formula shows the state transformation in the $\ket{11}$ subspace, i.e.\ conditional on getting one photon at each of the outputs. The probability of this event occurring decreases as $\eta \rightarrow \tfrac{1}{2}$. }
\label{fig12}
\end{center}
\end{figure}

\begin{figure}
\begin{center}
\includegraphics*[width=9cm]{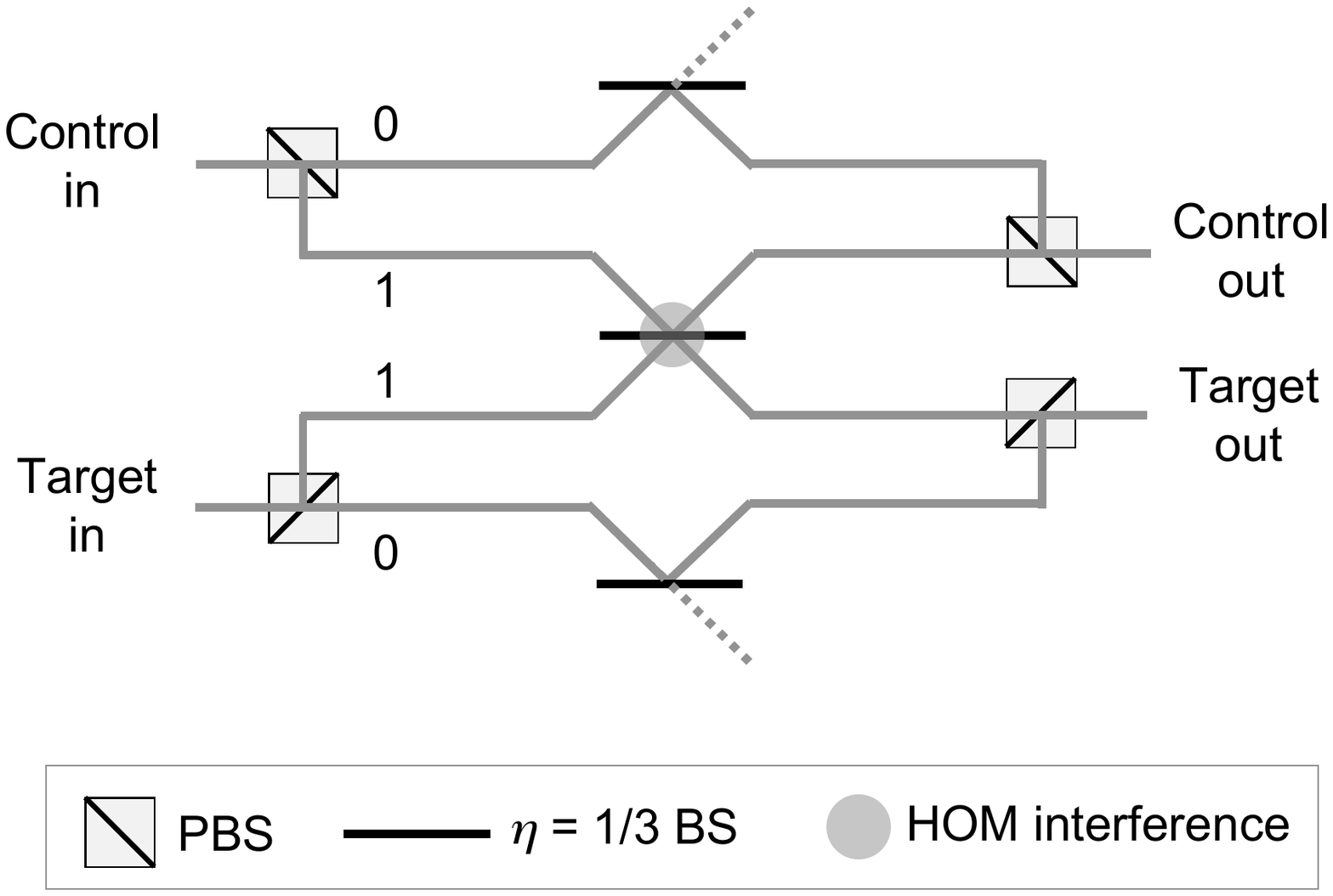}
\caption{ Conceptual diagram of a partial Hong-Ou-Mandel interference embedded in a polarizing interferometer. The phase flip only occurs on the logical $\ket{\bold1\bold1}$ term, leading to the CZ action of the circuit, conditional upon getting one photon in each of the control and target inputs. The beam splitters in the top and bottom arms balance the probability amplitudes for all logical basis states.}
\label{fig13}
\end{center}
\end{figure}

The basic operation of the scheme is as follows: a partial Hong-Ou-Mandel interference (Hong, Ou and Mandel (1987)), on a beam splitter of reflectivity $\eta$ can be configured such that the output state is a superposition of three possibilities. One of these possibilities is a phase flip of the $\ket{11}$ number state (Fig. \ref{fig12}). The remaining two possibilities correspond to final states with two photons in one mode---these lie outide the qubit Hilbert space we are interested in. If the number state $\ket{11}$ also represents the \textit{logical} $\ket{\bold1\bold 1}$ state, then the device can implement the nonlinearity required for an entangling two-qubit gate. Specifically, a beam splitter with an intensity reflectivity of $\eta=1/3$ acts upon a state with one photon in each mode (see Fig. \ref{fig12}) to produce the logical state transformation
$\ket{\bold1\bold 1}\rightarrow -\sqrt{\tfrac{1}{3}} \ket{\bold1\bold 1}$, conditional on the output containing one photon per mode (Ralph, Langford, Bell and White (2002)). Embedding this effect in a larger polarization interferometer (Fig. \ref{fig13}) produces the full controlled-SIGN (CS) logic
\begin{eqnarray}
\nonumber \ket{HH}&\rightarrow&\sqrt{\tfrac{1}{3}}\ket{HH}\\
\nonumber \ket{HV}&\rightarrow&\sqrt{\tfrac{1}{3}}\ket{HV}\\
\nonumber \ket{VH}&\rightarrow&\sqrt{\tfrac{1}{3}}\ket{VH}\\
\nonumber \ket{VV}&\rightarrow&-\sqrt{\tfrac{1}{3}}\ket{VV}
\end{eqnarray}
where only outputs in the two-qubit Hilbert space are considered---this is the effect of the measuement-induced nonlinearity. (We have used the notation $\ket{H(V)} \equiv \ket{\bold 0(\bold 1)}$ for a horizontally (vertically) polarized photon). The cost of this truncation of Hilbert space is a non-unit probability of operation; $P=1/9$ for this particular scheme.

An optical circuit implementing this scheme was constructed usng passively stable classical interferomters to ensure a robust experimental design (O'Brien, Pryde, White, Ralph, Branning (2003)), and achieved a high gate fidelity (O'Brien, Pryde, Gilchrist, James, Langford, Ralph, and White (2004)). It is evident that the key element of this scheme is high-quality nonclassical and classical interferences between modes of the optical circuit. Using SPDC sources and free space optics, nonclassical interences of $\approx 96\%$ (relative to the maximum achievable for the given beam splitter reflectivity) are typically achieveable, with classical interferences of $> 98\%$ common. In the original demonstrations (O'Brien, Pryde, White, Ralph, Branning (2003); O'Brien, Pryde, Gilchrist, James, Langford, Ralph, and White (2004)), slightly lower interference visibilities led to gate fidelities of $0.89-0.95$. Subsequent improvements in circuit design (described below) have led to entangling gate fidelities of up to 0.98 (Lanyon, Barbieri, Almeida, Jennewein, Ralph, Resch, Pryde, O'Brien, Gilchrist, and White (2009)).

How are these gate fidelities determined? The technique of quantum process tomography was developed in order to characterize quantum processes such as quantum gates (Chuang and Nielsen (1997); Poyatos, Cirac, and Zoller (1997))
% O'Brien, Pryde, Gilchrist, James, Langford, Ralph, and White [2004)). 
The basic idea of process tomography is to prepare, one at a time, a range of input states that span the space of allowed density matrices, and to perform measurements spanning the measurement space on identical copies of each input state. In so doing, information is obtained about how all possible input states map onto all possible output states. Full tomography provides a complete description of the process in the computational Hilbert space, so not only is it useful for determining gate fidelities 
%\footnote{Actually, it is possible to determine the process fidelity with a much smaller set of measurements than required to perform full tomography (O'Brien, Pryde, Gilchrist, James, Langford, Ralph, and White [2004]; White, Gilchrist, Pryde, O'Brien, Bremner, and Langford [2007])}
but it is also useful for determining the errors that are introduced by imperfect gate realizations. However, while the information obtained from tomography provides a mathematical description of the errors in the process, parallel modelling is generally required to obtain a physical insight into the errors in the quantum circuit (Rohde, Pryde, O'Brien, and Ralph (2005)).

The fidelity of these linear optics gates has been improved by simplifications of the optical ciruits and by better control of the optical modes. For example, the classical interferences that are required in a circuit like that in Fig. \ref{fig13} can be removed by using partially-polarizing beam splitters (Langford, Weinhold, Prevedel, Resch, Gilchrist, O'Brien, Pryde, and White (2005); Kiesel, Schmid, Weber, Ursin, and Weinfurter (2005); Okamoto, Hofmann, Takeuchi, and Sasaki (2005)). In this case, the polarization modes need not be separated for the nonclassical interference to work, because the beam splitter that enables the HOM works on only one polarization state. Not only does this allow the classical interference visibility to approach unity (relative to the maximum for the beam splitter reflectivity), it also significantly simplifies the optical setup. The tradeoff is that the HOM condition is now set by the physical reflectivity of the beam splitter, rather than being tuneable by a wave plate setting as it is in polarizing interferometers---it is therefore susceptible to manufacturing errors. However, it has been noted that the gate fidelity is relatively insensitive to imperfections in the BS reflectivity (Ralph, Langford, Bell and White (2002)). Additionally, the use of guided-mode optics at the input and output of the gates has improved nonclassical mode matching, a theme that will be explored further in subsection \ref{subsec:IntegOpt} below. 

Internal ancilla gates, where additional photons are used in the gate operation itself (more in line with the original KLM suggestion) have also been demonstrated. 
%These can also be connected back to simplifications of the original KLM scheme \cite{RAL01}. 
The most popular of these is the design of Pittman, Jacobs and Franson (2001) (also Franson, Donegan, Fitch, Jacobs, and Pittman (2002)), which is based on the teleportation-based quantum computing scheme of Gottesman and Chuang (1999) (see section \ref{sec:KLM}), and which uses a maximally-entangled photon pair as the ancilla resource. By harnessing the pre-exisiting entanglement in the ancilla, the gate is able to achieve a higher success probability of 1/4, with the trade-off of requiring both single photons and an entangled pair as inputs. In priniciple, the internal ancillas allow the gate to operate in a heralded fashion without detection of photons in the output modes, by the registration of particular photon number detections in the ancilla output ports. In practice, this requires high-efficiency number-resolving photon counters (see subsection \ref{subsec:Dets}), which are not yet available.

A simplified version of this gate, with the entangled ancilla replaced by a single photon, was demonstrated by Pittman, Fitch, Jacobs, and Franson (2003). Versions of the gate with entangled ancillas were demonstrated by Gasparoni, Pan, Walther, Rudolph, and Zeilinger (2004) and Zhao, Zhang, Chen, Zhang, Du, Yang, and Pan (2005). Although process tomography was not used to quantify the performance of these gates, they demonstrated logical basis fidelities of $\sim 80\%$. Later, a version of the gate was demonstrated that replaced the entangled ancilla with a nondeterministic Bell measurement (Bao, Chen, Zhang, Yang, Zhang, Yang, and Pan (2007))---this gate used the method of Hofmann (2005) to bound the gate fidelity in the range $0.78-0.88$.

As noted in section \ref{sec:LOQC}, one of the challenges of the KLM approach that applies to all of these gates is the unfavourable overheads in implementing deterministic gates from non-deterministic ones, even though the scheme is computationally ``efficient''. Several approaches have been described to help overcome this problem, amongst them the cluster state and parity state methods. 

In the cluster state method, a highly entangled state of many qubits is built up as a resource. Single qubit measurements on the state, coupled with feedforward operations onto the remaining qubits, suffice to perform universal quantum computing (see section \ref{CIL}). Demonstrations of the cluster state method have been performed for cluster states of up to six photons (Walther, Resch, Rudolph, Schenck, Weinfurter, Vedral, Aspelmeyer, and Zeilinger (2005); Prevedel, Tame, Stefanov, Paternostro, Kim, and Zeilinger (2007); Lu, Zhou, Guehne, Gao, Zhang, Yuan, Goebel, Yang, and Pan (2007); Chen, Li, Qiang, Chen, Goebel, Chen, Mair, and Pan (2007); Tokunaga, Kuwashiro, Yamamoto, Koashi, and Imoto (2008); Kiesel, Schmid, Weber, Toth, Guhne, Ursin, and Weinfurter (2005)), and also with cluster states of multiple qubits using more than one qubit per photon (by using multiple degrees of freedom on a single photon, for example---Vallone, Pomarico, Mataloni, De Martini, and Berardi (2007)). Although this latter approach is not scaleable, it may provide a platform for small-scale tests of information protocols beyond what can be achieved with single-qubit-per-photon encondings alone. As well as demonstrating the measurement operations on cluster states, feedforward operations have also been demonstrated (Prevedel, Walther, Tiefenbacher, Bohl, Kaltenbaek, Jennewein, and Zeilinger (2007); Vallone, Pomarico, De Martini, and Mataloni (2008)), with fidelities of $>95\%$ and feedforward operation times of order 150 ns---mostly limited by the speed of the detection electronics and the switching electronics, which has to switch the $\sim$ kV supply for Pockels cells. In future generations, one imagines using fiber or waveguide electro-optic modulators which can have sub-nanosecond switching times.

Experimentally demonstrating that the measurement and feedforward operations actually work is an important proof-of-principle verification of the scheme. Equally important is demonstrating the manufacture of cluster states. The most efficient scheme presently known is that of Browne and Rudolph (2005), which uses polarizing beam splitters and photon detection to forge links between entangled states of photons. Once these links have been established, the resulting state is larger than any of the input states. Present demonstrations of cluster state generation have not used this scheme in its fullest sense, instead relying on postselection of certain measurement results in quantum interference between entangled pairs (e.g. Walther, Resch, Rudolph, Schenck, Weinfurter, Vedral, Aspelmeyer, and Zeilinger (2005); Lu, Zhou, Guehne, Gao, Zhang, Yuan, Goebel, Yang, and Pan (2007); Kiesel, Schmid, Weber, Toth, Guhne, Ursin, and Weinfurter (2005)). Because both techniques rely on the same interference effect, the postselected technique is a good representation of the full model for cluster state generation. The standard cluster state growth technique requires the use of high-efficiency sources and detectors, which are presently still under development---see subsections \ref{subsec:Dets}, \ref{subsubsec:BellPS} below. 

Yet another class of linear optics gates uses pre-existing entanglement in a second degree of freedom to power the two-qubit operation. An example of this is the gate of Sanaka, Kawahara, and Kuga (2002), in which pre-existing time-energy entanglement from a spontaneous parametric downconversion source may be converted into polarization entanglement using an optical interferometer. By controlling the initial polarization states of the photons, a controlled-NOT operation between the polarization qubits was performed. This scheme is expandable, in principle, to multi-qubit gates (Gong, Guo, and Ralph 2008) or small scale circuits by generating multi-qubit entanglement in the other degree of freedom. 

\subsection{Circuits and protocols using linear optics gates}
\label{subsec:LOcircuits}

A variety of gates and protocols have been demonstrated using the various linear optics techniques discussed so far. These include: realizations of two-qubit entangling gates such as the CNOT (O'Brien, Pryde, White, Ralph, Branning (2003); Pittman, Fitch, Jacobs, and Franson (2003); Gasparoni, Pan, Walther, Rudolph, and Zeilinger (2004); Zhao, Zhang, Chen, Zhang, Du, Yang, and Pan (2005)) or CZ (Langford, Weinhold, Prevedel, Resch, Gilchrist, O'Brien, Pryde, and White (2005); Kiesel, Schmid, Weber, Ursin, and Weinfurter (2005); Okamoto, Hofmann, Takeuchi, and Sasaki (2005)); basic error encoding and detection (Prevedel, Tame, Stefanov, Paternostro, Kim, and Zeilinger (2007); Lu, Gao, Zhang, Zhou, Yang, and Pan (2008); O'Brien, Pryde, White, and Ralph (2005); Pittman, Jacobs, and Franson (2005)); realizations of simple quantum algorithms such as Deutsch's algorithm (Tame, Prevedel, Paternostro, Boehi, Kim, and Zeilinger (2007)); Grover's algorithm for small numbers of qubits (Kwiat, Mitchell, Schwindt, and White (2000); Walther, Resch, Rudolph, Schenck, Weinfurter, Vedral, Aspelmeyer, and Zeilinger (2005); Chen, Li, Qiang, Chen, Goebel, Chen, Mair, and Pan (2007)); a compiled version of Shor's algorithm for factoring 15 into its prime factors (Lanyon, Weinhold, Langford, Barbieri, James, Gilchrist, and White (2007); Lu, Browne, Yang, and Pan (2007)); a demonstration of the quantum phase estimation algorithm (QPEA) (Lanyon, Whiteld, Gillet, Goggin, Almeida, Kassal, Biamonte, Mohseni, Powell, Barbieri, Aspuru-Guzik and White (2009)); using polarization qubits, for applications in quantum chemistry (Aspuru-Guzik, Dutoi, Love and Head-Gordon (2005)) \footnote{The QPEA was also modified to perform photonic estimation of a classical optical phase (Higgins, Berry, Bartlett, Wiseman, and Pryde (2007))}; simulations of anyonic statistics and braiding (Lu, Gao, Guehne, Zhou, Chen, and Pan (2009); Pachos, Wieczorek, Schmid, Kiesel, Pohlner, and Weinfurter (2007)); investigations of quantum computing operations with no entanglement (Lanyon, Barbieri, Almeida, and White (2008)); quantum games (Prevedel, Stefanov, Walther, and Zeilinger (2007)); counterfactual quantum computation (Hosten, Rakher, Barreiro, Peters and Kwiat (2006)); demonstration of more complex logic gates (such as the three-qubit Toffoli gate and the controlled-arbitrary-unitary gate---Lanyon, Barbieri, Almeida, Jennewein, Ralph, Resch, Pryde, O'Brien, Gilchrist, and White (2009)); and more.

Additionally, linear optical CNOT gates and cluster states have enabled experimental investigations of fundamental quantum physics (Pryde, O'Brien, White, Ralph, and Wiseman (2005); Pryde, O'Brien, White, and Bartlett (2005); Walther, Aspelmeyer, Resch, and Zeilinger (2005)) and new quantum measurements (including quantum nondemolition measurements  and arbitarary strength measurements on flying qubits (Pryde, O'Brien, White, Bartlett, and Ralph (2004); Ralph, Bartlett, O'Brien, Pryde, and Wiseman (2006)), and entangling measurements (Langford, Weinhold, Prevedel, Resch, Gilchrist, O'Brien, Pryde, and White (2005); Walther and Zeilinger (2005))).

These results have demonstrated that optics is a suitable technology for performing quantum information processing applications. The outstanding challenge is to conquer the various challenges to scalability both in terms of resource usage to overcome non-determinism, and obtaining fault tolerance.

The current line of experimental research is to take a two-pronged approach: increasing the number of qubits and the sophistication of the optical circuits and algorithms; and improving the performance of auxilliary components (see section \ref{sec:AuxComp}). As the performance of these components improves, their incorporation into LOQC circuits will increase.

\subsection{Linear optics, field encodings}
\label{subsec:LOfields}

Our discussion so far has only considered \textit{photonic} LOQC. Although not yet as advanced, development has also been progressing in the area of linear optical quantum computing using field encodings. In particular, there has been significant emphasis on \textit{coherent state quantum computing}, largely focussing on the generation of the CSS (coherent state superposition) resource states.

Several groups have demonstrated close approximations to the CSS (see \ref{subsubsec:CSSsources}), based on subtraction of a photon from a squeezed vacuum state to generate a close approximation of an ``odd cat''---a CSS with only odd photon number terms in the number-state expansion. These types of CSS have been demonstrated for both continuous states of light (Neergaard-Nielsen, Nielsen, Hettich, Molmer, and Polzik (2006); Wakui, Takahashi, Furusawa, and Sasaki (2007)) and for fields in single short (typically $\sim$ picosecond) pulses (Ourjoumtsev, Tualle-Brouri, Laurat, and Grangier (2006)). Using these approaches, fidelities of $\sim 70\%$ (and upwards) with the ideal cat state have been achieved for states
%(Ourjoumtsev, Tualle-Brouri, Laurat, and Grangier (2006); Takahashi, Wakui, Suzuki, Takeoka, Hayasaka, Furusawa, and Sasaki [2008)) 
with an amplitude of $\sim 1$. Larger, squeezed cat states, with effective amplitudes of $\sim 1.6$ have been created using conditional homodyne detection of a two-photon state (Ourjoumtsev, Jeong, Tualle-Brouri and Grangier 2007). Other interesting states such as ``even cats'' have also been demonstrated (Takahashi, Wakui, Suzuki, Takeoka, Hayasaka, Furusawa, and Sasaki (2008)). Although actual gate operations employing these resource states have not yet been performed, preparation of necessary entangled states, such as the Bell-cat state and $\ket{Z_{\theta}}$ has been demonstrated (Ourjoumtsev, Ferreyrol, Tualle-Brouri, and Grangier (2009)).  
 
Cluster-state quantum computing has also been explored in the continuous-variables regime, with preliminary experiments based on use of QND-like beam splitter circuits for building cluster states (Yukawa, Ukai, Loock, and Furusawa (2008); Yoshikawa, Miwa, Huck, Andersen, Loock, and Furusawa (2008)), as well as plans to build frequency-comb cluster states using strongly squeezed light in multiple frequency modes (Pysher, Bloomer, Pfister, Kaleva, Roberts, and Battle (2008); Menicucci, Flammia, and Pfister (2008)). This latter field is in its early days. 

Although perhaps not the presently preferred scheme for optical quantum computing, the ability to make single-rail qubits in the $\ket{\textrm{photon~number}} \in \{\ket{0},\ket{1}\}$ subspace has also been demonstrated. Again, the generation of these field-encoded qubits progresses using a combination of techniques from continuous and discrete quantum optics---squeezing and photon counting (Babichev, Brezger, and Lvovsky (2004); Lvovsky and Mlynek (2002)). However, the full range of quantum operations has not yet been demonstrated with these states.

\subsection{Nonlinear optical quantum computing experiments}
\label{NOQCE}
To date, there have been very few demonstrations of optical quantum computing elements based on nonlinear interactions. Generally speaking, the approach employed for the existing experiments has been either to amplify the nonlinearity associated with a single atom (e.g. using a cavity), or to employ some weak nonlinearity amplified by the use of many atoms.

Perhaps the best-known experiment in the former category is that of Turchette, Hood, Lange, Mabuchi, and Kimble (1995), who used an atom in a cavity to demonstrate a conditional birefringent phase shift of $\sim 16^\circ$ per intracavity photon, with a weak coherent state probe. This experiment was a highly impressive feat of experimental optical quantum science, but it has proved challenging to increase the phase shift to larger values due to practical issues. Recently, the group of Vu\u{c}kovi\'{c} has demonstrated similar cavity QED experiments using the compact configuration of quantum dots in photonic crystal cavities (Englund, Fushman, Faraon, and Vu\u{c}kovi\'{c} (2009)). Using this technique, phase shifts of up to $0.16\pi$ radians have been observed (Fushman, Englund, Faraon, Stoltz, Petroff, and Vu\u{c}kovi\'{c} (2008)) at the single-photon level, also using a weak coherent beam. A host of other experiments have been performed addressing the nonlinearity available in the strong coupling regime. Some of these include: the observation of the photon blockade effect (Birnbaum, Boca, Miller, Boozer, Northup and Kimble (2005)); strong coupling with microtoroid cavities (Aoki, Dayan, Wilcut, Bowen, Parkins, Kippenberg, Vahala and Kimble 2006); two-photon dressed states (Schuster, Kubanek, Fuhrmanek, Puppe, Pinkse, Murr and Rempe (2008)); and strong coupling at microwave frequencies (Brune, Hagley, Dreyer, Maitre, Maali, Wunderlich, Raimond and Haroche 1996; Gleyzes, Kuhr, Guerlin, Bernu, Deleglise, Hoff, Brune, Raimond and Haroche 2007).

%Non-linearities close to those required can be realized in cavity quantum electro-dynamic (QED) situations featuring single emitters in cavities of extremely high finesse and small volume. This occurs in the so-called {\it strong coupling} regime, in which the dipole coupling between the cavity field and the emitter is significantly greater than the relaxation rates of both the cavity and the dipole. Strong coupling has been demonstrated at optical frequencies with single alkali atoms in Fabry Perot (Turchette, Hood,
%Lange, Mabuchi and Kimble 1995; Rempe?) and micro-toroid  cavities (Aoki, Dayan, Wilcut, Bowen, Parkins, Kippenberg, Vahala and Kimble 2006) and at microwave frequencies between Rydberg atoms (Brune, Hagley, Dreyer, Maitre, Maali, Wunderlich, Raimond
 %and Haroche 1996; Gleyzes, Kuhr, Guerlin, Bernu, Deleglise, Hoff, Brune, Raimond and Haroche 2007) .

Another proposed technique for generating large nonlinearities is electromagnetically induced transparency, where the application of a pump field can be used to cause a phase shift on another optical field (Schmidt and Imamoglu (1996)). Although such nonlinear phase shifts have not been demonstrated with single photons, promising low-field demonstrations have been made (Kang and Zhu (2003)). Even weak nonlinearities may be able to help realize quantum computer gates (Nemoto and Munro (2004)).

Recently, other nonlinear ideas have been explored. Franson et al. has shown theoretically that it is possible to use the two-photon absorption in a suitable atomic ensemble to provide a weak, Zeno-like measurement that suppresses the known failure mode of photonic linear optics approaches - moving out of the computational code space and generating states with two photons in the same mode (see section \ref{sec:zeno}). Although logic has not yet been demonstrated using this scheme, several indicative experiments have been performed, including low-intensity two-photon absorption experiments using rubidium vapour and guided-mode optics (Hendrickson, Pittman, and Franson (2009)).

\section{Auxiliary Components}
\label{sec:AuxComp}

\begin{figure}[htb]
\begin{center}
\includegraphics*[width=8cm]{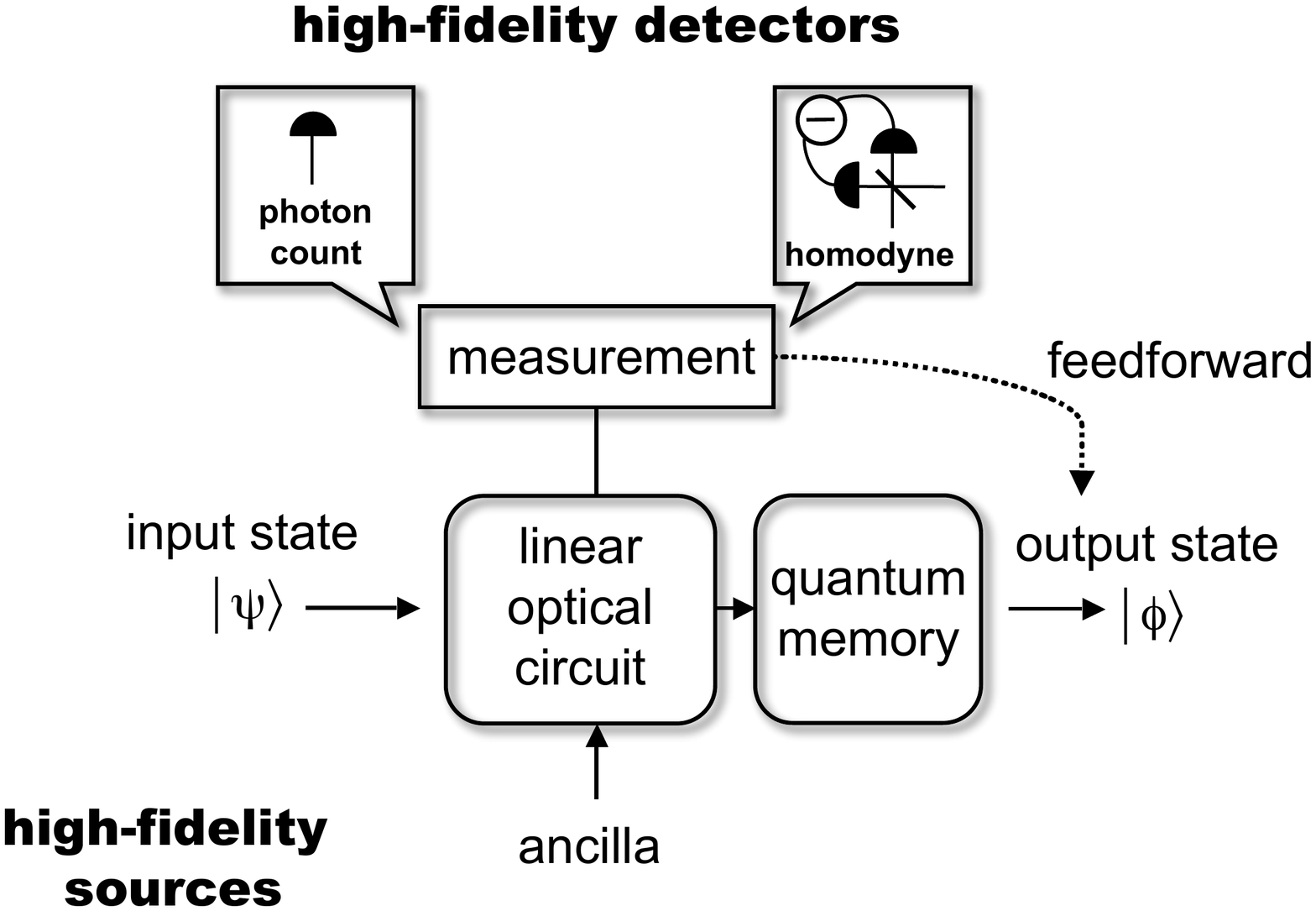}
\caption{Components of a measurement-induced-nonlinearity-based optical quantum computer. High fidelity sources of single photons, Bell pairs or CSS states are required for input states and ancillas, and high-fidelity detectors are required for state conditioning and final detection. A quantum memory allows states to be stored while other operations are occurring, and feedforward is used to update quantum states based on the results of previous detections. Photons interact interferometrically in a linear optics circuit. The architecture shown is conceptual, and a full-scale quantum computer would involve many repetitions of blocks such as these. }
\label{fig14}
\end{center}
\end{figure}

Demonstrations of optical quantum logic, and small algorithms, can be achieved with technology that is presently available. The ability to scale to larger problems will require efficient production and detection of single photon (or other highly quantum) states---a major challenge. In this section, we discuss the progress towards high-efficiency sources and detectors, as well as some of the other technologies which will likely play a part in larger-scale optical quantum computing implementations \footnote{There are a large and growing number of experimental results in this category, and it is not possible to provide exhaustive references --- we have chosen a representative sample}. Fig. \ref{fig14} provides a graphical overview of how some of the different components may fit together to form part of a large-scale optical quantum computer.

\subsection{Detectors}
\label{subsec:Dets}

The silicon avalanche photodiodes that serve as detectors in most present-day experiments have a quantum efficiency of about $50-75\%$, depending on wavelength and the manufacture of the individual device. Furthermore, they act as ``click'' or ``no-click detectors'', meaning that a positive signal indicates the detection of \textit{at least one} photon, but with no ability to assign photon number from a click.

Although the exact detector requirements vary from scheme to scheme, OQC protocols generally require that the efficiency be close to unity (see section \ref{sec:FT} for a discussion of just how close), and with at least some number-resolving capability (e.g.\ the ability to distinguish 0, 1 and 2 photon events from one another). Other desirable properties include high speed operation (in order to detect many photons per second) and low dark counts (in order to distinguish real events from thermal noise in the detector). Although the commonly used detectors do not perform especially favourably on the first two criteria, there are new detectors that are beginning to do so. 

In terms of quantum efficiency, the best developed single photon counting detector is the tungsten transition-edge superconducting (TES) sensor of Lita, Miller, and Nam (2008). It has reached a measured quantum efficiency of $\sim 95\%$, and it also possesses the capability to resolve photon numbers up to 7 per pulse. TES devices work by using a small, thermally isolated region of superconducting material that is temperature-stabilized such that the conduction is right at the boundary of superconducting and normal behaviour---a region of high thermal sensitivity. When photons are absorbed by the metal, the heat changes the its resistance in quantized steps, allowing for the photon number resolving capability. Because the device is bolomteric, the quantum efficiency is very high as long as the photon is absorbed and gives up its heat. 

Other high-efficiency, number resolving devices are also under development. Examples include the visible-light photon counter (Takeuchi, Kim, Yamamoto, and Hogue (1999); Kim, Takeuchi, Yamamoto, and Hogue (1999)), which has been demonstrated to have efficiencies of $>90\%$, as well as photon-number-resolving capability. Additionally, a wide variety of other technologies are under development, including other superconductor schemes (e.g.\ Rosfjord, Yang, Dauler, Kerman, Anant, Voronov, Gol'tsman, and Berggren (2006); Miki, Fujiwara, Sasaki, Baek, Miller, Hadfield, Nam, and Wang (2008)), single-photon upconversion schemes (VanDevender and Kwiat (2007); Thew, Zbinden, and Gisin (2008); Albota and Wong (2004); Roussev, Langrock, Kurz, and Fejer (2004)), and clever uses of existing technologies with new back-end electronics (Fujiwara and Sasaki (2006); Kardynal, Yuan, and Shields (2008)). Each of these schemes address one or both of the issues of detector quantum efficiency and number-resolving capability.

Continued development is required to achieve high efficiency, number resolving capability, low dark counts and high speed all in one device, but progress in detector technology is highly encouraging. 

The detectors just described are not yet commercially available, and in general are restricted to a few laboratories and projects. In parallel, a number of ideas are being implemented to increase the flexibility of existing avalanche photodiodes. Number resolving detectors can be simulated by multiplexing in time (Achilles, Silberhorn, Sliwa, Banaszek, Walmsley, Fitch, Jacobs, Pittman, and Franson (2004)) or space. Photons are split into different time bins or spatial modes using beamsplitters, and although this procedure is non-deterministic, the probability of having more than one photon per channel is low if the number of channels significantly exceeds the number of photons to be discriminated. Such detection schemes are usually employed with stardard single photon counting modules (SPCMs) based on avalanche photodiodes---devices that do not have a high intrinsic detection efficiency. If the single-unit quantum efficiency is $\eta$, then the chance of detecting $N$ photons (assuming we are in the limit where there is no more than one per channel) in a multichannel device is $\sim\eta^N$. With $\eta=0.5$ and $N=6$, this overall efficiency is less than $2\%$. Nevertheless, such schemes make possible experiments that may otherwise be out of reach, and in some cases, the use of tomographic detector characterization schemes (Lundeen, Feito, Coldenstrodt-Ronge, Pregnell, Silberhorn, Ralph, Eisert, Plenio, and Walmsley (2009); Achilles, Silberhorn, and Walmsley (2006)) may enable intial demonstrations even with significantly imperfect devices. 

%Producing and detecting single photon states efficiently is a major technological challenge. Currently the best single photon detectors have efficiencies around 90\% and the most efficient single photon sources are around 55\%, but typically in practical situations these numbers are much lower. This presents a major problem for single rail schemes where typically the loss of a photon results in a change to the qubit state and hence logical errors. In dual-rail schemes on the other hand, photon loss results in no qubit arriving (rather than the wrong qubit) and so can quite easily be filtered out of the data as we shall now describe. This is another reason why most optical quantum information demonstrations are currently based on dual-rail logic. 

\subsection{Sources}
\label{subsec:SOU}
High-efficiency, high-fidelity sources are another key requirement of building an optical quantum computer. Depending on the protocol, this may be an array of single photon sources, Bell-pair sources or CSS sources. In each case, the desired performance is that the source produces \textit{exactly} the required state \textit{every time}. Of course, a slight amount of imperfection should be tolerable, and it is an ongoing problem in the theory of error correction for optical schemes to determine how high these tolerances can be (see section \ref{sec:FT}). For this subsection, we will simply express the goals in terms of their absolute ideal values. 

\subsubsection{Single photon sources}
\label{subsubsec:SPS}
The original KLM proposal for LOQC harnessed the idea of the ideal single photon source, a device that, when triggered, produced one and only one photon into a desired mode with unit efficiency. Such a source does not yet exist. However, its development is the subject of a large and ongoing effort, encompassing several technological approaches. 

The initial step in developing a single photon source is identifying a quantum process that produces the single photon state, as opposed to some other common state such as a thermal state of light. Natural candidates include single atoms (Kuhn, Hennrich, and Rempe (2002); Darquie, Jones, Dingjan, Beugnon, Bergamini, Sortais, Messin, Browaeys, and Grangier (2005); McKeever, Boca, Boozer, Miller, Buck, Kuzmich, and Kimble (2004)), single ions (Diedrich and Walther (1987); Keller, Lange, Hayasaka, Lange, and Walther (2004)), single color centers (Kurtsiefer, Mayer, Zarda, and Weinfurter (2000); Brouri, Beveratos, Poizat, and Grangier (2000); Beveratos, Kuhn, Brouri, Gacoin, Poizat, and Grangier (2002)), single semiconductor quantum dots (Michler, Kiraz, Becher, Schoenfeld, Petroff, Zhang, Hu, and Imamoglu (2000); Noda, Chutinan and Imada (2000); Santori, Pelton, Solomon, Dale, and Yamamoto (2001); Moreau, Robert, Gerard, Abram, Manin, and Thierry-Mieg (2001); Santori, Fattal, Vu\u{c}kovi\'{c}, Solomon, and Yamamoto (2002); Yuan, Kardynal, Stevenson, Shields, Lobo, Cooper, Beattie, Ritchie, and Pepper (2002); Englund, Fushman, Faraon, and Vu\u{c}kovi\'{c} (2009)) and possibly other naturally quantized individual systems which will only emit one quantum of radiation after excitation. Additionally, ensemble or bulk-material approaches, conditioned on certain herald events, can be used as a source of single photons, albeit not a triggered one. Examples in this latter category include heralded spontaneous parametric downconversion (Hong and Mandel (1986); Pittman, Jacobs, and Franson (2002(b)); Pittman, Jacobs, and Franson (2005)) and four-wave mixing (Fulconis, Alibart, Wadsworth, and Rarity (2007); Li, Voss, Sharping, and Kumar (2005)), and the so-called DLCZ ensemble technique (Kuzmich, Bowen, Boozer, Boca, Chou, Duan, and Kimble (2003); Laurat, Riedmatten, Felinto, Chou, Schomburg, and Kimble (2006)).

Since producing more than one photon is an obvious form of error, the quantum optical second-order correlation function $g^{(2)}$---which can be used to identify states with more than one photon produced---is used to characterize the sources. The best single photon sources reported have $g^{(2)}(0)$ on the order of a percent, e.g.\ Vu\u{c}kovi\'{c}, Fattal, Santori, and Solomon (2003), which is substantially lower than the value $g^2(0)=1$ expected for a coherent state, for example. In many prototype single photon sources, much of the remaining 2-event signal may be attributed to contributions from the single photon and either an environmental background photon, or a thermal dark count in the detector. 

A more challenging problem for single photon sources is the issue of collection efficiency. Even if only one photon is generated, it must be reliably inserted into the desired spatial mode at each and every trigger event. In practice, this is remarkably difficult. Sources based on spontaneous emission suffer the problem that their natural emission is into $4\pi$ steradians, and collecting a sizeable fraction (ideally all) of this emission is challenging. A potential solution is to place the emitter in an optical cavity. Many cavity types have been employed, including Bragg stacks and photonic crystals for quantum dots, Fabry-Perot cavities for atoms and ions, and whispering-gallery-mode resonators in a variety of systems. Although these can, in principle, increase the collection efficiency to close to unity (Noda, Fujita and Asano (2007)), high degrees of coupling to the cavity (potentially close to $100\%$, e.g. Englund, Faraon, Zhang, Yamamoto, and Vu\u{c}kovi\'{c} (2007)) have not yet lead to high outcoupling efficiencies in practice. In the case of inefficient collection, one is faced with the problem that instead of having a single photon at each and every trigger event, the vacuum state is sometimes present instead. 

Heralded schemes such as spontaneous parametric downconversion and the DLCZ scheme are designed for more directional emission, so that the collection efficiency problem, in principle, is greatly reduced. Using these schemes, inferred single photon collection efficiencies of $83\%$ (Pittman, Jacobs, and Franson (2005)) and $50\%$ (Laurat, Riedmatten, Felinto, Chou, Schomburg, and Kimble (2006)) have been measured respectively. 

The final, and equally critical, criterion for single photon sources is the indistinguishability of the photons produced, both from a single source or from different elements in an array of SPSs. This indistinguishability is critical for achieving high-fidelity nonclassical interference, which is a key component of nearly all optical schemes. In some cases, moderate- to high-visibility quantum interference has already been measured for two photons from independent sources (e.g.\ (Beugnon, Jones, Dingjan, Darquie, Messin, Browaeys, and Grangier (2006); Kaltenbaek, Blauensteiner, Zukowski, Aspelmeyer, and Zeilinger (2006); Maunz, Moehring, Olmschenk, Younge, Matsukevich, and Monroe (2007); Mosley, Lundeen, Smith, Wasylczyk, U'Ren, Silberhorn, and Walmsley (2008); Sanaka, Pawlis, Ladd, Lischka, and Yamamoto (2009)). Distinguishability can be introduced in several ways---key examples include: the timing jitter (uncertain emission time) of spontaneous emission sources; different center frequencies of independent sources; spectal distinguishability introduced through mixture from non-transform limited sources; and the natural entanglement in energy-time of SPDC sources. It should be noted, however, that solutions to all of these problems exist in principle, and some have been demonstrated in practice. For instance, with careful spectral engineering it is possible to remove the frequency entanglement in SPDC sources without requiring filtering that would reduce efficiency (Mosley, Lundeen, Smith, Wasylczyk, U'Ren, Silberhorn, and Walmsley (2008)).

\subsubsection{Bell-pair sources}
\label{subsubsec:BellPS}

Some optical quantum computing schemes, such as certain cluster-state proposals (e.g.\ Browne and Rudolph (2005)), require maximally-entangled photonic qubit pairs (Bell pairs) rather than single photon sources. Having access to pre-existing entanglement of this kind simplifies the operations required in implementing LOQC. 

Generating \textit{triggered}, or even \textit{heralded}, entangled pairs is difficult. One tends to think of suitably phase-matched SPDC as naturally producing entangled pairs (Kwiat, Mattle, Weinfurter, Zeilinger, Sergienko, and Shih (1995)), but these are generated randomly and there is far from any guarantee that one and only one pair will be produced at each pump pulse---rather, the pairs are Poisson-distributed. Nevertheless, SPDC can produce entanglement with high fidelity, conditional on a pair being present---such sources are useful in coincidence experiments. A number of sources have been developed that produce highly entangled states with high flux (Altepeter, Jeffrey, and Kwiat (2005); Fedrizzi, Herbst, Poppe, Jennewein, and Zeilinger (2007)).

The most successful schemes to date for generating triggered entangled pairs are based on semiconductor quantum dots. This involves the production, through pumping, of a biexciton state which can decay via several indistinguishable paths to produce a pair of polarization-entangled photons (Akopian, Lindner, Poem, Berlatzky, Avron, Gershoni, Gerardot, and Petroff (2006); Young, Stevenson, Atkinson, Cooper, Ritchie, and Shields (2006)). The challenge of this approach is keeping the emitted photons from being distinguishable by other means, such as by slight difference in frequencies that arise due to splittings of the intermediate exciton levels. As with other photon sources, collection efficiency is still problematic, although similar solutions as those proposed for SPSs should also be viable.

Recently, a triggered Bell pair source using an atom in a cavity has been demonstrated, with the output state demonstrating a fidelity of $0.9$ with the singlet Bell state (Weber, Specht, Mueller, Bochmann, Muecke, Moehring and Rempe (2009)).
\subsubsection{CSS sources}
\label{subsubsec:CSSsources}

For CSQC, the required optical resources are coherent states and coherent state superpositions (CSSs). The former is easily generated on demand by using an appropriately stabilized laser or above-threshold parametric oscillator. The latter, being a fragile non-Gaussian state, is not surprisingly somewhat more difficult to generate. The generation of CSSs has already been covered in some detail in subsection \ref{subsec:LOfields} above. 

There are several areas for active research in developing CSS states. One limitation of present generation of sources is that they are \textit{heralded}, rather than triggered. The CSS is generated when the photon detector fires. However, because of both the inefficiency of photon counters and the uncertainty in splitting off a photon using the beam splitter, this event does not occur on every pulse (for example). Another problem to be overcome with CSS sources is the loss of fidelity due to inefficiency in the circuit. This introduces vacuum noise that decreases the fidelity of the state. 

% Although not yet demonstrated, there exist schemes to scale up the size of small CSS states to larger ones, and in particular to reach the ideal CSS size of \textbf{$|\alpha|\sim 1.4$?}. These schemes are similar to distillation, in that several (e.g. two) CSS states are used in order to generate a larger CSS. As with the generation of a small pseudo-CSS, this ``growing'' technique is non-deterministic and heralded, making it necessary to use more elaborate schemes to generate CSSs on demand.

\subsubsection{From heralded to triggered sources.}
\label{subsubsec:HeraldtoTrig}

Many of the schemes for generating single photon (or other resource) states for optical quantum computing suffer from being nondeterministic, either in principle or in practice. In the latter case, this is often a consequence of the fact that the outcoupling efficiency is not $100\%$. As long as the schemes are heralded, however, this does not have to be a show-stopper. Several groups are working on techniques to turn heralded sources into triggered, on-demand sources. 

The basic idea of most of these approaches is to use an array of parallel heralded sources and switching circuitry to bring the probability of getting a single event very close to unity (Migdall, Branning, and Castelletto (2002); Pittman, Jacobs, and Franson (2002(b))). Because the probability of not getting \textit{any} events decreases exponentially with the number of sources in parallel, the scheme can be very efficient in principle. In practice, sources of error such as dark counts in the heralding detectors, switch loss and timing considerations need to be addressed, and work is continuing in these areas. 

\subsection{Memory}
\label{subsec:Mem}

At times during a quantum computation, quantum information must be stored in memory. A simple example of this in LOQC is the requirement to hold a qubit while a feedforward operation on the state is triggered and implemented. The basic requirements of a memory are that it has high (ideally unit) fidelity, that the hold time be suitably long, and that the information can be switched in and out on demand. 

Photons---and other states useful for optical quantum computing---have spatiotemporal structure, and so it is necessary to design a memory with sufficient bandwidth and physical dimensions to accommodate the optical modes of interest. The simplest memory is the optical delay line, which consists of a free-space or fiber path length to store (or delay) the light by a fixed amount (Pittman, Jacobs, and Franson (2002); Prevedel, Walther, Tiefenbacher, Bohl, Kaltenbaek, Jennewein, and Zeilinger (2007)). While this is a simple, potentially high-bandwidth, and relatively high-fidelity solution, the basic delay line is also inflexible in that the memory is not switchable - for maximum flexibilty one wishes to be able to decide the length of time that the light is held in memory \textit{after} it is stored. 

A more sophisticated version of the delay line is the storage loop, in which light is sent around a loop such that once on each pass, it passes a switch which can be flipped to switch it back into the circuit (Pittman and Franson (2002)). In this case, the storage time is not arbitrary, but is rather a multiple of the loop's single round trip time. One of the major practical challenges of this type of memory is the loss through the switch, even when in the passive condition, which is experienced with each pass, making the chance of retaining a photon for many loop cycles small. 

Perhaps the most flexible memory, in principle at least, is that promised by atomic ensembles. A variety of techniques have been proposed to store light in these systems, but the feature common to all is that the optical state is transferred into a quantum state of a collection of atoms---often a spin-wave in the ensemble---which is later switched (on demand) back to the optical state.

Although several schemes exist for using atomic ensembles, perhaps the best-known are ``stopped light'' techniques\footnote{the \textit{slow light} phenomenon can also be used, in principle, to make an optical quantum delay line} using electromagnetically-induced transparency (EIT) (Chaneliere, Matsukevich, Jenkins, Lan, Kennedy, and Kuzmich (2005); Eisaman, Andre, Massou, Fleischhauer, Zibrov, and Lukin (2005); Appel, Figueroa, Korystov, Lobino, and Lvovsky (2008); Choi, Deng, Laurat, and Kimble (2008); Hetet, Buchler, Gloeckl, Hsu, Akulshin, Bachor, and Lam (2008)), and what is variously known as GEM (gradient echo memory) or CRIB (coherent reversible inhomogeneous broadening) (Hetet, Longdell, Alexander, Lam, and Sellars (2008); Tittel, Afzelius, Cone, Chaneli\'ere, Kr\"oll, Moiseev, and Sellars (2008)). 

EIT stopped light uses the slow light medium created by a narrow transparency window to compress the light into the medium, and an adiabatic switching of the control field captures the wavefunction in the atomic state of the ensemble.  GEM and CRIB work by absorptively creating an atomic coherence stored amongst the atoms in the medium. By globally manipulating the atomic coherence---specifically, by switching detunings---the optical quantum state can be recovered. The interaction of off-resonant light with spin-polarized ensembles has also been used to demonstrate quantum optical storage (Julsgaard, Sherson, Cirac, Fiurasek, and Polzik (2004)).
  
What is common to these techniques, in general, is that they are low bandwidth. For example, the EIT typically produces a delay for light in a $\sim$MHz bandwidth, a similar bandwidth as for a  preliminary CRIB demonstration (Hetet, Longdell, Alexander, Lam, and Sellars (2008)). This should be contrasted with the typical terahertz bandwidth of SPDC. However, as has been noted, SPDC may not be the ultimate choice for a SPS, and even if it is, it is possible to produce narrow bandwidth SPDC sources which might be more suitable for working with this type of memory (e.g.\ Haase, Piro, Eschner, and Mitchell (2009)). Additionally, improvements in materials development (such as in the solid-state systems) might allow for wider memory bandwidths. 

Two requirements for a memory are that one can achieve a suitably long delay time, and that the fidelity of the output state, with respect to the input, is high. This latter criterion is sometimes broken up into the storage efficiency and the conditional fidelity. In the case of a single photon, for example, this would be the probability of getting a photon out for each photon sent in, and the fidelity of that output photon's state with the input state. Storage efficiencies of $\sim 15\%$ been achieved in quantum memories (Choi, Deng, Laurat, and Kimble (2008); Appel, Figueroa, Korystov, Lobino, and Lvovsky (2008); Hetet, Longdell, Alexander, Lam, and Sellars (2008)), with the possibility of making this much higher---indeed, approaching unity in principle. Delay times are largely limited by coherence properties of the atomic media used for storage, and range from $\sim$ns to $\sim\mu$s, with the possibility of adapting second-timescale classical storage (Longdell, Fraval, Sellars, and Manson (2005)) to quantum memories in the future.

\subsection{Integrated Optics}
\label{subsec:IntegOpt}

In the long term, no one expects a large-scale optical quantum computer, comprising many thousands of gates, to be constructed from bulk components laid out on optical tables in a laboratory. As well as the alignment instability, such a quantum computer would take up an enormous amount of space! A more reasonable proposition is to have hardwired optical circuits, possibly on the micro-scale, using some form of integrated optics. As well as the obvious benefit in size and stability, such circuits should also simplify mode-matching operations that are central to achieving high-quality classical and non-classical interference. 

The two most obvious forms of guided optics are \textit{fiber optics} and \textit{integrated optics}---the latter is usually considered to mean planar waveguides or something written into or onto bulk material. Although there have been quite a number of examples of optical quantum information protocols in optical fiber, especially in regard to quantum key distribution (Gisin, Ribordy, Tittel, and Zbinden (2002)), there have not been many predominantly fiber implementations of quantum computing gates. Some notable demonstrations include ``plug and play'' fiber and free space modules (Pittman, Fitch, Jacobs, and Franson (2003)) and full fiber implementations (Clark, Fulconis, Rarity, Wadsworth, and O'Brien (2008)) that include a fiber-based photon source.

A recent development has been the use of integrated optics for quantum logic. The O'Brien group has demonstrated the nondeterministic CNOT gate of Ralph, Langford, Bell and White (2002) using planar waveguide circuits (Politi, Cryan, Rarity, Yu, and O'Brien (2008)). One of the properties of standard planar waveguides is that single-polarization operation is exceedingly preferred, so that a dual-rail spatial mode encoding is employed, where superpositions are between the occupancy of two spatial paths defined by waveguides. The key elements of such an encoding are beam splitters, phase shifters, and interferometers built from these components. To date, the ability to construct and utilise beam splitters with suitable ratios has been demonstrated, as has the ability to make tunable phase shifters using resistive heating of one of the interferometer arms. A future requirement will be fast phase shifters for feedforward operations (Prevedel, Walther, Tiefenbacher, Bohl, Kaltenbaek, Jennewein, and Zeilinger (2007)). The widespead use of electro-optic modulators in telecommunications waveguide technology suggest that this may be achievable in quantum optics devices as well. 

Another promising technique for building optical waveguide circuits is the \textit{direct write} technique, in which a high-power laser is used to write circular waveguide below the surface of some material, generally a glass (Marshall, Ams, and Withford (2006)). The advantage of this technique is that it can be used to generate three-dimensional circuits, which allows optical paths to easily cross. Preliminary demonstrations with this technology have achieved high visibility nonclassical interference for two-photon and multi-photon events (Marshall, Politi, Matthews, Dekker, Ams, Withford, and O'Brien (2009)).  

An outstanding problem for the application of waveguide technologies is the insertion loss of the waveguide circuits, and of specialized components (such as electro-optic devices). The first problem may be solved by introducing sources and detectors directly into the waveguide environment, and the development of fiber-based sources is already a step in that direction. Additionally, work is in progress towards the implementation of adaptive optics waveguide coupling schemes (Kumar, Kwiat, Migdall, Nam, Vu\u{c}kovi\'{c}, and Wong (2004)), in which the mode pattern from an SPDC source may be converted into the mode pattern corresponding to waveguide propagation. 

\section{Conclusion and Summary}

%The challenges involved in building a quantum computer of sufficient size that it can achieve useful outcomes are significant.
In this chapter, we have examined the potential of optics for realizing a quantum computer. Quantum information can be encoded
on light and can be carried, manipulated, entangled and measured with high precision. Both photonic encodings---where the information is carried on a binary degree of freedom such as polarization---and field encodings---with information encoded on different field states---are possible. As a sign of their utility in quantum information science, both types of encoding have been successfully deployed in quantum key distribution protocols (Poppe, Peev, Maurhart 2008).

Both linear and nonlinear strategies have been considered for implementing universal gate sets. In the non-linear approaches, very strong optical non-linearities are proposed for use in-line, to interact optical qubits with one another. In the linear approaches, off-line non-linearities, in the form of state production and measurement, are used to add measurement-induced non-linearity into a linear optical network. Although proposals exist for all combinations of strategies and encodings, the most advanced experimental gate demonstrations so far have been performed using the linear approach and photonic encoding.

It is possible, in principle, to obtain fault-tolerant operation with both particle and field encodings. Threshold estimates have been generated for several linear strategies, as well as one nonlinear one, using a generalized Steane code. A general trend can be identied whereby
higher thresholds tend to lead to higher resource overheads and vice versa, suggesting a trade-off between the precision of the physical operations and the number required. A new, high threshold, efficient fault tolerant protocol based on 3-D cluster states has recently
been developed by Raussendorf and Harrington (2007) that appears particularly well suited to optical architectures (Devitt, Fowler, Stephens, Greentree, Hollenberg, Munro, Nemoto 2008). Future work is likely to focus on the optimal encoding and gate types for these kinds of protocols.

A large-scale optical quantum computer will require specialized sources, detectors, switches and circuits, optical networks and memories. Each of these is presently under development, with many different approaches being pursued. While some of these optical technologies have already demonstrated performance levels close to the those required, significant improvements are still necessary in some areas to make large-scale quantum computing possible. Encouragingly, progress seems to be rapid and there are no in-principle reasons why the required performance levels cannot be achieved.

There are other directions in optics-related quantum computing that could not be covered in a chapter of this length. For example, hybrid optical/atomic or optical/solid-state systems may be contenders for quantum information processing applications. Optical "flying qubits" could act as a data bus, solving the connectivity problem in atomic or solid-state quantum computer architectures. Alternatively we might
use the "standing qubits" as memory, whilst processing the quantum information optically. An example of experimental progress in this direction is the generation of entanglement between distant ions using an optical quantum bus and LOQC processing (Matsukevich, Maunz,
Moehring, Olmschenk, and Monroe 2008). 
%These, and other emerging technologies, combined with achievements described in this review indicate a bright future for quantum information processing using optics.

Of the physical systems being considered for quantum computation, optics is perhaps the best understood in terms of the physics of the interactions and decoherence mechanisms. Optics has also demonstrated an outstanding precision in operations demonstrated to date. The gap between the theoretical requirements and experimental demonstrations, while large, has been shrinking consistently due to both advances in technology and improved protocols. We can conclude that there appear to be no fundamental barriers to optical quantum computation and although formidable practical barriers remain, current research suggests an optimistic outlook for overcoming them.

Acknowledgments: This work was supported by the Australian Research Council and the US Army Research Office run IARPA program on Quantum Computation.

\section{References}

%Lanyon (2009) B. P. Lanyon, M. Barbieri, M. P. Almeida, T. Jennewein, T. C. Ralph, K. J. Resch, G. Pryde, J. L. OBrien, A. Gilchrist and A. G. White, Nature Physics {\bf 5}, 134 (2009).
% GJP's REFs %
%%%%%%%%%%%%%%%%%%%%%%%%%%
% \\ \\
%\textbf{Geoff's Refs}
% \\ \\

Achilles, D., C. Silberhorn, C. Sliwa, K. Banaszek, I. A. Walmsley, M. J. Fitch, B. C. Jacobs, T. B. Pittman, and J. D. Franson, 2004, Jornal of Modern Optics \textbf{51}, 1499.

Achilles, D., C. Silberhorn, and I. A. Walmsley, 2006, Physical Review Letters \textbf{97}, 043602.

Akopian, N., N. H. Lindner, E. Poem, Y. Berlatzky, J. Avron, D. Gershoni, B. D. Gerardot, and P. M. Petroff, 2006, Physical Review Letters \textbf{96}, 130501.

Albota, M. A. and F. N. C. Wong, 2004, Optics Letters \textbf{29}, 1449.

Altepeter, J. B., E. R. Jeffrey, and P. G. Kwiat, 2005, Optics Express \textbf{13}, 8951.

Aoki, T., B. Dayan, E.Wilcut, W. P. Bowen, A. S. Parkins, T. J. Kippenberg, K. J. Vahala and H. J. Kimble, 2006, Nature {\bf 443} 671 (2006).

Appel, J., E. Figueroa, D. Korystov, M. Lobino, and A. I. Lvovsky, 2008, Physical Review Letters \textbf{100}, 093602.
%
%Bennett (1992)
%C.H. Bennett, S.J.Wiesner,
%Phys.Rev.Lett. {\bf 69}, 2881 (1992).

Babichev, S. A., B. Brezger, and A. I. Lvovsky, 2004, Physical Review Letters \textbf{92}, 047903.

Bachor, H-A., and T.C.Ralph, 2004, {\it A guide to experiments in quantum optics} (2nd Edition, Wiley-VCH, Weinheim).

Bao, X. H., T. Y. Chen, Q. Zhang, J. Yang, H. Zhang, T. Yang, and J. W. Pan, 2007, Physical Review Letters \textbf{98}, 170502.

%Braunstein, S.L., and H.J.Kimble, 1998, Phys.Rev.Lett. {\bf 80}, 869.

Braunstein, S.L., and A.Pati, Ed, 2003, {\it Continuous Variable Quantum Information}, (Kluwer Academic Publishers, The Netherlands).

Braunstein, S. L., and P. van Loock, 2005, Rev. Mod. Phys. {\bf 77}, 513.

Bennett, C.H., G.Brassard, C.Crepeau, R.Jozsa, A.Peres, W.K.Wootters, 1993,
Phys.Rev.Lett. {\bf 70}, 1895.

Beugnon, J., M. P. A. Jones, J. Dingjan, B. Darquie, G. Messin, A. Browaeys, and P. Grangier, 2006, Nature \textbf{440}, 779.

Beveratos, A., S. Kuhn, R. Brouri, T. Gacoin, J. P. Poizat, and P. Grangier, 2002, European Physical Journal D \textbf{18}, 191.

Brouri, R., A. Beveratos, J. P. Poizat, and P. Grangier, 2000, Optics Letters \textbf{25}, 1294.

Browne, D. E., and T. Rudolph, 2005, Physical Review Letters \textbf{95}, 010501.

Brune, M., E.~Hagley, J.~Dreyer, X.~Maitre, A.~Maali, C.~Wunderlich, J.~M.~Raimond
and S.~Haroche, 1996, Phys Rev Lett {\bf 77}, 4887.

Bullock, S.S, D.P.OÕLeary, and G.K.Brennen, 2005, Phys. Rev. Lett. {\bf 94}, 230502.

Chaneliere, T., D. N. Matsukevich, S. D. Jenkins, S. Y. Lan, T. A. B. Kennedy, and A. Kuzmich, 2005, Nature \textbf{438}, 833.

Chen, K., C. M. Li, Z. Qiang, Y. A. Chen, A. Goebel, S. Chen, A. Mair, and J. W. Pan, 2007, Physical Review Letters \textbf{99}, 120503.

Choi, K. S.,  H. Deng, J. Laurat, and H. J. Kimble, 2008, Nature \textbf{452}, 67.

Chuang, I. L. and M. A. Nielsen, 1997, Journal of Modern Optics \textbf{44}, 2455.

Clark, A.S., J. Fulconis, J. G. Rarity, W. J. Wadsworth, and J. L. O'Brien, 2008, arXiv:0802.1676.

Cochrane, P., G.~J.~Milburn and W.~J.~Munro, 1998, Phys Rev
A {\bf 59}, 2631.

Darquie, B., M. P. A. Jones, J. Dingjan, J. Beugnon, S. Bergamini, Y. Sortais, G. Messin, A. Browaeys, and P. Grangier, 2005, Science \textbf{309}, 454.

Dawson, C.M., H.L.Haselgrove, M.A.Nielsen, 2006 Phys. Rev. A {\bf 73}, 052306.

Deutsch, D., 1985, Proc.Roy.Soc.London {\bf A400}, 97.

Devitt, S.J., A.G.Fowler, A.M.Stephens, A.D.Greentree,L.C.L.Hollenberg, W.J.Munro, K.Nemoto, 2008, arXiv:0808.1782.

Diedrich, F. and H. Walther, 1987, Physical Review Letters \textbf{58}, 203.

Dirac, P.A.M., 1958, {\it The Principles of Quantum Mechanics}, 4th edition,
Oxford University Press, London.

Dodd, J. L., T. C. Ralph, and G. J. Milburn, 2003, Phys. Rev. A {\bf 68}, 042328.

Duan, L.-M., and H. J. Kimble, 2004, 
Phys. Rev. Lett. {\bf 92}, 127902.

Eisaman, M. D., A. Andre, F. Massou, M. Fleischhauer, A. S. Zibrov, and M. D. Lukin, 2005, Nature \textbf{438}, 837.

Englund, D., A. Faraon, B. Zhang, Y. Yamamoto, and J. Vu\u{c}kovi\'{c}, 2007, Optics Express \textbf{15}, 5550.

Englund, D., I. Fushman, A. Faraon, and J. Vu\u{c}kovi\'{c}, 2009, Photonics and Nanostructures, Fundamentals and Applications \textbf{7}, 56.

Enk, S.~J.~van, and O.~Hirota, \pra {\bf 64}, 2001,
022313.

Fedrizzi, A., T. Herbst, A. Poppe, T. Jennewein, and A. Zeilinger, 2007, Optics Express \textbf{15}, 15377.

Feynman, R.P., 1986, Foundations of Physics {\bf 16}, 507.

Franson, J. D., M. M. Donegan, M. J. Fitch, B. C. Jacobs, and T. B. Pittman, 2002, Physical Review Letters \textbf{89}, 137901

Franson, J.D., B.C. Jacobs, and T.B. Pittman, 2004, Phys. Rev. A \textbf{70}, 062302.

Fujiwara, M. and M. Sasaki, 2006, Optics Letters \textbf{31}, 691.

Fulconis, J., O. Alibart, W. J. Wadsworth, and J. G. Rarity, 2007, New Journal of Physics \textbf{9}, 276.

Fushman, I., D. Englund, A. Faraon, N. Stoltz, P. Petroff, and J. Vu\u{c}kovi\'{c}, 2008, Science \textbf{320}, 769.

Gasparoni, S., J. W. Pan, P. Walther, T. Rudolph, and A. Zeilinger, 2004, Physical Review Letters \textbf{93}, 020504.
Gaebel, T., M. Domhan, I. Popa, C. Wittmann, P. Neumann, F. Jelezko, J.R. Rabeau, N. Stavrias, A. D. Greentree, S. Prawer, J. Meijer, J. Twamley, P. R. Hemmer, J. Wrachtrup, 2006, Nature Physics {\bf 2}, 408.

Ghosh, R. and L. Mandel, 1987, Phys. Rev. Lett. {\bf 59}, 1903.

Gilchrist, A., A.J.F.Hayes and T.C.Ralph, 2007, Phys. Rev. A {\bf 75}, 052328.

Gisin, N., G. G. Ribordy, W. Tittel, and H. Zbinden, 2002, Reviews of Modern Physics \textbf{74}, 145.

Glancy, S., and E.Knill, 2006, Phys. Rev. A {\bf 73}, 012325.

Glauber, R.J., 1962, Phys.Rev. {\bf 131}, 2766.

Gleyzes, S., S. Kuhr, C. Guerlin, J. Bernu, S.Deleglise, U. B. Hoff,
M. Brune, J-M. Raimond and S. Haroche, 2007, Nature {\bf 446}, 297.

Gong, Y. X., G. Guo, and T. C. Ralph, 2008, Physical Review A \textbf{78}, 012305.

Gottesman,
D., I.L.Chuang, 1999, Nature {\bf 402}, 390.

Gottesman, D., A.~Kitaev, J.~Preskill, 2001, Phys Rev A {\bf
64},
012310.

Grover,  L.K., 1997, Phys.Rev.Lett. {\bf 79}, 325.

Haase, A., N. Piro, J. Eschner, and M. W. Mitchell, 2009, Optics Letters \textbf{34}, 55.

Haeffner, H., C.F. Roos,  R. Blatt, 2008, Phys. Rep. {\bf 469}, 155

Haselgrove, H.L., P. P. Rohde, 2008, Quantum Information and Computing, {\bf 8}, 399.

Hayes, A.J.F., A.Gilchrist, C.R.Myers and T.C.Ralph, 2004, J.Opt.B {\bf 6}, 533.

Hendrickson, S. M., T. B. Pittman, and J. D. Franson, 2009, Journal of the Optical Society of America B-Optical Physics \textbf{26}, 267.

Hetet, G., B. C. Buchler, O. Gloeckl, M. T. L. Hsu, A. M. Akulshin, H. A. Bachor, and P. K. Lam, 2008, Optics Express \textbf{16}, 7369-7381.

Hetet, G., J. J. Longdell, A. L. Alexander, P. K. Lam, and M. J. Sellars, 2008, Physical Review Letters \textbf{100}, 023601.

Higgins,, B. L., D. W. Berry, S. D. Bartlett, H. M. Wiseman, and G. J. Pryde, 2007, Nature \textbf{450}, 393.

%Hillery, M., 2000, Phys Rev A {\bf 61}, 022309.

Hofmann, H.F., and S.Takeuchi, 2002,
Phys. Rev. A {\bf 66}, 024308.

Hofmann, H. F. 2005, Physical Review Letters \textbf{94}, 160504.

Hong, C. K., and L. Mandel, 1986, Physical Review Letters \textbf{56}, 58.

Hong, C. K., Z.Y.Ou, L.Mandel, 1987, Phys. Rev. Lett.  {\bf 59}, 2044.

Huntington, E.H., and T.C.Ralph, 2004, Phys. Rev. A {\bf 69}, 042318.

Ivanovic, I. D., 1987, Phys. Lett. A {\bf 123}, 257.

Jeong, H., M.~S.~Kim, and J.~Lee, 2001,
Phys Rev A {\bf 64}, 052308.

Jeong, H., and T.C.Ralph, 2007, {\it Quantum Information with Continuous Variables of Atoms and Light}, Ed. N.Cerf, G.Leuchs and E.S.Polzik (Imperial College Press, London, 2007) 159.

Julsgaard, B., J. Sherson, J. I. Cirac, J. Fiurasek, and E. S. Polzik, 2004, Nature \textbf{432}, 482.

Kaltenbaek, R., B. Blauensteiner, M. Zukowski, M. Aspelmeyer, and A. Zeilinger, 2006, Physical Review Letters \textbf{96}, 240502.

Kang, H. S. and Y. F. Zhu, 2003, Physical Review Letters \textbf{91}, 093601.

Kardynal, B. E., Z. L. Yuan, and A. J. Shields, 2008, Nature Photonics \textbf{2}, 425.

Keller, M., B. Lange, K. Hayasaka, W. Lange, and H. Walther, 2004, Nature \textbf{431}, 1075.

Kieling, K., T. Rudolph, J. Eisert, 2007, Phys. Rev. Lett. {\bf 99}, 130501.

Kiesel, N., C. Schmid, U. Weber, G. Toth, O. Guhne, R. Ursin, and H. Weinfurter, 2005, Physical Review Letters \textbf{95}, 210502.

Kiesel, N., C. Schmid, U. Weber, R. Ursin, and H. Weinfurter, 2005, Physical Review Letters \textbf{95}, 210505.

Kim, J. S., S. Takeuchi, Y. Yamamoto, and H. H. Hogue, 1999, Applied Physics Letters \textbf{74}, 902.

Knill, E., R.Laflamme, G.J.Milburn, 2001,
Nature {\bf 409}, 46.

Knill, E., 2003, Phys. Rev. A {\bf 68}, 064303.

Kok, P., W.J. Munro, Kae Nemoto, T.C. Ralph, Jonathan P. Dowling, G.J. Milburn, 2007, Rev. Mod. Phys. {\bf 79}, 135.

Kuhn, A., M. Hennrich, and G. Rempe, 2002, Physical Review Letters \textbf{89}, 067901.

Kumar, P., P. Kwiat, A. L. Migdall, S. W. Nam, J. Vu\u{c}kovi\'{c}, and F. N. C. Wong, 2004, Quantum Information Processing \textbf{3}, 215.

Kurtsiefer, C., S. Mayer, P. Zarda, and H. Weinfurter, 2000, Physical Review Letters \textbf{85}, 290.

Kuzmich, A., W. P. Bowen, A. D. Boozer, A. Boca, C. W. Chou, L. M. Duan, and H. J. Kimble, 2003, Nature \textbf{423}, 731.

Kwiat, P. G., J. R. Mitchell, P. D. D. Schwindt, and A. G. White, 2000, Journal of Modern Optics \textbf{47}, 257.

Kwiat, P. G., K. Mattle, H. Weinfurter, A. Zeilinger, A. V. Sergienko, and Y. H. Shih, 1995, Physical Review Letters \textbf{75}, 4337.

Langford, N.K., T. J. Weinhold, R. Prevedel, K. J. Resch, A. Gilchrist, J. L. O'Brien, G. J. Pryde, and A. G. White, 2005, Physical Review Letters \textbf{95}, 210504.

Lanyon, B. P., T. J. Weinhold, N. K. Langford, M. Barbieri, D. F. V. James, A. Gilchrist, and A. G. White, 2007, Physical Review Letters \textbf{99}, 250505.

Lanyon, B. P., M. Barbieri, M. P. Almeida, and A. G. White, 2008, Physical Review Letters \textbf{101}, 200501.

Lanyon, B. P., M. Barbieri, M. P. Almeida, T. Jennewein, T. C. Ralph, K. J. Resch, G. J. Pryde, J. L. O'Brien, A. Gilchrist, and A. G. White, 2009, Nature Physics \textbf{5}, 134.

Laurat, J., H. de Riedmatten, D. Felinto, C. Chou, E. W. Schomburg, and H. J. Kimble, 2006, Optics Express \textbf{14}, 6912.

Leung, P.M.,  and T.C.Ralph, 2006, Phys. Rev. A {\bf 74}, 062325.

Leung. P.M., and T.C.Ralph, 2007, New J. Phys. {\bf 9} 224.

Li, X. Y., P. L. Voss, J. E. Sharping, and P. Kumar, 2005, Physical Review Letters \textbf{94}, 053601.

Lita, A. E., A. J. Miller, and S. W. Nam, 2008, Optics Express \textbf{16}, 3032.

Lloyd, S., and S. L. Braunstein, 1999, Phys. Rev. Lett. {\bf 82}, 1784.

Longdell, J. J., E. Fraval, M. J. Sellars, and N. B. Manson, 2005, Physical Review Letters \textbf{95}, 063601.

Lu, C., W. Gao, O. Guehne, X. Zhou, Z. Chen, and J. W. Pan, 2009, Physical Review Letters \textbf{102}, 030502.

Lu, C., D. E. Browne, T. Yang, and J. W. Pan, 2007, Physical Review Letters \textbf{99}, 250504.

Lu, C., X. Zhou, O. Guehne, W. Gao, J. Zhang, Z. Yuan, A. Goebel, T. Yang, and J. W. Pan, 2007, Nature Physics \textbf{3}, 91.

Lu, C., W. Gao, L. Zhang, X. Zhou, T. Yang, and J. W. Pan, 2008, Proceedings of the National Academy of Sciences of the United States of America \textbf{105}, 11050.

Lund, A. P., T. C. Ralph, 2002, Phys Rev A {\bf 66}, 032307.

Lund, A.P., T. C. Ralph, and H. L. Haselgrove, 2008, Phys. Rev. Lett. {\bf 100}, 030503.

Lundeen, J. S., A. Feito, H. Coldenstrodt-Ronge, K. L. Pregnell, C. Silberhorn, T. C. Ralph, J. Eisert, M. B. Plenio, and I. A. Walmsley, 2009, Nature Physics \textbf{5}, 27.

L\"utkenhaus N., J. Calsamiglia, and K.-A. Suominen, 1999,
Phys. Rev. A {\bf 59}, 3295.

Lvovsky, A. I., and J. Mlynek, 2002, Physical Review Letters \textbf{88}, 250401.

Marshall, G. D., A. Politi, J. C. F. Matthews, P. Dekker, M. Ams, M. J. Withford, and J. L. O'Brien, 2009, arXiv:0902.4357.

Marshall, G. D., M. Ams, and M. J. Withford, 2006, Optics Letters \textbf{31}, 2690.

Matsukevich, D. N., P. Maunz, D. L. Moehring, S. Olmschenk, and C. Monroe, 2008, Phys Rev Lett {\bf 100}, 150404.

Maunz, P., D. L. Moehring, S. Olmschenk, K. C. Younge, D. N. Matsukevich, and C. Monroe, 2007, Nature Physics \textbf{3}, 538.

McKeever, J., A. Boca, A. D. Boozer, R. Miller, J. R. Buck, A. Kuzmich, and H. J. Kimble, 2004, Science \textbf{303}, 1992.

Menicucci, N. C., P. van Loock, M. Gu, C. Weedbrook,
T. C. Ralph, and M. A. Nielsen, 2006,
Phys. Rev. Lett. {\bf 97}, 110501.

Menicucci, N. C., S. T. Flammia, and O. Pfister, 2008, Physical Review Letters \textbf{101}, 130501.

Michler, P., A. Kiraz, C. Becher, W. V. Schoenfeld, P. M. Petroff, L. D. Zhang, E. Hu, and A. Imamoglu, 2000, Science \textbf{290}, 2282.

Migdall, A. L., D. Branning, and S. Castelletto, 2002, Physical Review A \textbf{66}, 053805.

Miki, S., M. Fujiwara, M. Sasaki, B. Baek, A. J. Miller, R. H. Hadfield, S. W. Nam, and Z. Wang, 2008, Applied Physics Letters \textbf{92}, 061116.

Milburn,
G.J., 1989,
Phys.Rev.Lett. {\bf 62}, 2124.

Misra, B., E. C. G. Sudarshan, 1977, J.Math.Phys {\bf 18}, 756.

M{\o}lmer, K., 1997, Phys. Rev. A {\bf 55}, 3195.

Moreau, E., I. Robert, J. M. Gerard, I. Abram, L. Manin, and V. Thierry-Mieg, 2001, Applied Physics Letters \textbf{79}, 2865.

Mosley, P. J., J. S. Lundeen, B. J. Smith, P. Wasylczyk, A. B. U'Ren, C. Silberhorn, and I. A. Walmsley, 2008, Physical Review Letters \textbf{100}, 133601.

Neergaard-Nielsen, J. S., B. M. Nielsen, C. Hettich, K. Molmer, and E. S. Polzik, 2006, Physical Review Letters \textbf{97}, 083604.

Nemoto, K. and W. J. Munro, 2004, Physical Review Letters \textbf{93}, 250502.

Nielsen, M., and I.~Chuang, 2000,
{\it Quantum computation and quantum information}
(Cambridge University Press, Cambridge, UK).

Nielsen,
M., 2004, Phys. Rev. Lett. {\bf 93}, 040503.

Noda, S., A. Chutinan, M. Imada, 2000, Nature, {\bf 407}, 608.

Noda, S., M. Fujita, and T. Asano, 2007, Nature Photonics {\bf 1}, 449.

O'Brien, J. L., G.J.Pryde, A.G.White, T.C.Ralph, D.Branning, 2003, Nature {\bf 426}, 264.

O'Brien, J. L., G. J. Pryde, A. G. White, and T. C. Ralph, 2005, Physical Review A \textbf{71}, 060303.

O'Brien, J. L., G. J. Pryde, A. Gilchrist, D. F. V. James, N. K. Langford, T. C. Ralph, and A. G. White, 2004, Physical Review Letters \textbf{93}, 080502.

Okamoto, R., H. F. Hofmann, S. Takeuchi, and K. Sasaki, 2005, Physical Review Letters \textbf{95}, 210506.

Oort, J. H. and T. Walraven, 1956, Bulletin of the Astronomical Institutes of the Netherlands \textbf{12}, 285.

Ourjoumtsev, A., R. Tualle-Brouri, J. Laurat, and P. Grangier, 2006, Science \textbf{312}, 83.

Ourjoumtsev, A., A. Dantan, R. Tualle-Brouri, and P. Grangier, 2007, Physical Review Letters \textbf{98}, 030502.

Ourjoumtsev A., H.Jeong, R.Tualle-Brouri and P.Grangier, 2007, Nature {\bf 448}, 784.

Ourjoumtsev, A., F.Ferreyrol, R. Tualle-Brouri, and P. Grangier, 2009, Nature Physics, {\bf 5}, 189.

Pachos, J. K., W. Wieczorek, C. Schmid, N. Kiesel, R. Pohlner, and H. Weinfurter, 2007, arXiv:0710.0895.

Peters, N., J. Altepeter, E. Jeffrey, D. Branning, and P. G. Kwiat, 2003, Quantum Information \& Computation \textbf{3}, 503.

Pittman, T. B., M. J. Fitch, B. C. Jacobs, and J. D. Franson, 2003, Physical Review A \textbf{68}, 032316.

Pittman, T. B., and J. D. Franson, 2002, Physical Review A \textbf{66}, 062302.

Pittman, T. B., B. C. Jacobs, and J. D. Franson, 2001, Physical Review A \textbf{64}, 062311.

Pittman, T. B., B. C. Jacobs, and J. D. Franson, 2002a, Physical Review A \textbf{66}, 052305.

Pittman, T. B., B. C. Jacobs, and J. D. Franson, 2002b, Physical Review A \textbf{66}, 042303.

Pittman, T. B., B. C. Jacobs, and J. D. Franson, 2005, Physical Review A \textbf{71}, 052332.

Pittman, T. B., B. C. Jacobs, and J. D. Franson, 2005, Optics Communications \textbf{246}, 545.

Politi, A., M. J. Cryan, J. G. Rarity, S. Yu, and J. L. O'Brien, 2008, Science \textbf{320}, 646.

Poppe, A., M.Peev, O.Maurhart, 2008, Int J Quant Inf, {\bf 6}, 209.

Poyatos, J. F., J. I. Cirac, and P. Zoller, 1997, Physical Review Letters \textbf{78}, 390.

Prevedel, R., M. S. Tame, A. Stefanov, M. Paternostro, M. S. Kim, and A. Zeilinger, 2007, Physical Review Letters \textbf{99}, 250503.

Prevedel, R., A. Stefanov, P. Walther, and A. Zeilinger, 2007, New Journal of Physics \textbf{9}, 205.

Prevedel, R., P. Walther, F. Tiefenbacher, P. Bohl, R. Kaltenbaek, T. Jennewein, and A. Zeilinger, 2007, Nature \textbf{445}, 65.

Pryde, G. J., J. L. O'Brien, A. G. White, T. C. Ralph, and H. M. Wiseman, 2005, Physical Review Letters \textbf{94}, 220405.

Pryde, G. J., J. L. O'Brien, A. G. White, S. D. Bartlett, and T. C. Ralph, 2004, Physical Review Letters \textbf{92}, 190402.

Pryde, G. J., J. L. O'Brien, A. G. White, and S. D. Bartlett, 2005, Physical Review Letters \textbf{94}, 220406.

Pysher, M., R. Bloomer, O. Pfister, C. M. Kaleva, T. D. Roberts, and P. Battle, 2008, arXiv:0809.1611v1.

Ralph, T.C., 2006, Rep. Prog. Phys. {\bf 69}, 853.

Ralph, T. C., S. D. Bartlett, J. L. O'Brien, G. J. Pryde, and H. M. Wiseman, 2006, Physical Review A \textbf{73}, 012113.

Ralph, T.C., A.Gilchrist, G.J.Milburn, W.J.Munro and S.Glancy, 2003, Phys. Rev. A {\bf 68}, 042319.

Ralph, T.C., A.J.F.Hayes and A.Gilchrist, 2005, Phys. Rev. Lett. {\bf 95}, 100501.

Ralph, T.C., N.K.Langford, T.B.Bell, A.G.White, 2002,
Phys.Rev. A {\bf 65}, 062324.

Ralph, T.C., W.J.Munro and G.J.Milburn, 2002,
Proceedings of SPIE {\bf 4917}, 1.

Ralph, T.C., A.G.White, W.J.Munro, G.J.Milburn, 2002,
Phys.Rev. A {\bf 65}, 012314.
%Ralph (2007) T.C.Ralph, K.Resch and A.Gilchrist, Phys.Rev.A {\bf 75} 022313 (2007).

Raussendorf, R. and H.J.Briegel, 2001, Phys.Rev.Lett. {\bf 86}, 5188.

Raussendorf, R. and J.Harrington, 2007, Phys. Rev. Lett. {\bf 98}, 190504.

Reck, M., A.Zeilinger, H.J.Bernstein and P.Bertani, 1994, Phys. Rev. Lett. {\bf 73}, 58. 

Rohde, P. P., G. J. Pryde, J. L. O'Brien, and T. C. Ralph, 2005, Physical Review A \textbf{72}, 032306.

Rohde, P.P., T.C.Ralph, W.J.Munro, 2007, Phys.Rev.A {\bf 75}, 010302.

Rosfjord, K. M., J. K. W. Yang, E. A. Dauler, A. J. Kerman, V. Anant, B. M. Voronov, G. N. Gol'tsman, and K. K. Berggren, 2006, Optics Express \textbf{14}, 527.

Roussev, R. V., C. Langrock, J. R. Kurz, and M. M. Fejer, 2004, Optics Letters \textbf{29}, 1518.

Sakurai, J.J., 1985, {\it Modern Quantum Optics} (Addison-Wesley, California).

Sanaka, K., K. Kawahara, and T. Kuga, 2002, Physical Review A \textbf{66}, 040301.

Sanaka, K., A. Pawlis, T. Ladd, K. Lischka, and Y. Yamamoto, 2009, arXiv:0903.1849v1.

Santori, C., D. Fattal, J. Vu\u{c}kovi\'{c}, G. S. Solomon, and Y. Yamamoto, 2002, Nature \textbf{419}, 594.

Santori, C., M. Pelton, G. Solomon, Y. Dale, and E. Yamamoto, 2001, Physical Review Letters \textbf{86}, 1502.

Schmidt, H. and A. Imamoglu, 1996, Optics Letters \textbf{21}, 1936.

Schoelkopf, R. J., S. M. Girvin, 2008, Nature, {\bf 451}, 664.

Schumacher, B., 1995, Phys.Rev.A {\bf 51}, 2738.

Shapiro, J.H., 2006, Phys. Rev. A {\bf 73}, 062305.

Shor, P.W., 1994, {\it Proceedings of the 35th Annual
Symposium on the Foundations of Computer Science} (IEEE
Computer Society Press, Los Alamitos, California) 124. 

Shor, P., 1995, Phys Rev A {\bf 52}, 2493.

Silva, M., M.Roetteler, C.Zalka, 2005, Phys. Rev. A {\bf 72}, 032307.

Steane, A.M., 1996, Phys Rev Lett {\bf 77}, 793.

Stucki, D., N.Gisin, O.Guinnard, G.Ribordy, H.Zbinden, 2002,
New J.Phys. {\bf 4}, 41.

Takahashi, H., K. Wakui, S. Suzuki, M. Takeoka, K. Hayasaka, A. Furusawa, and M. Sasaki, 2008, Physical Review Letters \textbf{101}, 233605.

Takeuchi, S., J. Kim, Y. Yamamoto, and H. H. Hogue, 1999, Applied Physics Letters \textbf{74}, 1063.

Tame, M. S., R. Prevedel, M. Paternostro, P. Boehi, M. S. Kim, and A. Zeilinger, 2007, Physical Review Letters \textbf{98}, 140501.

Thew, R. T., H. Zbinden, and N. Gisin, 2008, Applied Physics Letters \textbf{93}, 071104.

Tittel, W., M. Afzelius, R. L. Cone, T. Chaneli\'ere, S. Kr\"oll, S. A. Moiseev, and M. Sellars, 2008, arXiv:0810.0172.

Tokunaga, Y., S. Kuwashiro, T. Yamamoto, M. Koashi, and N. Imoto, 2008, Physical Review Letters \textbf{100}, 210501.

Turchette, Q. A., C. J. Hood, W. Lange, H. Mabuchi, and H. J. Kimble, 1995, Physical Review Letters \textbf{75}, 4710.

Uskov, D. B., L.Kaplan, A. M. Smith, S. D. Huver, and J. P. Dowling, 2009, Phys Rev A {\bf 79}, 042326.

Vallone, G., E. Pomarico, F. De Martini, and P. Mataloni, 2008, Physical Review Letters \textbf{100}, 160502.

Vallone, G., E. Pomarico, P. Mataloni, F. De Martini, and V. Berardi, 2007, Physical Review Letters \textbf{98}, 180502.

Varnava, M., D.E.Browne and T.Rudolph, 2008, Phys. Rev. Lett. {\bf 100}, 060502.

VanDevender, A. P. and P. G. Kwiat, 2007, Journal of the Optical Society of America B-Optical Physics \textbf{24}, 295.

Vu\u{c}kovi\'{c}, J., D. Fattal, C. Santori, and G. S. Solomon, 2003, Applied Physics Letters \textbf{82}, 3596.

Wakui, K., H. Takahashi, A. Furusawa, and M. Sasaki, 2007, Optics Express \textbf{15}, 3568.%-3574

Walls D.F., and G.J.Milburn, 1994, {\it Quantum
Optics} (Springer-Verlag, Berlin).

%Wallraff (2004) A. Wallraff, D. I. Schuster, A. Blais, L. Frunzio, R.- S. Huang, J. Majer, S. Kumar, S. M. Girvin, R. J. Schoelkopf, Nature {\bf 431}, 162 (2004).
Walther, P., K. J. Resch, T. Rudolph, E. Schenck, H. Weinfurter, V. Vedral, M. Aspelmeyer, and A. Zeilinger, 2005, Nature \textbf{434}, 169.

Walther, P. and A. Zeilinger, 2005, Physical Review A \textbf{72}, 010302.

Walther, P., M. Aspelmeyer, K. J. Resch, and A. Zeilinger, 2005, Physical Review Letters \textbf{95}, 020403.

White, A. G., A. Gilchrist, G. J. Pryde, J. L. O'Brien, M. J. Bremner, and N. K. Langford, 2007, Journal of the Optical Society of America B-Optical Physics \textbf{24}, 172.

Wootters,
W.K., and W.H.Zurek, 1982, Nature {\bf 299}, 802.

Yoshikawa, J., Y. Miwa, A. Huck, U. L. Andersen, P. van Loock, and A. Furusawa, 2008, Physical Review Letters \textbf{101}, 250501.

Young, R. J., R. M. Stevenson, P. Atkinson, K. Cooper, D. A. Ritchie, and A. J. Shields, 2006, New Journal of Physics \textbf{8}, 29.

Yoran,
N., and B.Reznik, 2003, Phys. Rev. Lett. {\bf 91},
037903.

Yuan, Z. L., B. E. Kardynal, R. M. Stevenson, A. J. Shields, C. J. Lobo, K. Cooper, N. S. Beattie, D. A. Ritchie, and M. Pepper, 2002, Science \textbf{295}, 102.

Yukawa, M., R. Ukai, P. van Loock, and A. Furusawa, 2008, Physical Review A \textbf{78}, 012301.

Zhao, Z., A. N. Zhang, Y. A. Chen, H. Zhang, J. F. Du, T. Yang, and J. W. Pan, 2005, Physical Review Letters \textbf{94}, 030501.

\end{document}